\documentclass[a4paper,12pt, leqno]{article}
\usepackage{amsmath}
\usepackage{amssymb, latexsym}
\usepackage{amsfonts}
\usepackage{makeidx}
\usepackage{amssymb}

\date{}

\newtheorem{lem}{Lemma}[section]
\newtheorem{thm}{Theorem}[section]
\newtheorem{prop}{Proposition}[section]
\newtheorem{cor}{Corollary}[section]
\newtheorem{definition}{Definition}[section]

\numberwithin{equation}{section}
\newcommand{\dbar}{d\!\!\!{\lower-0.6ex\hbox{$-$}}\!}
\newcommand{\dslash}{d\!\!\!{\lower-0.6ex\hbox{$-$}}}
\newcommand{\e}{\varepsilon}
\newcommand{\h}{\hbar}
\newcommand{\ott}{\lower-0.4ex\hbox{${\scriptscriptstyle{\otimes}}$}}
\newcommand{\btt}{\lower-0.2ex\hbox{${\scriptscriptstyle{\bullet}}$}}
\newcommand{\ctt}{\lower-0.2ex\hbox{${\scriptscriptstyle{\circ}}$}}
\newcommand{\dtt}{\lower-0.2ex\hbox{${\scriptscriptstyle{\diamond}}$}}
\newcommand{\odt}{\lower-0.4ex\hbox{${\scriptscriptstyle{\odot}}$}}

\begin{document}

\title{Deformation Expression for Elements of Algebras (III)\\
--Generic product formula for ${*}$-exponentials\\
 \qquad\quad  of quadratic forms--}


\author{
     Hideki Omori\thanks{ Department of Mathematics,
             Faculty of Sciences and Technology,
        Tokyo University of Science, 2641, Noda, Chiba, 278-8510, Japan,
         email: omori@ma.noda.tus.ac.jp}
        \\Tokyo University of Science
\and  Yoshiaki Maeda\thanks{Department of Mathematics,
                Faculty of Science and Technology,
                Keio University, 3-14-1, Hiyoshi, Yokohama,223-8522, Japan,
                email: maeda@math.keio.ac.jp}
          \\Keio University
\and  Naoya Miyazaki\thanks{ Department of Mathematics, Faculty of
Economics, Keio University,  4-1-1, Hiyoshi, Yokohama, 223-8521, Japan,
        email: miyazaki@hc.keio.ac.jp}
        \\Keio University
\and  Akira Yoshioka \thanks{ Department of Mathematics,
          Faculty of Science, Tokyo University of Science,
         1-3, Kagurazaka, Tokyo, 102-8601, Japan,
         email: yoshioka@rs.kagu.tus.ac.jp}
           \\Tokyo University of Science
  }

\maketitle

\thispagestyle{empty}

\tableofcontents

\pagestyle{plain}



\setcounter{equation}{0}

\bigskip 
In a noncommutative algebra there is no canonical way to express 
elements in univalent way, which is often called 
``ordering problem''. In this note we give product formula of the 
Weyl algebra in generic ordered expression. 
In particular, the generic product formula of
$*$-exponential functions of quadratic forms will be given. 

In differential geometry, it is widely accepted 
that geometrical notion should have coordinate free expression. 
Obviously, algebraic structure of 
$({\mathbb C}[{\pmb u}], {*}_{\Lambda})$ 
depends only on the skew part of ${\Lambda}$.  
It seems reasonable to accept the independence  of ordering principle as 
a basic principle that the physical implication should be 
independent of ordered expressions.

In the last section, we mention the independence of ordering principle
(IOP), and  how this principle breaks down in the system containing 
$*$-exponential functions of quadratic forms.
 
As a result, we obtain a kind of ``double covering group'' of 
$Sp(m;{\mathbb C})$ which is simply connected, but 
this contains the double covering group (meta-plectic group) of 
$Sp(m;{\mathbb R})$. Several extraordinary properties of 
$*$-exponential functions of quadratic forms will be given. 

In these calculus, we found peculiar elements, called polar elements, 
each of which has infinitely many square roots. 

\section{General product formula and intertwiners}\label{PBW-theorem} 

We start with a little general setting as follows:
Let ${\mathfrak S}(n)$ and ${\mathfrak A}(n)$ be 
the spaces of complex symmetric matrices and skew-symmetric 
matrices respectively, and 
${\mathfrak M}(n){=}
{\mathfrak S}(n)\oplus{\mathfrak A}(n)$.
For an arbitrary fixed $n{\times}n$-complex matrix 
$\Lambda{\in}{\mathfrak M}(n)$, we 
define a product ${*}_{_{\Lambda}}$ on the space of polynomials   
${\mathbb C}[\pmb u]$ by the formula 
\begin{equation}
 \label{eq:KK}
 f*_{_{\Lambda}}g=fe^{\frac{i\h}{2}
(\sum\overleftarrow{\partial_{u_i}}
{\Lambda}{}^{ij}\overrightarrow{\partial_{u_j}})}g
=\sum_{k}\frac{(i\h)^k}{k!2^k}
{\Lambda}^{i_1j_1}\!{\cdots}{\Lambda}^{i_kj_k}
\partial_{u_{i_1}}\!{\cdots}\partial_{u_{i_k}}f\,\,
\partial_{u_{j_1}}\!{\cdots}\partial_{u_{j_k}}g.   
\end{equation}
It is known and not hard to prove that 
$({\mathbb C}[\pmb u],*_{_{\Lambda}})$ is an associative algebra. 
Clearly, if $\Lambda$ is symmetric, then the algebra obtained
is commutative and is isomorphic  
to the standard polynomial algebra with $\h$. 

\medskip
For every $\Lambda$, $\partial_{u_{i}}$ acts as a 
derivation of the algebra 
$({\mathbb C}[\pmb u],*_{_{\Lambda}})$. 
Noting this, we define 
for any other constant symmetric matrix $K$ 
a new product $*_{_{\Lambda,K}}$ by the formula  
$$  
\begin{aligned}
f*_{_{\Lambda{,}K}}g=&f
e^{\frac{i\h}{2}
(\sum\overleftarrow{\partial_{u_i}}
{K}^{ij}{*_{_{\Lambda}}}\overrightarrow{\partial_{u_j}})}g\\
=&\sum_{k}\frac{(i\h)^k}{k!2^k}
{K}^{i_1j_1}\cdots{K}^{i_kj_k}
(\partial_{u_{i_1}}\cdots\partial_{u_{i_k}}f){*_{_{\Lambda}}}
(\partial_{u_{j_1}}\cdots\partial_{u_{j_k}}g). 
\end{aligned}
$$
This is  also an associative algebra 
$({\mathbb C}[\pmb u],*_{_{\Lambda,K}})$. 
Since $\Lambda$, $K$ are constant matrices 
and the non-commutativity of matrix algebra 
is not used in the calculation of the product formula, 
the new product formula can be rewritten as 
$$
f*_{_{\Lambda,K}}g=
\sum_{k}\frac{(i\h)^k}{k!2^k}
{(\Lambda{+}K)}^{i_1j_1}
\cdots{(\Lambda{+}K)}^{i_kj_k}
\partial_{u_{i_1}}\cdots\partial_{u_{i_k}}f
\partial_{u_{j_1}}\cdots\partial_{u_{j_k}}g
$$ 
by noting that the exchanging indexes of 
$\partial_{u_{i_1}\cdots u_{i_k}}$ is permitted. 
That is, ${*}_{_{\Lambda,K}}={*}_{_{\Lambda+K}}$.

This formula may be written as
\begin{equation}
 \label{eq:prprod}
fe^{\frac{i\h}{2}(\sum\overleftarrow{\partial_{u_i}}
{(\Lambda{+}K)}^{ij}\overrightarrow{\partial_{u_j}})}g
=
fe^{\frac{i\h}{2}(\sum\overleftarrow{\partial_{u_i}}
{K}^{ij}
{e^{\frac{i\h}{2}(\sum\overleftarrow{\partial_{u_k}}
{\Lambda}^{kl}\overrightarrow{\partial_{u_k}})}}
  \overrightarrow{\partial_{u_j}})}g.
\end{equation}

Using a symmetric matrix $K$, we compute 
$\frac{1}{k!}
(\frac{i\h}{4}\sum{K}^{ij}
\partial_{u_i}\partial_{u_j})^k(f{*}_{_K}g)$ 
by noting that this is written as follows:
$$
\begin{aligned}
\sum_{p{+}q{+}r=k}
\frac{(i\h)^r}{r!2^r}
{K}^{i_1j_1}\cdots&{K}^{i_rj_r}
\partial_{u_{i_1}}\cdots\partial_{u_{i_r}}
\frac{1}{p!}(\frac{i\h}{4}
\sum{K}^{ij}\partial_{u_i}\partial_{u_j})^pf\\
&\quad\times\partial_{u_{j_1}}\cdots\partial_{u_{j_r}}
\frac{1}{q!}
(\frac{i\h}{4}\sum{K}^{ij}\partial_{u_i}\partial_{u_j})^qg.
\end{aligned}
$$
Using this formula, we have the following formula:
\begin{equation}
  \label{eq:Hochsch}
  \begin{aligned}
e^{\frac{i\h}{4}\sum{K}^{ij}\partial_{u_i}\partial_{u_j}}
\Big(&\big(e^{-\frac{i\h}{4}\sum{K}^{ij}
\partial_{u_i}\partial_{u_j}}f\big)
{*_{_\Lambda}}
\big(e^{-\frac{i\h}{4}\sum{K}^{ij}\partial_{u_i}\partial_{u_j}}g\big)\Big)\\
=&fe^{\frac{i\h}{2}(\sum\overleftarrow{\partial_{u_i}}
{*_{\Lambda}}{K}^{ij}{*_{_\Lambda}}
\overrightarrow{\partial_{u_j}})}g= f{*}_{_{\Lambda{+}K}}g.
 \end{aligned}
\end{equation}

Set $\Lambda=K{+}J$ where $K$, $J$ are the symmetric part and 
the skew-part of $\Lambda$ respectively. 
Since the commutator 
$[u_i,u_j]={i\h}J^{ij}$ is given by the skew-part of $\Lambda$, 
the algebraic structure of 
$({\mathbb C}[\pmb u], *_{_\Lambda})$ depends only on $J$, whose 
isomorphism class may be denoted by $({\mathbb C}[\pmb u], *_{_J})$
or simply by $({\mathbb C}[\pmb u], *)$ by noticing this class consists 
of a {\it single} algebra.   

This is confirmed directly by the formula \eqref{eq:Hochsch}. 
Namely, we see the following:
\begin{cor}
Let 
$I_0^{^K}(f)=
e^{\frac{i\h}{4}\sum{K}^{ij}\partial_{u_i}\partial_{u_j}}$, 
and $I_{_K}^0(f)=
e^{-\frac{i\h}{4}\sum{K}^{ij}\partial_{u_i}\partial_{u_j}}$. 
Then $I_0^{^K}$ is an isomorphism of 
$({\mathbb C}[\pmb u];{*}_{_{\Lambda}})$ onto 
$({\mathbb C}[\pmb u];{*}_{_{\Lambda+K}})$.
\end{cor}

It is clear that the product $f{*_{_{\Lambda}}}g$ 
is defined if one of $f, g$ is 
a polynomial and another is a smooth function.

Let $H{\!o}l({\mathbb C}^n)$ be the space of all 
holomorphic functions on the complex $n$-plane ${\mathbb C}^n$ with 
the uniform convergence topology on each compact domain. 
The next one gives a useful remark:
\begin{lem}\label{frechet} 
$H{\!o}l({\mathbb C}^n)$ with the topology above 
is a Fr{\'e}chet space defined by a countable family of seminorms. 
\end{lem}

\begin{prop}\label{extholom} 
For every $p(\pmb u)\in{\mathbb C}[\pmb u]$, 
the left-multiplication 
$f\to p(\pmb u)*_{_{\Lambda}}f$ and the right-multiplication 
$f\to f*_{_{\Lambda}}p(\pmb u)$ are both continuous 
linear mapping of $H\!ol({\mathbb C}^n)$ into itself. 

If two of $f, g, h$ are polynomials, then associativity 
$(f{*_{_{\Lambda}}}g{*_{_{\Lambda}}})h=
f{*_{_{\Lambda}}}(g{*_{_{\Lambda}}}h)$
holds.
\end{prop}

\subsection{Expression parameters and intertwiners}
\label{Expinter}
In what follows we treat the case of $2m$ variables, and 
we use notations 
\begin{equation}\label{Weyl}
{\pmb u}=(u_1,u_2,\cdots,u_{2m})=
(\tilde{\pmb u},\tilde{\pmb v}),\quad  
\tilde{\pmb u}=(\tilde{u}_1,\cdots,\tilde{u}_m),\,\,
\tilde{\pmb v}=(\tilde{v}_1,\cdots,\tilde{v}_m).
\end{equation}
The skew part $J$ is fixed to be the standard skew-symmetric 
matrix 
$J=\left[
{\footnotesize 
{\begin{matrix} 
   0 & {-}I\\ 
   I & 0 
 \end{matrix}}} 
\right]$. The algebra is called 
the {\bf Weyl algebra} and the isomorphism class is 
denoted by $W_{\h}(2m)$. 

We use sometimes notations $(u_1,\cdots,u_m, v_1,\cdots,v_m)$ 
instead of
$(\tilde{u}_1,\cdots,\tilde{u}_m,\tilde{v}_1,\cdots,\tilde{v}_m)$ 
when no confusion is suspected.

For the case of a universal enveloping algebra of a Lie algebra, 
Poincar{\'e}-Birkhoff-Witt theorem ensures
that this is realized on the space of ordinary 
polynomials by giving a new associative product. 
However, there is no standard way of 
unique expressing elements of algebra. 

\medskip
Note that if the generator system is fixed, then  
Proposition\,\ref{extholom} gives a representation 
of the algebra. The product formula \eqref{eq:KK} 
gives also the unique expression of elements of 
this algebra by the usual polynomials. 
For instance, computing $u^i{*}u^j{*}u^k$ by using 
\eqref{eq:KK} gives the
expression of $u^i{*}u^j{*}u^k$ as a polynomial. Thus, 
the product formula
\eqref{eq:KK} will be referred to $K$-{\it ordered expression} 
(or ${K}$-{\it ordering}), 
i.e. if generators are fixed, giving an ordering expression 
is nothing but giving a product formula on the space of 
polynomials which defines the Weyl algebra $W_{\h}$.

\medskip
By this formulation of orderings, the intertwiner between
$K$-ordered expression and $K'$-ordered expression 
is explicitly given as follows:
\begin{prop}
\label{intwn}
For every $K, K'\in{\mathfrak S}(n)$, the intertwiner
is defined by  
\begin{equation}
\label{intertwiner}
I_{_K}^{^{K'}}(f)=
\exp\Big(\frac{i\h}{4}\sum_{i,j}(K^{'ij}{-}K^{ij})
\partial_{u_i}\partial_{u_j}\Big)f \,\,
(=I_{0}^{^{K'}}(I_{0}^{^{K}})^{-1}(f)), 
\end{equation}
and by \eqref{eq:Hochsch} it gives an isomorphism 
$I_{_K}^{^{K'}}:({\mathbb C}[{\pmb u}]; *_{_{K+J}})\rightarrow 
({\mathbb C}[{\pmb u}]; *_{_{K'+J}})$.
Namely, the following identity holds for any 
$f,g \in {\mathbb C}[{\pmb u}]:$ 
\begin{equation}\label{intertwiner2}
I_{_K}^{^{K'}}(f*_{_\Lambda}g)=
I_{_K}^{^{K'}}(f)*_{_{\Lambda'}}I_{_K}^{^{K'}}(g),
\end{equation}
where $\Lambda=K{+}J$, $\Lambda'=K'{+}J$.
\end{prop}
Intertwiners do not change the algebraic structure $*$, 
but do change the expression of elements by the ordinary 
commutative structure.  

\begin{center}
\fbox
{If the skew part $J$ is fixed, 
we often use notation $*_{_K}$ instead of $*_{_\Lambda}$}
\end{center}
%
%
In what follows, we use the notation $*_{_K}$ instead of 
$*_{_\Lambda}$, since the skew-part $J$ is fixed as the 
standard skew-matrix.
We use notations 
$$
{\pmb u}=(u_1,u_2,\cdots,u_{2m})=
(\tilde{\pmb u},\tilde{\pmb v}),\quad  
\tilde{\pmb u}=(\tilde{u}_1,\cdots,\tilde{u}_m),\,\,
\tilde{\pmb v}=(\tilde{v}_1,\cdots,\tilde{v}_m).
$$

As in the case of one variable, infinitesimal intertwiner 
$$
dI_{_K}(K')=\frac{d}{dt}\Big|_{t=0}I_{_K}^{^{K{+}tK'}}=
\frac{i\h}{4} {K'}_{ij}\partial_{u_i}\partial_{u_j}
$$
is viewed as a flat connection on the trivial bundle 
$\coprod_{K\in{\mathfrak S}(n)}
H{\!o}l({\Bbb C}^{n})$. The equation of parallel translation 
along a curve $K(t)$ is given by 
\begin{equation}\label{parallel}
\frac{d}{dt}f_t=
dI_{_{\dot K}(t)}(\dot K(t))f_t, \quad 
\dot K(t)=\frac{d}{dt}K(t),
\end{equation}
but this may not have a solution for some initial function.

Note that according to the choice of $K=0, K_0, {-}K_0$, $I$,  
where 
$$
(0,\,\, K_0, {-}K_0, I)= 
\left(
 \left[
{\footnotesize
{\begin{matrix}
   0 & 0\\
   0 & 0
\end{matrix}}}
 \right],\,\,
\left[
{\footnotesize
{\begin{matrix}
   0 & I\\
   I & 0
 \end{matrix}}}
\right],\,\,
\left[
{\footnotesize
{\begin{matrix}
   0 &\!\!{-}I\\
   {-}I&\!\!0
 \end{matrix}}}\right],\,\,
\left[
{\footnotesize
{\begin{matrix}
   I&\!\!0\\
   0&\!\!I
 \end{matrix}}}\right]
\right),
$$
\begin{tabular}{l|l} 
Choice of $K$ & (name of ordering)\\ \hline
$K=0$         & Weyl ordered expression\\ \hline 
$K_0=\left[
{\footnotesize{
\begin{matrix}
   0 & I\\
   I & 0
 \end{matrix}}}
\right]$  & Normal ordered expression \\ \hline
$-K_0$    & Anti-normal ordered expression \\ \hline
$\left[{\footnotesize{
\begin{matrix}
   I& 0\\
   0 & I
 \end{matrix}}}
\right]$  & Unit ordered expression \\ 
\hline
General  $K$  & $K$-ordered expression
\end{tabular}
\hfill\parbox[c]{.4\linewidth}
{the product formulas \eqref{eq:KK} give the Weyl ordered 
expression and the normal ordered expression, 
the antinormal ordered expression respectively, but 
the unit ordered expression is not so familiar in physics.

For each ordered expression, the  product formulas are given  
respectively by the following formula:}
\begin{equation}\label{ppformula}
 \begin{aligned}[c]
f({\pmb u}){*{_{_0}}}g({\pmb u})=&
f\exp 
\frac{\h i}{2}\{\overleftarrow{\partial_{v}} 
     {\wedge}\overrightarrow{\partial_{u}}\}g,
     \quad{\text{(Moyal product formula)}}\\
f({\pmb u}){*{_{_{K_0}}}}g({\pmb u})=&
f\exp {\h i}\{\overleftarrow{\partial_{v}}\,\, 
       \overrightarrow{\partial_{u}}\}g,
       \qquad{\text{($\Psi$DO product formula)}}  \\
f({\pmb u}){*{_{_{{-}K_0}}}}g({\pmb u})=&
f\exp{-\h i}\{\overleftarrow{\partial_{u}}\,\, 
       \overrightarrow{\partial_{v}}\}g,
       \quad{\text{($\overline{\Psi}$DO product formula)}}   
 \end{aligned}
\end{equation} 
where 
$\overleftarrow{\partial_{v}}{\wedge}
\overrightarrow{\partial_{u}}
=\sum_i(\overleftarrow{\partial_{\tilde{v}_i}}
\overrightarrow{\partial_{\tilde{u}_i}}
-\overleftarrow{\partial_{\tilde{u}_i}}
\overrightarrow{\partial_{\tilde{v}_i}})$ and 
$\overleftarrow{\partial_{v}}\,\, 
       \overrightarrow{\partial_{u}}
=\sum_i\overleftarrow{\partial_{\tilde{v}_i}}\,\, 
       \overrightarrow{\partial_{\tilde{u}_i}}$.

The product formula for the unit ordered expression 
is a bit complicated to write down, 
but it is easy to obtain. For instance 
$$
u_{*_{_I}}^2{=}u^2{+}\frac{i\h}{2}, \quad 
u{*_{_I}}e^{-\frac{1}{i\h}u^2}{=}0{=}
e^{-\frac{1}{i\h}u^2}{*_{_I}}u 
\quad e.t.c.
$$
while the Weyl ordered expression gives 
$$ 
v{*_{_0}}e^{-\frac{2}{i\h}uv}{=}
0{=}e^{-\frac{2}{i\h}uv}{*_{_0}}u. 
$$

The next one is trivial, but an important remark:
\begin{prop}\label{trivrmk} 
Every entire function $f(u,v)=\sum a_{kl}u^kv^l$ can be viewed as a
$K$-ordered expression of an element of extended Weyl algebra.  
\end{prop}

The relations between two different expressions are given 
by intertwiners, but computations in the algebra 
can be done by using only the associativity and the fundamental 
commutation relations. Note for instance that $u{*}v{-}v{*}u=-i\h$ 
give for every polynomial $p(v{*}v)$ of $v{*}u$ that  
$$
u{*}p(v{*}u)=p(u{*}v){*}u,\quad (\text{bumping identity}).
$$ 
Let $u{\ctt}v=\frac{1}{2}(u{*}v{+}v{*}u)$; the 
symmetric product. The bumping identity gives  
$$
u{*}(u{*}v){*}v=u{*}(u{\ctt}v{+}\frac{1}{2}i\h){*}v
=(u{\ctt}v{+}\frac{1}{2}i\h){*}(u{\ctt}v{+}\frac{3}{2}i\h).
$$
Throughout this series, we use notation ${:}{\bullet}{:}_{_K}$ to indicate
the expression parameter for elements of $W_{\h}$. For instance, we write 
$$
{:}u_i{*}u_j{:}_{_K}{=}u^2{+}\frac{i\h}{2}(K{+}J)_{ij},\quad 
{:}u_j{*}u_j{:}_{_I}=u_j^2{+}\frac{i\h}{2}\quad etc.
$$
A remarkable feature of the first three formulas of
\eqref{ppformula} is seen as follows:
$$
\begin{aligned}
&{:}{\tilde u}_j{*}{\tilde v}_j{:}_0={\tilde u}_j{\tilde v}_j{-}\frac{1}{2}i\h, \quad 
 {:}{\tilde v}_j{*}{\tilde u}_j{:}_0={\tilde u}_j{\tilde v}_j{+}\frac{1}{2}i\h,\quad(\text{Weyl ordering})\\ 
&{:}\sum a_{kl}\tilde u_*^k{*}\tilde v_*^l{:}_{_{K_0}}=\sum a_{kl}\tilde u^k\tilde v^l, \qquad
(\text{normal ordering}), \\
&{:}\sum a_{kl}\tilde v_*^k{*}\tilde u_*^l{:}_{_{{-}K_0}}=\sum a_{kl}\tilde u^k\tilde v^l,\quad 
(\text{anti-normal ordering}),\\ 
\end{aligned}
$$
but concrete product formulas will be used to extend the algebra transcendentally.

\bigskip
\noindent
{\bf Weyl ordered expression}. \,\, 
In general, define  $w_*(u_kv_l)$ by $\frac{1}{(k{+}l)!}\sum x_1{*}x_2{*}\cdots{*}x_{k{+}l}$,
where $x_i$ is ${\tilde u}_k$ or ${\tilde v}_l$ and the summation runs through
all possible rearrangement of $u^k{*}v^l$. 
\begin{equation}\label{nikou}
(\tilde u{+}\tilde v)_*^n=\sum_k \,\,{}_nC_kw_*(\tilde u^k\tilde v^{n{-}k}), 
\quad {:}w_*(\tilde u^k\tilde v^l){:}_{0}=\tilde u^k\tilde v^{l}.
\end{equation}

\noindent
{\bf Special expression parameter $K_s$}.\,\,
In \cite{OMMY3}, we introduced the special ordered expression
$K_s$ to control the distribution of singular points of
$*$-exponential functions of quadratic forms.

By $K_s$-product formula, we see that 
$(H\!ol({\mathbb C}^n),*_{_{K_s}})$ contains a subalgebra which is 
isomorphic to the Clifford algebra $Clif\!f(m)$. 

\bigskip
\noindent
{\bf Siegel class of expression parameters}.\,\,
In \cite{OMMY3}, we introduced the class  ${\mathfrak S}_+({\mathbb  R}^n)$ 
of expression parameters and gave several remarks. This is 
$$
{\mathfrak S}_+({\mathbb R}^n)=
\{K; {\rm{Re}}\,\frac{1}{\h}\langle \xi(iK),\xi\rangle\geq c_{K}|\xi|^2, 
\quad \exists c_k>0,~\forall\xi\in {\mathbb R}^n\}
$$
which  will be called 
the {\it imaginary positive definite class} or the {\it  Siegel class}.
${\mathfrak S}_+({\mathbb R}^n)$ is $G\!L(n,{\mathbb  R})$-invariant.  
Expressions in this class is easy to treat up to $*$-exponential
functions of linear forms and their integrals.  
Further remarks will be given in the last section.

\subsubsection{Linear change of generators}

Next, we consider the effect of linear changes of generators such as 
$$
{u'}_i=\sum u_kS_i^k, \quad S\in G\!L(n,{\mathbb C}),\quad 
 ({\pmb u}'={\pmb u}S).
$$
Since $\partial_{u_i}=\sum S_i^k\partial_{u'_k}$, the product formula is rewritten 
by using new generators as 
\begin{equation} \label{eq:KK2}
 f*_{_{\Lambda}}g=fe^{\frac{i\h}{2}
(\sum\overleftarrow{\partial_{u'_i}}
({}^t\!S{\Lambda}S)^{ij}\overrightarrow{\partial_{u'_j}})}g.
\end{equation}
Hence the notation $*_{_\Lambda}$ is better to be replaced  
$*_{_{\Lambda'}}$ where $\Lambda'{=}{}^t\!S{\Lambda}S$.

Therefore the algebraic structure of 
$({\mathbb C}[\pmb u], *_{_\Lambda})$ 
depends only on the conjugate class of the skew part $J$.
If ${}^t\!SJS=J$, that is, $S$ is a symplectic linear change of
generators such as 
$$
{u'}_i=\sum u_kS_i^k, \quad S\in Sp(m,{\mathbb C}), 
$$
the mapping ${\pmb u}\to {\pmb u}'$ 
does not change the algebraic structure. 
Change of generators are viewed often as coordinate transformations,
but note here that $I_{_K}^{^{{}^t\!SKS}}$  
is something like the ``square root'' of 
symplectic  coordinate transformations.  

Since $\det S=1$ for $S\in Sp(m,{\mathbb C})$, we see 
$\det{}^t\!SKS=\det K$, hence the isomorphic 
change by the intertwiner $I_{_K}^{^{K'}}$ can not be  
covered by a coordinate transformation if 
$\det K\not=\det K'$.

\subsection{Star-exponential functions of linear functions}
For ${\pmb a}, {\pmb b}\in{\mathbb C}^{2m}$, we set 
$\langle{\pmb a}{\varGamma},{\pmb b}\rangle
=\sum_{ij=1}^{2m}{\varGamma}^{ij}a_ib_j$, 
$\langle{\pmb a},{\pmb u}\rangle=\sum_{i=1}^{2m} a_iu_i$. These 
will be denoted also by 
${\pmb a}{\varGamma}\,{}^t{\pmb b}$ and 
$\langle{\pmb a},{\pmb u}\rangle={\pmb a}\,{}^t\!{\pmb u}$.  
For $f(\pmb u)\in H{\!o}l({\mathbb C}^{2m})$, 
the direct calculation via the product formula 
\eqref{eq:KK} by using Taylor expansion gives the following:

\begin{equation}\label{extend}
\begin{aligned}
&e^{s\frac{1}{i\h}\langle{\pmb a},{\pmb u}\rangle}
{*_{_K}}f({\pmb u})=
e^{s\frac{1}{i\h}\langle{\pmb a},{\pmb u}\rangle}
f({\pmb u}{+}\frac{s}{2}{\pmb a}(K{+}J)),\\ 
& 
f({\pmb u}){*_{_K}}
e^{-s\frac{1}{i\h}\langle{\pmb a},{\pmb u}\rangle}=
f({\pmb u}{+}\frac{s}{2}{\pmb a}(-K{+}J))
e^{-s\frac{1}{i\h}\langle{\pmb a},{\pmb u}\rangle}
\end{aligned} 
\end{equation} 
as natural extension of the product formula. This gives also the 
associativity 
\begin{equation}\label{adjointlike}
\Big(e^{s\frac{1}{i\h}\langle{\pmb a},{\pmb u}\rangle}
{*_{_K}}f({\pmb u})\Big){*_{_K}}e^{t\frac{1}{i\h}\langle{\pmb b},{\pmb u}\rangle}
=
e^{s\frac{1}{i\h}\langle{\pmb a},{\pmb u}\rangle}
{*_{_K}}\Big(f({\pmb u}){*_{_K}}e^{t\frac{1}{i\h}\langle{\pmb b},{\pmb u}\rangle}\Big),
\quad f(\pmb u)\in H{\!o}l({\mathbb C}^{2m}).
\end{equation}

By a direct calculation of intertwiner, we see that     
\begin{equation}
  \label{eq:intwin}
I_{_K}^{^{K'}}(e^{\frac{1}{i\h}\langle{\pmb a},{\pmb u}\rangle})
=e^{\frac{1}{4i\h}\langle{\pmb a}(K'{-}K),{\pmb a}\rangle}
e^{\frac{1}{i\h}\langle{\pmb a},{\pmb u}\rangle}.   
\end{equation}
Hence, $\{e^{\frac{1}{4i\h}\langle{\pmb a}K,{\pmb a}\rangle}
e^{\frac{1}{i\h}\langle{\pmb a},{\pmb u}\rangle}; 
K\in{\mathfrak S}_{\mathbb C}(2m)\}$ is a parallel section of 
$\coprod_{K\in{\mathfrak S}_{\mathbb C}(2m)}
{H{\!o}l}({\Bbb C}^{2m}).$

We denote this element symbolically by 
$e_*^{\frac{1}{i\h}\langle{\pmb a},{\pmb u}\rangle}$. 
Namely we denote   
\begin{equation}
  \label{eq:tempexp}
:e_*^{\frac{1}{i\h}\langle{\pmb a},{\pmb u}\rangle}:_{_K}
=e^{\frac{1}{4i\h}\langle{\pmb a}K,{\pmb a}\rangle}
e^{\frac{1}{i\h}\langle{\pmb a},{\pmb u}\rangle}
=e^{\frac{1}{4i\h}\langle{\pmb a}K,{\pmb a}\rangle
{+}\frac{1}{i\h}\langle{\pmb a},{\pmb u}\rangle}.  
\end{equation}
By using the product formula in $K$-ordered expression, 
we have easily the exponential law  
$$
{:}e_*^{s\frac{1}{i\h}\langle{\pmb a},{\pmb u}\rangle}
{:}_{_{K}}{*_{_{K}}}
{:}e_*^{t\frac{1}{i\h}\langle{\pmb a},{\pmb u}\rangle}
{:}_{_{K}}=
{:}e_*^{(s{+}t)\frac{1}{i\h}\langle{\pmb a},{\pmb u}\rangle}
{:}_{_{K}},\,\,
\forall K\in {\mathfrak S}(2m).
$$
The exponential law may be written by omitting the suffix $K$ as  
$$
e_*^{s\frac{1}{i\h}\langle{\pmb a},{\pmb u}\rangle}{*}
e_*^{t\frac{1}{i\h}\langle{\pmb a},{\pmb u}\rangle}=
e_*^{(s{+}t)\frac{1}{i\h}\langle{\pmb a},{\pmb u}\rangle},\quad 
e^{s}e_*^{t\frac{1}{i\h}\langle{\pmb a},{\pmb u}\rangle}=
e_*^{s{+}t\frac{1}{i\h}\langle{\pmb a},{\pmb u}\rangle}
$$
together with the exponential law with the 
ordinary exponential functions.

\bigskip

Furthermore note also that 
${:}\langle{\pmb a},{\pmb u}\rangle{:}_{_{K}}=
\langle{\pmb a},{\pmb u}\rangle$ 
for every $K$, 
and 
$e_*^{\frac{s}{i\h}\langle{\pmb a},{\pmb u}\rangle}$  
is the solution of the evolution equation 
$$
\frac{d}{dt}{:}e_*^{\frac{t}{i\h}\langle{\pmb a},{\pmb u}\rangle}
{:}_{_{K}}
=\frac{1}{i\h}{:}\langle{\pmb a},{\pmb u}\rangle{:}_{_{K}}{*_{_K}}
{:}e_*^{\frac{t}{i\h}\langle{\pmb a},{\pmb u}\rangle}{:}_{_{K}}
\,\,{\text{with initial data}}\,\, {:}e_*^{\frac{0}{i\h}\langle{\pmb a},{\pmb u}\rangle}{:}_{_{K}}=1.
$$ 

$e_*^{s\frac{1}{i\h}\langle{\pmb a},{\pmb u}\rangle}=
\{e^{s^2\frac{1}{4i\h}\langle{\pmb a}K\,{\pmb a}\rangle}
e^{s\frac{1}{i\h}\langle{\pmb a},{\pmb u}\rangle}; 
K\in{\mathfrak S}(2m)\}$ forms a one parameter 
group of parallel sections. 

\bigskip
By applying \eqref{extend} to 
${:}e_*^{\pm s\frac{1}{i\h}\langle{\pmb a},{\pmb u}\rangle}{:}_{_K}$, 
we have for every $f\in H\!ol({\mathbb C}^n)$ the associativity   
\begin{equation}\label{adjointlike}
{:}(e_*^{s\frac{1}{i\h}
\langle{\pmb a},{\pmb u}\rangle}{*}f_*({\pmb u}))
{*}e_*^{-s\frac{1}{i\h}\langle{\pmb a},{\pmb u}\rangle}{:}_{_K}
={:}f_*({\pmb u}{+}{s}{\pmb a}J){:}_{_K}=
{:}e_*^{s\frac{1}{i\h}
\langle{\pmb a},{\pmb u}\rangle}{*}(f_*({\pmb u})
{*}e_*^{-s\frac{1}{i\h}\langle{\pmb a},{\pmb u}\rangle}){:}_{_K}.
\end{equation}
This gives also the real analyticity of 
$e_*^{s\frac{1}{i\h}
\langle{\pmb a},{\pmb u}\rangle}{*}f_*({\pmb u})
{*}e_*^{-s\frac{1}{i\h}\langle{\pmb a},{\pmb u}\rangle}$ in $s$.

It is remarkable that if $K=0$, then 
$:e_*^{\frac{1}{i\h}\langle{\pmb a},{\pmb u}\rangle}:_{_K}=
e^{\frac{1}{i\h}\langle{\pmb a},{\pmb u}\rangle}$, that is,
$*$-exponential functions of linear functions are 
ordinary exponential functions in Weyl ordered expression. 
On the other hand, if $K\in{\mathfrak S}_+({\mathbb R}^n)$ then 
${:}e_*^{\pm s\frac{1}{i\h}\langle{\pmb a},{\pmb u}\rangle}{:}_{_K}$ 
has a very strong property that 
$$
{:}e_*^{\pm s\frac{1}{i\h}\langle{\pmb a},{\pmb u}\rangle}{:}_{_K}
=e^{\frac{s^2}{4i\h}\langle{\pmb a}K,{\pmb a}\rangle}
e^{\pm\frac{s}{i\h}\langle{\pmb a},{\pmb u}\rangle}
$$
is rapidly decreasing in $s\in{\mathbb R}$.

\subsubsection{Extension of products}\label{Extprodpp}

For every positive real number $p$, we set 
\begin{equation}
\label{sysnorm1}
  {\mathcal E}_p({\Bbb C}^{n})=
\{f \in H\!ol({\Bbb C}^{n})\, ;\, 
 \|f\|_{p,s}=\sup\, |f|\, e^{-s|\xi|^p}<\infty,\,\,\forall s>0\}
\end{equation}
where $|\xi|=(\sum_i|u_i|^2)^{1/2}$. The family of seminorms 
$\{||\,\cdot\,||_{p,s}\}_{s>0}$ induces a topology 
on $ {\mathcal E}_p({\Bbb C}^{n} )$ and 
$({\mathcal E}_p({\Bbb C}^{n} ),\cdot)$ is 
an associative commutative Fr\'{e}chet algebra, 
where the dot $\cdot$ is 
the ordinary product for functions in 
${\mathcal E}_p({\Bbb C}^{n})$. 
It is easily seen that for $0<p<p'$, there is a 
continuous embedding 
\begin{equation}
  {\mathcal E}_p({\Bbb C}^{n} )\subset 
{\mathcal E}_{p'}({\Bbb C}^{n} ) 
\end{equation}   
as commutative Fr\'{e}chet algebras (cf.\,\cite{GS}), and that 
${\mathcal E}_p({\Bbb C}^{n})$ is $G\!L(n,{\mathbb C})$-invariant. 

We denote  
\begin{equation}
{\mathcal E}_{p+}({\Bbb C}^{n})
=\bigcap_{p'>p}{\mathcal E}_{p'}({\Bbb C}^{n}),\quad 
(\text{with the intersection topology})
\end{equation}
It is obvious that every polynomial is contained in 
${\mathcal E}_{p}({\Bbb C}^{n})$, that is 
$p({\pmb u})\in {\mathcal E}_{0+}({\Bbb C}^{n})$, 
 and $\Bbb C[{\pmb u}]$ is dense 
in ${\mathcal E}_{p}({\Bbb C}^{n})$ for any $p>0$ 
in the Fr{\'e}chet topology defined by the family of 
seminorms $\{||\,\,||_{p,s}\}_{s>0}$. 

\bigskip
We easily see that 
$e^{\frac{1}{i\h}\langle{\pmb a},{\pmb u}\rangle}
\in{\mathcal E}_{1+}({\Bbb C}^{n})$. Moreover, it is not 
difficult to show that an exponential 
function $e^{p({\pmb u})}$ of a polynomial of degree $d$ is 
contained in ${\mathcal E}_{d+}({\Bbb C}^{n})$, but not in 
${\mathcal E}_{d}({\Bbb C}^{n})$.
 
Theorems \ref{main01} and \ref{main02} stated below give  
basic tools to study $*$-functions (cf. \cite{ommy} for the proof), 
although most of the concrete formulas can be obtained without these 
theorems. 

\begin{thm}\label{main01}
For $0<p\leq 2$, the product formula \eqref{eq:KK} 
extends to give the following{:}

\noindent
$(1)$ The space $({\mathcal E}_{p}({\Bbb C}^{n}),*_{_{K}})$ 
       forms a complete noncommutative topological associative
       algebra over $\mathbb C$.

\noindent
$(2)$ The intertwiner $I_{_K}^{^{K'}}$ extends to give an 
isomorphism of  $({\mathcal E}_{p}({\Bbb C}^{n}),*_{_{K}})$ 
onto $({\mathcal E}_{p}({\Bbb C}^{n}),*_{_{K'}})$.
\end{thm}

\medskip
\noindent
{\bf Remark}\,\,For the second statement, it is enough to prove that 
$I_{_K}^{^{K'}}$ extends to give a linear isomorphism of 
${\mathcal E}_{p}({\Bbb C}^{n})$ onto itself. 
The property (2) shows that if $p\leq 2$, $\coprod_{K\in{\mathfrak S}(n)}
{\mathcal E}_p({\Bbb C}^{n})$ is a trivial subbundle, and 
this is in fact an algebra bundle 
$$
\coprod_{K\in{\mathfrak S}(n)}
({\mathcal E}_{p}({\Bbb C}^{n}),*_{_K}). 
\quad (0<p\leq 2)
$$
The equation of parallel translation \eqref{parallel} has a unique solution
for the initial function $f$ is in  
${\mathcal E}_{p}({\Bbb C}^{n})$, $0<p\leq 2$.

It is easily seen that the following identities hold on 
${\mathcal E}_{p}({\Bbb C}^{n})$, $p\leq 2$
\begin{equation}
  \label{eq:intid}
I_{_{K'}}^{^{K}}I_{_{K}}^{^{K'}}=1,\quad
I_{_{K'}}^{^{K''}}I_{_K}^{^{K'}}=I_{_K}^{^{K''}}.  
\end{equation}
Hence, for every $f\in {\mathcal E}_{p}({\Bbb C}^{n})$, the set 
$f_*({\pmb u})=
\{I_{0}^{^{K}}(f); K\in{\mathfrak S}_{\mathbb C}(n)\}$ 
is a globally defined parallel section. 

\medskip
For $p>2$, we note the following: 
\begin{thm}
\label{main02}
For $p>2$, the product formula \eqref{eq:KK} 
gives continuous bi-linear mappings of 
\begin{equation}
{\mathcal E}_{p}({\Bbb C}^{n})\times
{\mathcal E}_{p'}({\Bbb C}^{n})\rightarrow 
{\mathcal E}_{p}({\Bbb C}^{n}),\quad 
{\mathcal E}_{p'}({\Bbb C}^{n})\times
{\mathcal E}_{p}({\Bbb C}^{n})\rightarrow 
{\mathcal E}_{p}({\Bbb C}^{n}),  
\end{equation}
for $\forall p'$ such that $\frac{1}{p}+\frac{1}{p'}\geq 1$.

For $f, g, h\in{\mathcal E}_{p}({\Bbb C}^{n})$ $(p>2)$,
 the associativity 
$(f{*}_{_{K}}g){*}_{_{K}}h=f{*}_{_{K}}(g{*}_{_{K}}h)$  
holds if two of $f, g, h$ are in 
${\mathcal E}_{p'}({\Bbb C}^{n})$ such that 
$\frac{1}{p}+\frac{1}{p'}\geq 1$. 
\end{thm}

%
\bigskip
Note that the linear change of coordinate 
${\pmb u}'={\pmb u}S$ by $S\in G\!L(n,{\mathbb C})$ 
gives naturally the topological linear isomorphism 
$\Phi_S:{\mathcal E}_{p}({\Bbb C}^{n})\to 
{\mathcal E}_{p}({\Bbb C}^{n}),\,\, p \geq 0,$
and this is an isomorphism as ${\mathbb C}[{\pmb u}]$-bi-modules 
for every $p\geq 0$; 
$$
\Phi_S:({\mathcal E}_{p}({\Bbb C}^{n}); *_{_\Lambda})\to 
({\mathcal E}_{p}({\Bbb C}^{n}; *_{_{{}^t\!S\Lambda S}}).
$$
This is not an automorphism, but an {\it outer isomorphism}.  

\subsubsection{Remarks on elements obtained by integrals}

Suppose $f(x)$ is a continuous mapping of a compact domain $D$ into 
${\mathcal E}_{p}({\Bbb C}^{n})$. As $\|f(x)\|_{p,s}$ is bounded on
$D$, its integral over $D$ is bounded. Hence  we see 
$\int_Df(x)dx \in {\mathcal E}_{p}({\Bbb C}^{n}).$
This will be used to compute Fourier series.

\begin{lem}
  \label{safe}
For every compact interval $I$, the integral 
$\int_{I}
{:}e_*^{t\frac{1}{i\h}\langle{\pmb a},{\pmb u}\rangle}
{:}_{_{K}}dt$
gives an element of ${\mathcal E}_{1+}({\mathbb C}^{n})$ 
for every $K\in {\mathfrak S}(n)$, and 
$$
I_{_K}^{^{K'}}\Big(\int_{I}
{:}e_*^{t\frac{1}{i\h}
\langle{\pmb a},{\pmb u}\rangle}{:}_{_{K}}dt\Big)
=\int_{I}
{:}e_*^{t\frac{1}{i\h}
\langle{\pmb a},{\pmb u}\rangle}{:}_{_{K'}}dt
$$  
$\{\int_{I}{:}
e_*^{t\frac{1}{i\h}\langle{\pmb a},{\pmb u}\rangle}{:}_{_{K}}dt;
K\in {\mathfrak S}(n)\}$ 
is a parallel section which may be denoted by 
$\int_{I}
e_*^{t\frac{1}{i\h}\langle{\pmb a},{\pmb u}\rangle}dt$ 
without showing expression parameters.
\end{lem}

\bigskip
Since 
$e^{\frac{1}{i\h}\langle{\pmb a},{\pmb u}\rangle}
\in{\mathcal E}_{1+}({\Bbb C}^{n})$,
Theorem \ref{main02} shows that 
$e^{\frac{1}{i\h}\langle{\pmb a},{\pmb u}\rangle}{*_{_K}}f$, 
$f{*_{K}}e^{\frac{1}{i\h}\langle{\pmb a},{\pmb u}\rangle}$ 
are defined for every 
$f\in \bigcup_{p>2}{\mathcal E}_{p}({\Bbb C}^{n})$, but 
in fact these are defined for every $f\in H\!ol({\mathbb C}^n)$ by 
\eqref{extend}.

As we have seen above, ${*}$-product integrals of exponential
functions of linear functions are remained in the class.  
${\mathcal E}_{1+}({\mathbb C}^{2m})$. 
Note that usual integral can be defined for elements of
 ${\mathcal E}_{1+}({\mathbb C}^{2m})$. 
However, we have to be careful to use the integral on 
noncompact domain, for such integrals often give elements 
outside the domain where the integrand is considered.  
Here, we give a typical example. 

Suppose 
${\rm{Re}}(\frac{1}{i\h}\langle{\pmb a}K,{\pmb a}\rangle)<0$, that is
$K$ is in the Siegel class. 
Then, the integral $\int_{-\infty}^{\infty}
{:}e_*^{t\frac{1}{i\h}
\langle{\pmb a},{\pmb u}\rangle}{:}_{_K}dt$
converges.  The formula of Fourier transform gives 
$$
\int_{-\infty}^{\infty}
{:}e_*^{t\frac{1}{i\h}
\langle{\pmb a},{\pmb u}\rangle}{:}_{_K}dt=
\int_{\mathbb R}e^{\frac{t^2}{4i\h}\langle{\pmb a}K,{\pmb a}\rangle}e^{t\frac{1}{i\h}
\langle{\pmb a},{\pmb u}\rangle}dt=
2(\frac{-i\h\pi}{\langle{\pmb a}K,{\pmb a}\rangle})^{1/2}
\,\,e^{-\frac{1}{i\h}\frac{1}{\langle{\pmb a}K,{\pmb a}\rangle}
\langle{\pmb a},{\pmb u}\rangle^2}.
$$ 
Since 
$\frac{1}{i\h}\langle{\pmb a},{\pmb u}\rangle{*}: 
H\!ol({\mathbb C}^{2m})\to H\!ol({\mathbb C}^{2m})$ 
is continuous, we see that
\begin{equation}
  \label{eq:confirm}
\begin{aligned}
\frac{1}{i\h}\langle{\pmb a},{\pmb u}\rangle{*}
\int_{-\infty}^{\infty}
e_*^{t\frac{1}{i\h}\langle{\pmb a},{\pmb u}\rangle}dt=
\lim_{N,N'\to\infty}
\int_{-N}^{N'}\frac{1}{i\h}\langle{\pmb a},{\pmb u}\rangle{*}
e_*^{t\frac{1}{i\h}\langle{\pmb a},{\pmb u}\rangle}dt
=\int_{-\infty}^{\infty}
\frac{d}{dt}
e_*^{t\frac{1}{i\h}\langle{\pmb a},{\pmb u}\rangle}dt=0.
\end{aligned}  
\end{equation}
Since 
$\langle{\pmb a},{\pmb u}\rangle{*_{_K}}
f(\langle{\pmb a},{\pmb u}\rangle)
=\langle{\pmb a},{\pmb u}\rangle 
f(\langle{\pmb a},{\pmb u}\rangle)
{+}\frac{i\h}{2}\langle{\pmb a}K,{\pmb a}\rangle
f'(\langle{\pmb a},{\pmb u}\rangle)$, 
the direct calculation also gives 
$$
\langle{\pmb a},{\pmb u}\rangle{*_{_K}}
e^{-\frac{1}{i\h}\frac{1}{\langle{\pmb a}K,{\pmb a}\rangle}
\langle{\pmb a},{\pmb u}\rangle^2}=0.
$$ 
%

\medskip 
Under the condition ${\rm{Re}}(\frac{1}{i\h}\langle{\pmb a}K,{\pmb a}\rangle)<0$, we denote 
as in \cite{OMMY3} 
$$
\frac{1}{2\pi\h}\int_{-\infty}^{\infty}
e_*^{t\frac{1}{i\h}\langle{\pmb a},{\pmb u}\rangle}dt
=\delta_*(\langle{\pmb a},{\pmb u}\rangle)
$$
Moreover, we see that integrals 
$\int_{-\infty}^{0}
{:}e_*^{t\frac{1}{i\h}\langle{\pmb a},{\pmb u}\rangle}{:}_{_K}dt,
\,\, 
{-}\!\int_{0}^{\infty}
{:}e_*^{t\frac{1}{i\h}\langle{\pmb a},{\pmb u}\rangle}{:}_{_K}dt$
are both inverses of 
$\frac{1}{i\h}\langle{\pmb a},{\pmb  u}\rangle$. We denote these by 
$$
(\frac{1}{i\h}\langle{\pmb a},{\pmb u}\rangle)_{*+}^{-1}
=\int_{-\infty}^{0}
{:}e_*^{t\frac{1}{i\h}\langle{\pmb a},{\pmb u}\rangle}{:}_{_K}dt,
\quad 
(\frac{1}{i\h}\langle{\pmb a},{\pmb u}\rangle)_{*-}^{-1}=
-\int_{0}^{\infty}
{:}e_*^{t\frac{1}{i\h}\langle{\pmb a},{\pmb u}\rangle}{:}_{_K}dt.
$$
This apparently breaks associativity 
$$
\Big((\frac{1}{i\h}\langle{\pmb a},{\pmb u}\rangle)_{*+}^{-1}{*_{_K}}
\frac{1}{i\h}\langle{\pmb a},{\pmb u}\rangle\Big){*_{_K}}
(\frac{1}{i\h}\langle{\pmb a},{\pmb u}\rangle)_{*-}^{-1}
\not=
(\frac{1}{i\h}\langle{\pmb a},{\pmb u}\rangle)_{*+}^{-1}{*_{_K}}
\Big(\frac{1}{i\h}\langle{\pmb a},{\pmb u}\rangle{*_{_K}}
(\frac{1}{i\h}\langle{\pmb a},{\pmb u}\rangle)_{*-}^{-1}\Big).
$$

\subsubsection{Remarks on real analyticity and on associativity}

A mapping $f: U\to F$ from an open subset $U$ of $\mathbb R$ 
into a Fr{\'e}chet space $F$ is called to be   
{\bf real analytic}, if for every $a\in U$ there is $\e(a)>0$
such that $f$ is written in the form  
$$
f(a+s)=\sum_k \frac{1}{k!}a_k s^k, \quad  a_k\in F, \quad 
|s|<\e(a). 
$$
$a_k$ is given by $a_k=\partial_s^kf|_{s=0}$.

If $F$ is a Banach space and $\sum_k\frac{1}{k!}\|a_k\||s|^k$
converges, then the power series 
$\sum_k a_k s^k$ is called to 
{converge absolutely} under the norm. 

If a Fr{\'e}chet space $F$ is defined by a countable 
family of seminorms  
$\{\|f\|_m; m=1,2,3\cdots\}$, then replace this part by 
the absolute convergence of 
$\sum_k\frac{1}{k!}\|a_k\|_m|s|^k$ 
w.r.t. seminorms $\|\cdot\|_m$. 
 A power series $\sum_k a_k s^k$ converges if this 
converges absolutely under every seminorms. 

\noindent
{\bf Radius of convergence}\,\,Suppose a 
Fr{\'e}chet space $F$ is defined by a countable family of 
seminorms $\{\|f\|_m; m=1,2,3\cdots\}$.

\begin{lem}
  \label{powerser}
For a power series $\sum_k a_k s^k$, $a_k\in F$,
there exists uniquely a real number $R$ 
$(0\leq R\leq \infty)$ satisfying $(1)$ 
and $(2)$ below:
\begin{description}
\item[(1)] If $|s|<R$, then the power series 
$\sum_k a_k s^k$ converges 
absolutely under every seminorm $\|\cdot\|_m$.
 
\item[(2)] If $|s|>R$, then $\sum_k a_k s^k$ does not converge 
for some seminorm. 
\end{description}
\end{lem}

\medskip\noindent
{\bf Proof}\,\,Suppose $\sum_k a_k s_0^k$ converges  
at $s_0$. Then $a_k s_0^k$ is bounded 
under every seminorm $\|\cdot\|_m$. Set 
$\sup_k\|a_k s_0^k\|_m\leq M_m$. Then for every 
$s$ such that $|s|<|s_0|$ 
$$
\sum_k\|a_ks^k\|_m\leq \sum_k M_m|s/s_0|^k 
= M_m\frac{1}{1-|s/s_0|}<\infty.
$$
It follows the convergence of $\sum_ka_ks^k$. 
${}$\hfill$\square$

\normalsize
\begin{lem}
  \label{powbibun}
$\sum_{k\geq 0}a_k s^k$ and 
$\sum_{k\geq 1}ka_k s^{k-1}$ have same radius of convergence. 
\end{lem}

Real analyticity is left invariant under every continuous 
linear transformation:
\begin{lem}
  \label{linsbl}
Let $F, G$ be Fr{\'e}chet spaces and 
$\varphi : F\to G$ be a continuous linear mapping.  
If $f: U\to F$ is real analytic, then 
$\varphi f: U\to G$ is also real analytic. 
\end{lem}
Since  $X\to p({\pmb u}){*}X{*}q({\pmb u})$ is a continuous 
linear mapping, Lemma\,\ref{frechet} gives the  following:
\begin{lem}\label{realanal}
Let $U$ be an connected open neighborhood of $0$ of 
${\mathbb R}^{\ell}$ 
Suppose $\psi: U\to H{\!o}l({\mathbb C}^n)$ be a
real analytic mapping. Then 
$x\to p({\pmb u}){*}\psi(x){*}q({\pmb u})$ is also a 
real analytic on $U$ for every polynomial 
$p({\pmb u}),\,q({\pmb u})$. 
\end{lem}

\noindent
{\bf Remarks on the associativity}

Products of exponential functions of quadratic forms 
may not be defined, and even if the product is defined 
the associativity may not hold, since these are elements of 
${\mathcal E}_{2+}({\Bbb C}^{n})$. In general, we do not 
have the associativity even for a polynomial $p({\pmb u})$
$$
(e^{H({\pmb u})}{*}p({\pmb u})){*}e^{K({\pmb u})},
   \quad e^{H({\pmb u})}{*}(p({\pmb u}){*}e^{K({\pmb u})}), 
$$
since $p(u)$ has two different $*$-inverses in general.

\medskip

However, if we can treat elements in 
$({\mathbb C}[\pmb u][[\h]], {*_{_K}})$, 
the space of formal power series of $\h$,
then $*_{_\Lambda}$-product is always defined by the 
product formula \eqref{eq:KK} and the associativity holds. 

Elements of ${\mathcal E}_{2+}({\Bbb C}^{n})$ are often 
given as a real analytic function of $\h$ defined on 
certain interval containing $\h=0$.  The following 
is easy to see:

\begin{thm}\label{assocthm}
Suppose $f(\h,{\pmb u})$,  
$g(\h,{\pmb u})$ and $h(\h,{\pmb u})$
are given as real analytic function of $\h$ in some interval 
$[0,H]$. If 
$$
f(\h,{\pmb u}){*_{_K}}g(\h,{\pmb u}),\,\, 
(f(\h,{\pmb u}){*_{_K}}g(\h,{\pmb u})){*_{_K}}h(\h,{\pmb u}),\,\, 
g(\h,{\pmb u}){*_{_K}}h(\h,{\pmb u}),\,\, 
f(\h,{\pmb u}){*_{_K}}(g(\h,{\pmb u}){*_{_K}}h(\h,{\pmb u}))
$$
are defined as real analytic functions on $[0,H]$, then 
the associativity 
$$
(f(\h,{\pmb u}){*_{_K}}g(\h,{\pmb u})){*_{_K}}h(\h,{\pmb u})
=f(\h,{\pmb u}){*_{_K}}(g(\h,{\pmb u}){*_{_K}}h(\h,{\pmb u}))
$$
holds. 
\end{thm}
We refer this theorem to the 
{\bf formal associativity theorem}.

\noindent 
{\bf Remark 1}. 
In what follows, elements are often given 
in the form $f(\frac{1}{i\h}\varphi(t),{\pmb u})$ 
by using a real analytic function $f(t,{\pmb u})$, $t{\in}[0,T]$, 
where $\varphi(t)$ is a real analytic function such that 
$\varphi(0){=}0$. 
(Cf.\eqref{eq:tempexp}, \eqref{eq:prodfmla}

In such a case, 
replacing $t$ by $s\h$ gives a real analytic function 
of $\h$, and such an element is embedded in 
$({\mathbb C}[\pmb u][[\h]], {*_{_K}})$. 
Thus, we can apply the above theorem. We call such elements 
{\bf classical elements}. 
However, there are many elements in 
${\mathcal E}_{2+}({\Bbb C}^{n})$ 
written in the form $f(\frac{1}{i\h}\varphi(t),{\pmb u})$ 
such that $\varphi(0){\not=}0$.

\section{Blurred covering group of $Sp(m,{\mathbb C})$}

In this section we first treat the infinitesimal 
$*$-action of quadratic forms 
on the space of exponential functions of quadratic forms. 
We treat this in general expressions by using intertwiners.  
Since the space of quadratic forms is isomorphic to the 
Lie algebra of $Sp(m,\mathbb C)$, i.e. 
$$
\{\langle{\pmb u}A,{\pmb u}\rangle; A\in{\mathfrak S}(2m)\}
\cong {\mathfrak{sp}}(m,{\mathbb C})=
\{\alpha; \alpha J{+}J\,{}^t\!\alpha=0\}
$$ 
as Lie algebra, the natural ${*}$-action of quadratic forms
can be viewed as the 
{\bf infinitesimal action} of the Lie group $Sp(m,\mathbb C)$. 

In contrast with that infinitesimal intertwiners are viewed 
as a flat connection on the trivial bundle 
$\coprod_{K{\in}{\mathfrak S}(2m)}{\mathbb C}
e^{{\mathfrak S}(2m)}$, 
whose fiber is the space  of exponential functions of 
quadratic forms ${\mathbb C}e^{{\mathfrak S}(2m)}$, 
all possible infinitesimal 
actions of quadratic forms gives a tangential 
distribution on each fiber  
$\coprod_{K{\in}{\mathfrak S}(2m)}{\mathbb C}
e^{{\mathfrak S}(2m)}$.  
 
\subsection{Infinitesimal actions of quadratic forms}
\label{infaction}

On every fiber at $K$, consider left multiplication   
$$ 
{:}\langle{\pmb u}A,{\pmb u}\rangle{:}_{_K}{*_{_K}} : 
{\mathbb C}e^{{\mathfrak S}(2m)}
  \to {\mathbb C}e^{{\mathfrak S}(2m)}
$$
Since $\frac{1}{i\hbar}{:}\langle{\pmb u}A,{\pmb u}\rangle{:}_{_K}=
\frac{1}{i\hbar}\langle{\pmb u}A,{\pmb u}\rangle
+\frac{1}{2}{\rm Tr}(A\!K)$, 
we see 
\begin{equation}\label{bulletone}
\frac{1}{i\hbar}{:}\langle{\pmb u}A,{\pmb u}\rangle{:}_{_K}{*_{_K}}
(ge^{\frac{1}{i\hbar}\langle{\pmb u}Q,{\pmb u}\rangle})
=\Big(\frac{1}{2}{\rm{Tr}}\big((K{-}J)A(K{+}J)Q+A\!K\big) 
{+}\frac{1}{i\hbar}\langle{\pmb u}{Q'},{\pmb u}\rangle\Big)
ge^{\frac{1}{i\hbar}\langle{\pmb u}{Q}),{\pmb u}\rangle}  
\end{equation}
where 
${Q'}=
{A}+{A}(K{+}J){Q}+{Q}(K{-}J){A}
+{Q}(K{-}J){A}(K{+}J){Q}$, and $A\in{\mathfrak S}(2m)$.  
The term 
$\frac{1}{i\hbar}\langle{\pmb u}{Q'},{\pmb u}\rangle$ 
will be called the {\it infinitesimal phase} part, and 
$\frac{1}{2}{\rm{Tr}}\big((K{-}J)A(K{+}J)Q+A\!K\big)$ will be called 
the {\it infinitesimal amplitude} part.

Moving  $A\in{\mathfrak S}(2m)$ at every fixed 
$ge^{\frac{1}{i\hbar}\langle{\pmb u}Q,{\pmb u}\rangle}$, we have 
a linear subspace of the tangent space of 
${\mathbb C}e^{{\mathfrak S}(2m)}$ at 
$ge^{\frac{1}{i\hbar}\langle{\pmb u}Q,{\pmb u}\rangle}$. 
We call this the singular distribution of infinitesimal 
actions of quadratic forms.

\bigskip
On the other hand, there is a natural 
correspondence between  ${\mathfrak{sp}}(m;\mathbb C)$ 
and ${\mathfrak S}(2m)$.
$$
{\mathfrak{sp}}(m;\mathbb C)\cong {\mathfrak S}(2m)\quad 
{\text{via}}\quad
\alpha\in {\mathfrak{sp}}(m;\mathbb C)
\Leftrightarrow \alpha J\in{\mathfrak S}(2m),  
\quad 
J=
\begin{bmatrix} 
0&{-}1\\
1&0
\end{bmatrix}.
$$
We make the correspondence as follows: 
$$
 A \Leftrightarrow\alpha{=}{-}AJ,\qquad 
 Q \Leftrightarrow\xi{=}{-}QJ. 
$$
We set also ${\kappa'}{=}J\!K'$, $\kappa{=}J\!K$ in 
${\mathfrak{sp}}(m;\mathbb C)$.
Through these, intertwiners $I_{_K}^{^{K'}}$ defined on 
$\coprod_{K\in{\mathfrak S}(2m)}{\mathbb C}
e^{{\mathfrak S}(2m)}$ in \S\,3.1 of \cite{OMMY4}    
is easily translated on 
$\coprod_{\kappa\in{\mathfrak{sp}}(m;\mathbb C)}
{\mathbb C}e^{{\mathfrak{sp}}(m;\mathbb C)J}$ as 
$$
I_{\kappa}^{\kappa'}(ge^{\langle{\pmb u}(\frac{1}{i\h}\alpha J),{\pmb u}\rangle})
=\frac{g}{\sqrt{\det(I{-}\alpha(\kappa'{-}\kappa))}}
e^{\langle{\pmb u}(\frac{1}{i\h}\frac{1}{I{-}\alpha(\kappa'{-}\kappa)}\alpha J),
{\pmb u}\rangle} . 
$$
These intertwiners may be viewed 
as coordinate transformations:    
$I_{\kappa}^{\kappa'}:{\mathbb C}
e^{{\mathfrak{sp}}(m;\mathbb C)J}\to
{\mathbb C}e^{{\mathfrak{sp}}(m;\mathbb C)J}$. 
For the precise treatment of patching by intertwiners, we set
$$
I_{\kappa}^{\kappa'}\frac{1}{\sqrt{\det(I{-}\alpha\kappa))}}
e^{\frac{1}{i\h}\langle{\pmb u}
(\frac{1}{I{-}\alpha\kappa}\alpha J),{\pmb u}\rangle}=
\frac{1}{\sqrt{\det(I{-}\alpha\kappa'))}}
e^{\frac{1}{i\h}\langle{\pmb u}
(\frac{1}{I{-}\alpha\kappa'}\alpha J),{\pmb u}\rangle}
$$
$$
\widetilde{\mathcal D}_{\kappa}
=\{
\frac{1}{\sqrt{\det(I{-}\alpha\kappa))}}
e^{\frac{1}{i\h}\langle{\pmb u}
(\frac{1}{I{-}\alpha\kappa}\alpha J),{\pmb u}\rangle}; \alpha\in 
{\mathcal D}_{\kappa}\},\quad 
{\mathcal D}_{\kappa}{=}\{\alpha; \det ({I{-}\alpha\kappa) }\not=0\}.
$$
As $\kappa$ moves in the whole space ${sp}(m,{\mathbb C})$, and 
for any $\alpha$, we can find $\kappa$ such that 
$\det(I{-}\alpha\kappa){\not=}0$,  
we easily see that 
\begin{equation}\label{spspsp}
\bigcup_{\kappa}{\mathcal D}_{\kappa}=
{\mathfrak{sp}}(m,{\mathbb C}), \quad 
\bigcap_{\kappa}{\mathcal D}_{\kappa}=\{0\}.
\end{equation}

$\widetilde{\mathcal D}_{\kappa}$ is a double cover of 
${\mathcal D}_{\kappa}$. 
(Recall we set $\sqrt{1}=\{\pm 1\}$ in the case 
$\kappa=0$.)  
Let $\pi:\widetilde{\mathcal D}_{\kappa}\to{\mathcal D}_{\kappa}$ be the 
natural projection. 
As it was seen in in \S\,3.1 of \cite{OMMY4}
%
%
intertwiners $I_{\kappa}^{\kappa'}$ give isomorphisms 
$$
\begin{matrix}
\widetilde{\mathcal D}_{\kappa}& \supset & 
\pi^{-1}({\mathcal D}_{\kappa}{\cap}{\mathcal D}_{\kappa'})&
\overset{I_{\kappa}^{\kappa'}}{\longrightarrow}&
\pi^{-1}({\mathcal D}_{\kappa'}{\cap}{\mathcal D}_{\kappa})&
\subset &\widetilde{\mathcal D}_{\kappa'}\\
\downarrow\,\pi&  &\downarrow\,\pi&  &\downarrow\,\pi&  
&\downarrow\,\pi \\
{\mathcal D}_{\kappa}& \supset & 
{\mathcal D}_{\kappa}{\cap}{\mathcal D}_{\kappa'}&
=\!=&
{\mathcal D}_{\kappa'}{\cap}{\mathcal D}_{\kappa}&
\subset&{\mathcal D}_{\kappa'}
\end{matrix}   
$$
However intertwiners 
$I_{\kappa}^{\kappa'}$ are 2-to-2 mappings. 
Thus, the union 
$\bigcup_{\kappa}\widetilde{\mathcal D}_{\kappa}$ 
is a manifold-like object glued by 2-to-2 coordinate 
transformations. 

\bigskip

Set ${\alpha'}{=}{-}Q'J$, ${\xi}{=}{-}QJ$, 
${\alpha}{=}{-}AJ$, $\kappa{=}J\!K$. These are 
$\in {\mathfrak{sp}}(m;\mathbb C)$. 
We want to translate the equality $\eqref{bulletone}$ by these replacement.   
First, the infinitesimal phase part is rewritten as 
\begin{equation*}
\begin{aligned}[t]
{\alpha'} =&{\alpha}-{\alpha}(I{-}\kappa){\xi}+{\xi}(I{+}\kappa){\alpha}
-{\xi}(I{+}\kappa){\alpha}(I{-}\kappa){\xi}\\
=&\big(I+{\xi}(I{+}\kappa)\big){\alpha}\big(I-(I{-}\kappa){\xi}\big),
\end{aligned}
\end{equation*}
and it is easy to see 
$\big(I+{\xi}(I{+}\kappa)\big){\alpha}\big(I-(I{-}\kappa){\xi}\big)\in
{\mathfrak{sp}}(m;\mathbb C)$ by moving 
$J$ in the l.h.s of \\
$J\big(I+{\xi}(I{+}\kappa)\big){\alpha}\big(I-(I{-}\kappa){\xi}\big)$
to the r.h.s. \,\,
Hence, the equality $\eqref{bulletone}$ is translated into 
\begin{equation}\label{diamond}
\begin{aligned}
(\diamondsuit)\quad \frac{1}{i\hbar}
{:}\langle{\pmb u}(\alpha J),{\pmb u}\rangle{:}_{_K}{*_{_K}}
(ge^{\frac{1}{i\hbar}\langle{\pmb u}\,\xi J,{\pmb u}\rangle})
&{=}
\Big(\frac{1}{2}{\rm{Tr}}\big((\kappa{+}I)\alpha(\kappa{-}I)\xi{+}\alpha\kappa\big)
{+}\langle{\pmb u}(\alpha'J),{\pmb u}\rangle\Big)
ge^{\frac{1}{i\hbar}\langle{\pmb u}\,{{\xi}J}),{\pmb u}\rangle}\\
\alpha'&{=}(I+{\xi}(I{+}\kappa)){\alpha}(I-(I{-}\kappa){\xi}).   
\end{aligned}
\end{equation}

\medskip
\noindent 
By moving $\alpha{\in}{\mathfrak{sp}}(m,{\mathbb C})$ we make a 
 subspace $D_{(\kappa,ge^{\xi})}$ of the tangent space 
 $T_{ge^{\xi}}{\mathbb C}e^{{\mathfrak S}(2m)}$ of 
${\mathbb C}e^{{\mathfrak S}(2m)}$ at  
$ge^{\frac{1}{i\hbar}\langle{\pmb u}\,\xi J,{\pmb u}\rangle}$. 
We make also a distribution (a singular subbundle):    
\begin{equation}\label{distr}
\begin{aligned}
D_{(\kappa,ge^{\xi})}=&
\Big\{\Big(\frac{1}{2}
{\rm{Tr}}\big((\kappa{+}I)\alpha(\kappa{-}I)\xi{+}\alpha\kappa\big)
{+}\frac{1}{i\h}
\langle{\pmb u}\,{{\alpha'}\!J},{\pmb u}\rangle,\,\,\alpha\Big); 
 \alpha\in {sp}(m,{\mathbb C})\Big\}\\
&\qquad\qquad{\text{where }} \,\,{\alpha'}=
\big(I{+}{\xi}(I{+}\kappa)\big){\alpha}
\big(I{-}(I{-}\kappa){\xi}\big). 
\end{aligned}
\end{equation}
on the space ${\mathbb C}e^{{\mathfrak S}(2m)}$.
Note the following:
\begin{lem}\label{lemma0}
 $\det(I{+}{\xi}(I{+}\kappa))
{=}\det(I{-}(I{-}\kappa)\xi)
{=}\det(I{-}{\xi}(I{-}\kappa))$.
\end{lem}

\noindent
{\bf Proof}\, For the first 
equality, use $\det J{=}1$ and   
$$
\det(J(I{+}{\xi}(I{+}\kappa))){=}
\det((I{-}{}^t\!{\xi}(I{-}{}^t\!\kappa))J)=
\det(I{-}(I{-}\kappa){\xi}).
$$ 
For the second, we use the 
standard trick 
$$
\det(I{-}(I{-}\kappa){\xi})=
\det({\xi}^{-1}{\xi}{-}(I{-}\kappa){\xi})
=\det({\xi}{\xi}^{-1}{-}{\xi}(I{-}\kappa))
$$
via an appropriate approximation of $\xi$ by nonsingular 
element.\hfill$\Box$
\bigskip


\medskip
First of all, we consider open subsets where the rank of distribution is constant:
\begin{lem}
$\alpha\to{\alpha'}$ is a bijection of 
${\mathfrak{sp}}(m;\mathbb C)$ onto itself,
  if and only if 
$\det(I{+}{\xi}(I{+}\kappa))\not=0.$
In this case, Lemma\,\ref{lemma0} gives 
$$
\alpha=(I{+}{\xi}(I{+}\kappa))^{-1}\alpha'(I{-}(I{-}\kappa)\xi)^{-1}.
$$
That is in $K$-ordered expression, the infinitesimal action 
$\{\langle{\pmb u}(\alpha J),{\pmb u}\rangle{*};\,\,  
\alpha{\in}{sp}(m;\mathbb C)\}$ degenerates only at the point 
$\xi$ such that $\det(I{+}{\xi}(I{+}\kappa))=0$. 
\end{lem}

\medskip

Let 
${\mathcal O}_{\kappa}
=\{\xi\in {\mathfrak{sp}}(m;\mathbb C);
\det(I{+}\xi(I{+}\kappa))\not=0\}$ for every 
${\kappa}{\in}{\mathfrak{sp}}(m;{\mathbb C})$.    
Since this distribution is given by the infinitesimal action 
of a Lie group, we have 
\begin{prop}
The distribution $D_{(\kappa,ge^{\xi})}$ 
is constant corank one and involutive on ${\mathcal O}_{\kappa}$. 
\end{prop}

The goal of this section is as follows:
\begin{thm}\label{goal}
Maximal integral submanifold through $g\in {\mathbb C}_{\times}$ over ${\mathcal O}_{\kappa}$
 is given by  
\begin{equation}
 \label{eq:kyokumen1}
\{g\sqrt{\det(I{+}(I{{+}}\kappa){\alpha})}
e^{\langle{\pmb u}(\frac{1}{i\hbar}{\alpha}J), {\pmb u}\rangle}; 
\alpha\in{\mathcal O}_{\kappa}\}  
\end{equation}
This is a nontrivial double cover of ${\mathcal O}_{\kappa}$. 
\end{thm}
The proof is given in several steps. 
Note first that the maximal integral submanifold through $1$ must be closed
under $*_{\kappa}$-product.

\medskip
\noindent
{\bf Step 1}
First note that the phase part of the distribution takes arbitrary 
element. Thus consider elements 
$g(t)e^{\frac{1}{i\h}\langle{\pmb u}(t\tilde\alpha J),{\pmb u}\rangle}$
by fixing $\tilde\alpha$. 
We want to make the tangent vectors of this curve are always in the distribution.
Taking derivative, we have 
$$
\Big(\frac{d}{dt}g(t){+}
\frac{1}{i\h}\langle{\pmb u}(\tilde\alpha J),{\pmb u}\rangle g(t)\Big)
e^{\frac{1}{i\h}\langle{\pmb u}(t\tilde\alpha J),{\pmb u}\rangle}.
$$
Comparing this with \eqref{diamond} at $\xi=t\tilde\alpha$, we take 
$\alpha(t)$ so that 
$$
\tilde\alpha=
(I{+}t\tilde\alpha(I{+}\kappa))\alpha(t)(I{-}t\tilde\alpha(I{-}\kappa)).
$$ 
Then, the infinitesimal action by 
$\langle{\pmb u}(\alpha(t)J),{\pmb u}\rangle$ satisfies
$$
\begin{aligned}
&{:}\frac{1}{i\h}\langle{\pmb u}(\alpha(t)J),{\pmb u}\rangle{:}_{_K}{*_{_K}}
(g(t)e^{\frac{1}{i\h}\langle{\pmb u}(t\tilde\alpha J,{\pmb u}\rangle})\\
&=
\left\{
\frac{1}{2}{\rm{Tr}}
\big((\kappa{+}I){\alpha(t)}(\kappa{-}I)t\tilde\alpha{+}\alpha(t)\kappa\big)
{+}\frac{1}{i\h}\langle{\pmb u}(\tilde\alpha J),{\pmb u}\rangle
\right\}g(t)e^{\frac{1}{i\h}\langle{\pmb u}(t\tilde\alpha J),{\pmb u}\rangle}\\
&=\Big(\frac{d}{dt}g(t){+}
\frac{1}{i\h}\langle{\pmb u}(\tilde\alpha J),{\pmb u}\rangle g(t)\Big)
e^{\frac{1}{i\h}\langle{\pmb u}(t\tilde\alpha J),{\pmb u}\rangle}
\end{aligned}.
$$
Plugging in 
$\alpha(t)=(I{+}{t\tilde\alpha}(I{+}\kappa))^{-1}
\tilde\alpha(I{-}(I{-}\kappa)t\tilde\alpha)^{-1}$ into the above,  
$g(t)$ is obtained by solving 
$$
\frac{d}{dt}g(t)=
\frac{1}{2}{\rm{Tr}}
\big((\kappa{+}I){\alpha(t)}(\kappa{-}I)t\tilde\alpha{+}\alpha(t)\kappa\big)g(t),\quad 
g(0)=g.
$$
\medskip
\noindent
{\bf Step 2}
To solve this equation, we first solve it in the case
$\kappa{=}0$. The equation becomes
$$
\frac{d}{dt}\log g(t)=
\frac{1}{2}
{\rm{Tr}}\frac{t\tilde\alpha^2}{1-(t\tilde\alpha)^2}=
\frac{1}{4}\frac{d}{dt}{\rm{Tr}}\log(1{-}(t\tilde\alpha)^2).
$$
It follows that 
$$
g(t)=e^{{\rm{Tr}}\log(1{-}(t\tilde\alpha)^2)^{\frac{1}{4}}}
=\sqrt[4]{\det(1{-}(t\tilde\alpha)^2)}.
$$

On the other hand, since $\det(1{-}t\tilde\alpha){=}\det(1{+}t\tilde\alpha)$, 
we have 
$g(t)=\sqrt{\det(1{+}t\tilde\alpha)},$
that is, 
\begin{lem}\label{integsug}
$\sqrt{\det(1{+}t\tilde\alpha)}\,\,
e^{\frac{1}{i\h}\langle{\pmb u}(t\tilde\alpha J),{\pmb u}\rangle}$
is in an integral submanifold. 
\end{lem}

\medskip
\noindent
{\bf Step 3}
The integral manifold for the general $\kappa$ is obtained by 
the intertwiner $I_0^{\kappa}$. We have 
$$
I_0^{\kappa}\Big(\sqrt{\det(1{+}\tilde\alpha)}
e^{\frac{1}{i\h}\langle{\pmb u}(t\tilde\alpha J),{\pmb u}\rangle}\Big)
=\frac{\sqrt{\det(I{+}\tilde\alpha)}}{\sqrt{\det(I{-}\tilde\alpha\kappa)}}
e^{\frac{1}{i\h}
\langle{\pmb u}(\frac{1}{I{-}\tilde\alpha}\tilde\alpha J),{\pmb u}\rangle}.
$$
Replacing $\frac{1}{I{-}\tilde\alpha}\tilde\alpha=\alpha$ gives 
$$
\tilde\alpha=\alpha\frac{1}{1+\alpha\kappa}{=}\frac{1}{I{+}\alpha\kappa}\alpha.
$$
Plugging this and using the algebraic calculation such that 
$\frac{\sqrt{x}}{\sqrt{x}}=1$, we have the following:
\begin{prop}\label{integmanif}
In $\kappa$-ordered expression, the maximal integral submanifold is 
given by 
$$
c{\widetilde{\mathcal O}}_{\kappa}=
\{
c\sqrt{\det(I{+}\alpha(I{+}\kappa))}\,\,
e^{\frac{1}{i\h}\langle{\pmb u}(\alpha J),{\pmb u}\rangle}; \alpha\in 
{\mathcal O}_{\kappa}\}
$$
where ${\mathcal O}_{\kappa}=
\{\alpha\in {sp}(m,{\mathbb C}); \det(I{+}\alpha(I{+}\kappa))\not=0\}$.
\end{prop} 

Note that Proposition\,\ref{integmanif} shows that we have only to know the 
phase part to know the integral submanifold. 
By definition ${\widetilde{\mathcal O}}_{\kappa}$ is 
the maximal integral submanifold through 
$(1,0)\in {\mathbb C}{\times}{sp}(m;\mathbb C)$.  
Setting $c=1$ in Proposition\ref{integmanif}, we see that 
\begin{equation}\label{integralmanifmax}
\pi_{\kappa}: {\widetilde{\mathcal O}}_{\kappa}\to{\mathcal O}_{\kappa}
\end{equation}
is a nontrivial double cover, which is just the forgetful mapping of
the amplitude part. The significance of the set 
${\mathcal O}_{\kappa}$ will be explained in the next section by the
Cayley transform. 

\subsubsection{Integral submanifolds and twisted 
Cayley transforms}\label{Cayley}

The Cayley transform  $C_0(X)=\frac{I-X}{I+X}$ 
has following properties:
For $X\in {\mathfrak{sp}}(m;\mathbb C)$ with $\det(I+X)\not=0$, we see 
$C_0(X)\in Sp(m;\mathbb C)$ and  
$\det(I+C_0(X))=(\det(I+X))^{-1}$. 
\begin{equation}
\label{eq:Cayley}
X\in {\mathfrak{sp}}(m;\mathbb C)\Leftrightarrow C_0(X)\in Sp(m;\mathbb C),
\quad C_0^2(X)=X.
\end{equation}

Let ${\mathcal O}_0=\{X\in {\mathfrak{sp}}(m;\mathbb C); \det(I{+}X)\not=0\}$.  
$C_0:{\mathcal O}_0\to Sp(m;\mathbb C)$ is viewed as a local 
coordinate system $Sp(m;\mathbb C)$, which covers an open dense subset of 
$Sp(m;\mathbb C)$.

Let ${\mathcal O}_\kappa=
\{\alpha\in {\mathfrak{sp}}(m;\mathbb C); \det(I{+}(I{+}\kappa)\alpha)\not=0\}$, and 
define
\begin{equation}
  \label{eq:cayley22}
\begin{aligned}[c]
C_{\kappa}(\alpha)=
(I-(I{-}{\kappa})\alpha)&\frac{1}{I+(I{+}{\kappa})\alpha}
=\frac{1}{I{+}\alpha(I{+}{\kappa})}(I{-}\alpha(I{-}{\kappa})), \\
(C_{\kappa})^{-1}(Y)=
&\frac{1}{I{-}{\kappa}{+}Y(I{+}{\kappa})}(I{-}Y)=
(I{-}Y)\frac{1}{I{-}{\kappa}{+}(I{+}{\kappa})Y}.
\end{aligned}
\end{equation}
$C_{\kappa}:{\mathcal O}_{\kappa}\to Sp(m;\mathbb C)$ gives
also a local coordinate system of $Sp(m;\mathbb C)$. We call 
\eqref{eq:cayley22} the {\it twisted Cayley transform}. 
Since  
$C_{\kappa}: {\mathcal O}_{\kappa}\to 
C_{\kappa}({\mathcal O}_{\kappa})$ is 
a diffeomorphism, we often identify 
${\mathcal O}_{\kappa}$ with 
$C_{\kappa}({\mathcal O}_{\kappa})$ through the 
twisted Cayley transform $C_{\kappa}$. 
The following Lemma is crucial to our purpose:
\begin{lem}\label{crucial00}
$\bigcup\{C_{\kappa}({\mathcal O}_{\kappa}); 
\kappa\in {\mathfrak{sp}}(m,{\mathbb C})\} 
=Sp(m,{\mathbb C})$. 
On the other hand, if $\alpha\not\in {\mathcal O}_0$, then 
$\frac{1}{I{-}\alpha\kappa}\alpha
\not\in{\mathcal O}_{\kappa}$ for every 
$\kappa\in {\mathcal D}_{\alpha}{=}
\{\kappa{\in}{\mathfrak{sp}}(m;{\mathbb C}); 
\det(I-\alpha\kappa)\not=0\}$.   
\end{lem}

\noindent 
{\bf Proof}\, Suppose there is a $Y{\in}Sp(m,{\mathbb C})$ 
such that $\det(I{-}{\kappa}{+}Y(I{+}{\kappa})){=}0$ for  
every $\kappa\in {\mathfrak{sp}}(m,{\mathbb C})$. Then such a $Y$ must satisfy 
$\det(\frac{1{-}\kappa}{I{+}\kappa}{+}Y){=}0$. 
Since $\frac{1{-}\kappa}{I{+}\kappa}$ moves in an open dense domain of   
$Sp(m,{\mathbb C})$, it follows $\det(X{+}Y){=}0$ for every 
$X\in Sp(m,{\mathbb C})$. Set $X{=}Y$ to get a contradiction. 
Thus we see that for every $Y$ there is $\kappa$ such that 
$\det(I{-}{\kappa}{+}Y(I{+}{\kappa})){\not=}0.$ 
Hence $C_{\kappa}^{-1}(Y)$ exists. 
The rest of Lemma follows easily. \hfill $\Box$

\bigskip

Define $T_{\kappa'{-}\kappa}(\alpha)=\frac{1}{I-\alpha(\kappa'{-}\kappa)}\alpha$. 
Then  
$T^{-1}_{\kappa'{-}\kappa}(\alpha)=\frac{1}{I{+}\alpha(\kappa'{-}\kappa)}\alpha$
$$
T_{\kappa'{-}\kappa}(\alpha)\in {\mathfrak{sp}}(m;\mathbb C)
\Longleftrightarrow \alpha \in {\mathfrak{sp}}(m;\mathbb C).
$$  
It is easy to see 
$$
\frac{1}{I{-}\alpha(\kappa'{-}\kappa)}(I{+}\alpha(I{+}{\kappa}))= 
I{+}T_{\kappa'{-}\kappa}(\alpha)(I{+}{\kappa'}).
$$
Hence, we have the following:

\begin{center}
\fbox{\parbox[c]{.8\linewidth}{
${}$\hfill{The Cayley transform gives the phase part of intertwiners.}\hfill${}$\\
$T_{-\kappa}\sim C_0^{-1}C_{\kappa},\quad T_{\kappa'-\kappa}\sim
C_{\kappa'}^{-1}C_{\kappa}$ \,\,({\text{$\sim$ means equality in algebraic calculations}})}}
\end{center}

\bigskip\noindent
On $Sp(m,{\mathbb C})$, the coordinate transformations are given by
the phase part of the intertwiners.

Hence, by setting 
${\mathcal O}_{\kappa\kappa'}={\mathcal O}_{\kappa}\cap{\mathcal O}_{\kappa'}$, intertwiners give 2-to-2 mappings 
\begin{equation}
\label{eq:coodtrns}
I_{\kappa}^{\kappa'}: \pi_{\kappa}^{-1}({\mathcal O}_{\kappa\kappa'})\to 
            \pi_{\kappa'}^{-1}({\mathcal O}_{\kappa'\kappa}) 
\end{equation}
just as in Proposition 3.1\,in \S\,3.1 of \cite{OMMY4}.
%
%
Since $0{\in}\bigcap_{\kappa}{\mathcal O}_{\kappa}$, and 
$C_{\kappa}(0)=I$, 
 we can consider $\bigcup_{\kappa}{\widetilde{\mathcal O}}_{\kappa}$ as 
an object patched by the intertwiners $I_{\kappa}^{\kappa'}$ as a 
bundle-like object over 
$\bigcup_{\kappa}C_{\kappa}({\mathcal O}_{\kappa})=Sp(m;{\mathbb C})$.

\medskip
Computing the derivative of the twisted Cayley transform, we have 
$$
\begin{aligned}[t]
(dC_{\kappa})_{\xi}({\alpha}){=}
-(I{-}\kappa){\alpha}\frac{1}{I{+}(I{+}\kappa){\xi}}
{=}
-(I{-}(I{-}\kappa){\xi})\frac{1}{I{+}(I{+}\kappa){\xi}}
(I{+}\kappa){\alpha}\frac{1}{I{+}(I+\kappa){\xi}}.
\end{aligned}
$$
Using the bumping identity  
$\frac{1}{I{+}(I{+}\kappa){\xi}}(I{+}\kappa)=
(I{+}\kappa)\frac{1}{I{+}{\xi}(I{+}\kappa)}$, we easily see 
\begin{equation} 
  \label{eq:keykey}
(dC_{\kappa})_{{\xi}}((I{+}{\xi}(I{+}\kappa))\alpha 
(1{-}(I{-}\kappa){\xi}))=-2\alpha C_{\kappa}({\xi}).  
\end{equation}
Thus, the phase part of the  distribution $D_{\kappa}$ is translated by 
$C_{\kappa}$ into the right invariant tangential  distribution
on $Sp(m;\mathbb C)$.    
Recall that 
$(I{+}{\xi}(I{+}\kappa))\alpha(1{-}(I{-}\kappa){\xi})$ appeared 
already in \eqref{diamond}, $(\diamondsuit)$ as the infinitesimal phase part. 
\begin{prop}\label{infactionquad}
The infinitesimal phase part of the infinitesimal action 
${:}\frac{1}{i\h}\langle{\pmb u}\alpha J,{\pmb u}\rangle{:}_{\kappa}{*_\kappa}$ 
is translated by the twisted Cayley transform 
$C_{\kappa}$ into the right invariant distribution by 
$\alpha\in {\mathfrak{sp}}(m;\mathbb C)$ on $Sp(m;\mathbb C)$.
\end{prop}

As the distribution in the previous section is defined by the infinitesimal 
action of $*$-exponential functions 
$e_*^{t\frac{1}{i\h}\langle{\pmb u}A,{\pmb u}\rangle}$, 
the maximal integral submanifold must be closed by the left 
multiplication $e_*^{t\frac{1}{i\h}\langle{\pmb u}A,{\pmb u}\rangle}{*}$. 

Therefore, the joint object 
$\{\widetilde{\mathcal O}_{\kappa};\kappa\in {sp}(m,{\mathbb C})\}$  
must have certain ``Lie group-like'' properties with manifold-like 
properties patched by $2$-to-$2$ coordinate transformations.  
A general product formula will be given in the next section.

\bigskip

On the other hand, recall the second statement of
Lemma\,\ref{crucial00} gives  
\begin{prop}\label{vacpolar} 
If $\det(I{+}\alpha)=0$, then the infinitesimal 
$*$-action of the quadratic forms 
to the parallel section 
$$
\Big\{\frac{1}{\sqrt{\det(I{-}\alpha\kappa)}}
e^{\frac{1}{i\h}\langle{\pmb u}\frac{1}{I{-}\alpha\kappa}\alpha,{\pmb u}\rangle};
\kappa\in {\mathfrak{sp}}(m;{\mathbb C}) \Big\}
$$
degenerates at every $\kappa$. Namely, 
$(\frac{1}{\sqrt{\det(I{-}\alpha\kappa)}}; 
\frac{1}{I{-}\alpha\kappa}\alpha)\not\in 
{\widetilde{\mathcal O}}_\kappa$ for every 
$\kappa\in {\mathcal D}_{\alpha}$, where ${\mathcal D}_{\alpha}=$\\
$\{\kappa{\in}{\mathfrak{sp}}(m;{\mathbb C}); 
\det(I-\alpha\kappa)\not=0\}$. 
\end{prop}

The most degenerate (minimal rank) orbit will be called  
the orbit of {\bf vacuums}. An example is given by $\alpha=
\begin{bmatrix}
1&0\\
0&{-}1
\end{bmatrix}$ 
in the case $m=1$. These elements will be used to make 
matrix representations in the later chapter.

\subsection{General product formula}\label{productformula}

It is rather hard to construct general manifold theory patched 
together by $2$-to-$2$ coordinate transformations, 
for such objects do not have underlying topological spaces. 
In spite of this, there is no difficulty forming  
a local/classical differential geometry. 
Hence a certain general theory is easy to construct 
for a Lie group-like object by using infinitesimal 
algebraic notions other than point set pictures. 
It is natural to think this gives an intuitive concrete 
object of ``gerbes''. 

Proposition\,\ref{infactionquad} shows that if one concerns 
only the phase parts of the $*$-product, then  
one can compute these via the group structure 
of $Sp(m;{\mathbb C})$ through twisted Cayley transform.

By this observation, we first investigate 
the product $*_0$ defined on 
${\mathbb C}_{\times}\times{\mathcal O}_{0}$ as follows:   
\begin{center}
  $(g; a){*}_{0}(g';b)
=\left(gg'\Big(\frac{1}{\sqrt{\det(1{+}ab)}}\Big); 
C_0^{-1}(C_0(a)C_0(b))\right)$ \\
$C_0^{-1}(C_0(a)C_0(b))=
\frac{1}{1{+}a}(a{+}b)\frac{1}{1{+}ab}(1{+}a)$  
  \end{center}

Note first the following general identity: 
\begin{lem}
$\frac{1}{1{+}a}(a{+}b)\frac{1}{1{+}ab}(1{+}a)\sim
(1{+}b)\frac{1}{1{+}ab}(a{+}b)\frac{1}{1{+}b}$, where 
the reason of the notation $\sim$ instead of $=$ 
is that algebraic calculation such as 
$(1{+}a)\frac{1}{1{+}a}=1$ is used in the proof. 
Hence, one may replace 
$\frac{1}{1+a}(a+b)\frac{1}{1+ab}(1+a)$ by 
$(1{+}b)\frac{1}{1{+}ab}(a{+}b)\frac{1}{1{+}b}$. 
\end{lem}

\noindent
{\bf Proof}\,\, follows immediately by the identity 
$(a{+}b)\big(1{+}\frac{1}{1{+}ab}(a{+}b)\big)=
\big(1{+}(a{+}b)\frac{1}{1{+}ab}\big)(a{+}b)$.
${}$ \hfill $\Box$

\medskip
As far as concerning the phase part $C_0(a)$, and  
forgetting about the singularity, this gives a group which is 
isomorphic to $Sp(m;{\mathbb C})$. 
  
To consider the amplitude part, we define 
$$
(g;\alpha){*}_{\kappa}(g';\beta)\sim 
I_0^{\kappa}
\big(I^0_{\kappa}(g;\alpha){*}_0I^0_{\kappa}(g';\beta)\big).
$$
Since 
$I^0_{\kappa}(g;\alpha)
=\big( g(\det(I{+}\alpha\kappa))^{-\frac{1}{2}};T_{-\kappa}(\alpha)\big)$,
the definition of $C_{\kappa}$ gives that
\begin{equation}
\label{prodkappa}
(g;\alpha)*_{{\kappa}}(g';\beta)
=\Big(gg'(\frac{\det(P{+}Q(I{+}\kappa))}
{\det(P(I{+}\alpha(I{+}\kappa))(I{+}\beta(I{+}\kappa)))})^{\frac{1}{2}}
;C_{\kappa}^{-1}(C_{\kappa}(\alpha)C_{\kappa}(\beta))\Big)    
\end{equation}
where $P=I{+}\alpha(I{-}\kappa)\beta(I{+}\kappa)$,\,\, 
$Q=\alpha{+}\beta{+}2\alpha\kappa\beta$, \,\,
and    
\begin{equation}
C_{\kappa}^{-1}(C_{\kappa}(\alpha)C_{\kappa}(\beta))
=(I{+}\beta(I{+}{\kappa}))\,\frac{1}{P}\,Q \,\,\frac{1}{I{+}(I{+}{\kappa})\beta}.
\end{equation}
We easily see that 
$\det(P{+}Q(I{+}\kappa))
=\det(I{+}\alpha(I{+}\kappa))(I{+}\beta(I{+}\kappa))$.
Hence, the first component of the r.h.s of \eqref{prodkappa} is 
$gg'(\frac{1}{\det P})^{\frac{1}{2}}$. Hence, we obtain 
\begin{equation}
  \label{eq:prodfmla}
(g;\alpha)*_{{\kappa}}(g';\beta){=}
\Big(gg'(\frac{1}{\det P})^{\frac{1}{2}}; 
(I{+}\beta(I{+}{\kappa}))\,\frac{1}{P}\,Q \,\,
\frac{1}{I{+}(I{+}{\kappa})\beta}\Big).
\end{equation}
The product formula \eqref{eq:prodfmla} works only for 
$\alpha, \beta$ such that $\det P\not=0$, and 
$\det(I{+}(I{+}{\kappa})\beta)\not=0$. But, one can 
choose the expression parameter $\kappa$ so that these 
conditions are satisfied. 

\begin{center}
\fbox{\parbox[c]{.85\linewidth}
{${}$\hfill{The product formula is classical}\hfill${}$\\
Singularities move when $\kappa$ moves.  
For every $\forall (g;a),(g';b)$, the product 
$I_0^{\kappa}(g;a)*_{\kappa}I_0^{\kappa}(g';b)$ is 
defined in a generic (open dense) expression $\kappa$. 
By this algebraic trick, the product 
is defined for every pair, which  will be denoted by 
$(g;a)*(g';b)$. \\
It is remarkable that the product formula does not 
involve $\h$.}} 
\end{center}

This follows from that we treat elements written in the form 
$e^{\frac{1}{i\h}Q({\pmb u})}$. Therefore, for elements written 
in the form $e^{Q({\pmb u})}$ the product must be written 
in the form $e^{(i\h)^2R({\pmb u})}$, and hence the product 
formula is real analytic in $\h\geq 0$. Hence one can apply the 
formal associativity Theorem\,\ref{assocthm}. 

\begin{prop}\label{blurred}
Associativity holds with $\pm$ ambiguity. 
\end{prop}

We call this object a {\it blurred Lie group},
and denote it by $(Sp^{(\frac{1}{2})}_{\mathbb C}(m);{*})$. 
This is not an object in which $\pm a$ is treated 
simply as a single point, since they can be locally 
distinguished.

\medskip
For instance, we first note the following: 
\begin{center}
 $(1;0)$ is the identity with respect to 
${*}_{\kappa}$-product for $\forall\kappa$
\end{center}
In particular, in the Weyl ordered expression, 
the integral manifold through $(1;0)$ is 
$$
{\widetilde{\mathcal O}}_0=
\{\sqrt{(\det(1{+}a)};a); a\in {\mathcal O}_0\}.
$$ 

Although the sign ambiguity remains,  
we obtain the following:
\begin{prop}\label{doublevalinv00}
\begin{description}
\item[(a)] If $A, B\in {\widetilde{\mathcal O}}_0$, 
 and if $A{*}_0B$ is defined, then 
$A{*}_0B\in {\widetilde{\mathcal O}}_0$. 
\item[(b)] $(1;0)$ is the identity.
\item[(c)] The inverse $(\sqrt{\det(1{+}a)};a)^{-1}$ 
is given by $(\sqrt{\det(1{-}a)}; -a)$. 
\end{description}
$$
(\sqrt{\det(1{+}a)}; a)*_{0}(\sqrt{\det(1{-}a)}; -a)
=(\sqrt{1}; 0).
$$
\end{prop}
In general, $\sqrt{1}$ must be treated as $\pm 1$, 
but concerning the inverse, this 
should be $1$ by {\bf continuous tracing} from the 
identity $(1;0)$ to the point $(\sqrt{\det(1{+}a)}; a)$.  

\bigskip 
Since ${\widetilde{\mathcal O}}_{\kappa}$ is 
a local Lie group with the identity and 
${\mathfrak{sp}}(m;{\mathbb C})$ as its tangent space, 
we have the following: 
\begin{prop}\label{Liealgebra}  
${\mathfrak{sp}}(m;{\mathbb C})$ is the Lie algebra of 
$Sp^{(\frac{1}{2})}_{\mathbb C}(m)$.
\end{prop}
By $A=\alpha J$, ${\mathfrak{sp}}(m;{\mathbb C})$ is naturally
identified with the space of expression parameters 
${\mathfrak S}(2m)$. 

Since the element $(1; 0)$ may be viewed as the identity of  
the blurred Lie group $Sp^{(\frac{1}{2})}_{\mathbb C}(m)$, 
the tangent space of $Sp^{(\frac{1}{2})}_{\mathbb C}(m)$ 
at $(1;0)$ is naturally identified with 
${\mathfrak{sp}}(m,{\mathbb C})$.

\medskip
In the next section,  we define one parameter subgroups of 
$(Sp^{(\frac{1}{2})}_{\mathbb C}(m);{*})$, and the $*$-exponential 
mapping
$$
\exp_*: {\mathfrak{sp}}(m;{\mathbb C})\to Sp^{(\frac{1}{2})}_{\mathbb C}(m), 
$$
and we show that every one parameter subgroup has discrete branched 
singular points in generic ordered expression.  

\medskip

\subsection{Abstract definition of blurred Lie groups}

Here we give a {\it tentative} abstract definition of blurred Lie
(covering) groups. As we do not have enough concrete examples, it
seems to be too early to give the notion of isomorphisms or the  
general theory. 

%

Let $G$ be a locally simply arcwise connected topological 
group and let 
$\{{\mathcal O}_{\alpha}; \alpha\in I\}$ be 
an open covering of $G$.

\medskip
\noindent
(a) For every $\alpha\in I$, ${\mathcal O}_{\alpha}$ contains 
the identity $e$. ${\mathcal O}_{\alpha}$ is called an 
{\bf abstract expression 

space}, and $\alpha$ is called an expression parameter.

\noindent
(b) For every $\alpha\in I$, ${\mathcal O}_{\alpha}$ is open, 
dense and connected, but it may not be simply connected.

\noindent
(c) For every $\alpha, \beta\in I$, there is a homeomorphism 
$\phi_{\alpha}^{\beta}: {\mathcal O}_{\alpha}\to {\mathcal O}_{\beta}$.

\noindent
(d) For every $g, h\in G$, there is $\alpha\in I$ and 
continuous path  $g(t), h(t)\in G$, $t\in [0,1]$,
such that 

$g(0)=h(0)=e$, $g(1)=g, h(1)=h$ and 
$g(t), h(t), g(t)h(t)$ are in ${\mathcal O}_{\alpha}$ 
for every $t\in [0,1]$.

\medskip
An open covering $\{{\mathcal O}_{\alpha}; \alpha\in I\}$ 
is called {\bf natural covering} of $G$ if it satisfies 
$(a){\sim}(d)$. The condition $(c)$ shows that there is an 
abstract topological space $X$ homeomorphic to every 
${\mathcal O}_{\alpha}$. 
We consider a connected covering space $\pi:\tilde X\to X$.    
This is same to say we consider a connected covering 
$\pi_{\alpha}: \widetilde{\mathcal O}_{\alpha}\to 
{\mathcal  O}_{\alpha}$ 
for each $\alpha$.
It is  easy see that $\pi_{\alpha}^{-1}(e)$ is a 
group given as a quotient group of the fundamental group 
of ${\mathcal O}_{\alpha}$. As $G$ is locally simply connected,
$\pi_{\alpha}^{-1}(e)$ forms a discrete group, and 
 $\phi_{\alpha}^{\beta}$ lifts to an isomorphism  
$\tilde\phi_{\alpha}^{\beta}: \pi^{-1}_{\alpha}(e)\to\pi^{-1}_{\beta}(e)$. 
We denote  $\pi^{-1}_{\alpha}(e)=\varGamma_{\alpha}$, and 
the isomorphism class is denoted by ${\varGamma}$.  

Choose $\tilde{e}_{\alpha}\in\pi_{\alpha}^{-1}(e)$ and 
call $\tilde{e}_{\alpha}$ a tentative identity. For any continuous
path $g(t)$ in ${\mathcal O}_{\alpha}$ such that $g(0)=g(1)=e$, 
the continuous tracing among the set $\pi^{-1}(g(t))$ starting at 
$\tilde{e}_{\alpha}$ gives a group element 
$\gamma \in \varGamma_{\alpha}$.
  
By a standard argument, it is easy to make 
$\widetilde{\mathcal O}_{\alpha}$ a local group 
such that $\pi_{\alpha}$ is a homomorphism: 
We define first that 
$\tilde e_{\alpha}\tilde e_{\alpha}=\tilde e_{\alpha}$.
For paths
$g(t), h(t), g(t)h(t)$ such that they are in ${\mathcal O}_{\alpha}$ for every 
$t\in [0,1]$ and $g(0)=h(0)=e$, 
we define the product 
by a continuous tracing among the set-to-set mapping 
$$
\pi^{-1}_{\alpha}(g(t))\pi^{-1}_{\alpha}(h(t))=\pi^{-1}_{\alpha}(g(t)h(t)).
$$
We set 
${\mathcal O}_{\alpha\beta}=
{\mathcal O}_{\alpha}\cap {\mathcal O}_{\beta},\quad 
{\mathcal O}_{\alpha\beta\gamma}=
{\mathcal O}_{\alpha}\cap {\mathcal O}_{\beta}\cap 
{\mathcal O}_{\gamma}$ for simplicity.

As $G$ is locally simply connected, the full inverse 
$\pi_{\alpha}^{-1}V$ of a simply connected neighborhood 
$V\subset{\mathcal O}_{\alpha}$ of the identity 
$e\in G$ is the disjoint union 
$\coprod_{\lambda}{\tilde V}_{\lambda}$, each member 
${\tilde X}_{\lambda}$ of which is homeomorphic to $V$.     
Moreover $\pi_{\alpha}^{-1}{\mathcal O}_{\alpha\beta}$ is 
also a local group for every $\beta$.

\medskip
\noindent
{\bf Isomorphisms modulo $\varGamma$, Controlled discontinuity} 

For every $\alpha,\beta$, we define the notion of 
``isomorphism'' $I_{\alpha}^{\beta}$ 
of local groups, which corresponds to the notion of 
intertwiners in the previous section:  
$$
\begin{matrix}
\widetilde{\mathcal O}_{\alpha}&\supset&
\pi_{\alpha}^{-1}{\mathcal O}_{\alpha\beta}&
\overset{I_{\alpha}^{\beta}}\longrightarrow&
\pi_{\beta}^{-1}{\mathcal O}_{\beta\alpha}&\subset&
\widetilde{\mathcal O}_{\alpha}\\
\downarrow\pi_{\alpha}&{ }&{ }&{ }&{ }&{ }&
\downarrow\pi_{\beta}\\
{\mathcal O}_{\alpha}&\supset&
{\mathcal O}_{\alpha\beta}&
=\!=\!=&
{\mathcal O}_{\beta\alpha}&\subset&{\mathcal O}_{\beta}
\end{matrix}
$$
such that $I_{\beta}^{\alpha}=(I_{\alpha}^{\beta})^{-1}$, but 
the cocycle condition 
$I_{\alpha}^{\beta}I_{\beta}^{\gamma}I_{\gamma}^{\alpha}=1$
is not required for ${\mathcal O}_{\alpha\beta\gamma}$. 

Since the correspondence $I_{\alpha}^{\beta}$ 
does not make sense as a point set mapping, we should 
be careful for the definition.

Note that $I_{\alpha}^{\beta}$ is a collection of $1$-to-$1$ mapping 
$I_{\alpha}^{\beta}(g): \pi_{\alpha}^{-1}(g)\to
\pi_{\beta}^{-1}(g)$ for every 
$g\in{\mathcal O}_{\alpha\beta}=
{\mathcal O}_{\beta\alpha}$, which may not be continuous in 
$g$.   
  
For each $g$ there is a neighborhood 
$V_g$ of the identity $e$ such that 
$V_gg\subset {\mathcal O}_{\alpha\beta}$ and  the local trivialization  
$\pi_{\alpha}^{-1}(V_gg)=V_gg{\times}\pi_{\alpha}^{-1}(g)$.  
Thus $I_{\alpha}^{\beta}(g)$ extends to 
the correspondence  
$$
{\tilde I}_{\alpha}^{\beta}(h,g): 
\pi_{\alpha}^{-1}(hg)\to
\pi_{\beta}^{-1}(hg),\quad h\in V_g 
$$
which commutes with the local deck transformations. 

\begin{definition}
The collection $I_{\alpha}^{\beta}{=}\{I_{\alpha}^{\beta}(g); g\in{\mathcal O}_{\alpha\beta}\}$ 
is called an isomorphism 
modulo $\varGamma$, if the product 
$I_{\beta}^{\alpha}(hg){\tilde I}_{\alpha}^{\beta}(h,g)$ 
is in the group $\varGamma$ for every 
$g\in {\mathcal O}_{\alpha\beta}$ and $h\in V_g$. $($It follows the
continuity of $I_{\alpha}^{\beta}(hg)$ w.r.t. $h$.$)$
\end{definition}
The condition given by this definition means roughly 
that $I_{\alpha}^{\beta}(g)$ has discontinuity in $g$ only 
in the group $\varGamma$.

\medskip
${\widetilde G}=\{\widetilde{\mathcal O}_{\alpha}, 
\pi_{\alpha}, I_{\alpha}^{\beta}; \alpha,\beta\in I\}$ 
is called a {\bf blurred covering group} of $G$ 
if each $\widetilde{\mathcal O}_{\alpha}$ is a covering local 
group of ${\mathcal O}_{\alpha}$, where
$\{{\mathcal O}_{\alpha};\alpha\in I\}$
is a natural open covering of a locally simply arcwise 
connected topological group $G$ and   
$I_{\alpha}^{\beta}$ are isomorphisms 
modulo $\varGamma$. 

\medskip
Because of the failure of the cocycle condition, this 
object does neither form a covering group, nor a topological 
point set. However, this object looks like a covering group. 

\medskip 
For $g$, let $I_g$ be the set of expression parameters 
involving $g$; 
$I_g=\{\alpha\in I; {\mathcal O}_{\alpha}\ni g\}$. 
For every  $\alpha\in I(g,h,gh)=I_g\cap I_h\cap I_{gh}$,  
we easily see that 
$\pi_{\alpha}^{-1}(g)\pi_{\alpha}^{-1}(h)=
\pi_{\alpha}^{-1}(gh)$. In general, this is viewed as 
set-to-set correspondence, but if $g$ or $h$ is in a 
small neighborhood of the identity, we can make 
these correspondence a genuine point set mapping. 
Hence, we have the notion of indefinite small action 
or ``infinitesimal left/right action''  
of small elements to the object. This corresponds to  
the infinitesimal action 
$w_*^2{*}$ or ${*}w_*^2$ in the previous section.

\medskip
Next, we choose an element 
$\tilde{e}_{\alpha}\in\pi_{\alpha}^{-1}(e)$, and call it  
a local identity. On the other hand,  
$\pi_{\alpha}^{-1}(e)$ is called {\it the set of local identities} 
of ${\widetilde{G}}$. 
The failure of the cocycle condition gives that 
${\mathfrak M}_{\alpha}\tilde{e}_{\alpha}$ may not be a 
single point set, but forms a discrete abelian group. 
Hence an identity of our object is always a 
{\it local identity}.

Since $G$ is a locally simply connected, there is an open 
simply connected neighborhood $V_{\beta}$ of $e$ contained 
in ${\mathcal O}_{\beta}$. Hence, there is the unique lift 
$\tilde{V}_{\beta}$ through $\tilde{e}_{\beta}$. 
Setting 
$\tilde{V}_{\beta\gamma}=
\tilde{V}_{\beta}\cap\tilde{V}_{\gamma}$ e.t.c.,
 we see easily 
$I_{\beta}^{\gamma}(\tilde{V}_{\beta\gamma})
=\tilde{V}_{\gamma\beta}.$

\medskip
The $\{{\tilde g}_{\alpha}\in \tilde{\mathcal O}_{\alpha};\alpha\in I\}$
may be viewed as an element of ${\widetilde{G}}$ if 
$I_{\alpha}^{\beta}{\tilde g}_{\alpha}={\tilde g}_{\beta}$, 
but this is not a single point set by the same reason. 
In spite of this, one can distinguish individual points within 
a small local area.

\section{Star-exponential functions of quadratic forms}
\label{star-exponential2}

For an element $H_*$ of the algebra, we define the $*$-exponential function $e_*^{tH_*}$ as the 
real analytic solution of 
\begin{equation}\label{leftevol}
\frac{d}{dt}f_*(t)=H_*{*}f_*(t), \quad f(0){=}1,
\end{equation} 
provided the solution exists.   
More precisely, we define $e_*^{tH_*}$ as the family 
$\{f_t(K)\}$ of univalent solutions of the evolution equation 
\begin{equation}\label{starexp}
\frac{d}{dt}f_t(K)={:}H_*{:}_{_K}{*_{_K}}f_t(K), 
\end{equation}
with the initial condition $f_0(K)=1$. 
We think of $f_t(K)$ as the $K$-ordered expression of 
$e_*^{tH_*}$, and denote it by ${:}e_*^{tH_*}{:}_{_K}=f_t(K)$. 
Uniqueness is ensured if we consider 
only real analytic solutions. 
\eqref{leftevol} is called the {\it left evolution  equation}. 
The {\it right evolution equation} is defined similarly, but this is
not used except when otherwise mentioned.  

\medskip
If $H_*$ is a $*$-polynomial,   
\eqref{leftevol} can be rewritten as a partial differential 
evolution equation.    
If the equation 
$\frac{d}{dt}{:}f_*(t){:}_{_K}={:}H_*{:}{*}f_*(t){:}_{_K}$ 
has a unique solution for the initial element 
$f_*(0){=}g_*$, then 
the solution will be denoted by ${:}e_*^{tH_*}{*}g_*{:}_{_K}$.

As it was seen in 
\S\,\ref{Expinter}, a star exponential function 
$e_*^{\frac{1}{i\h}\langle{\pmb\xi}, {\pmb u}\rangle}$ 
of a linear form $\langle{\pmb\xi}, {\pmb u}\rangle$, 
 was welldefined as the family 
$\{e^{\frac{1}{4i\h}
\langle{\pmb\xi}K,{\pmb\xi}\rangle}
e^{\frac{1}{i\h}\langle{\pmb\xi}, {\pmb u}\rangle}\}$ for all 
$K{\in}{\mathfrak S}(n)$.  
Provided ${:}e_*^{sH_*}{:}_{_{K}}$ exists for every 
$s{\in}{\mathbb C}$,  they form a 
complex one parameter subgroup, for the exponential 
law holds by the uniqueness of real analytic solutions. 

\medskip

Here we give several general remarks on $*$-exponential functions of
quadratic forms.

\noindent
(1) If $H_*$ is a quadratic form,  ${:}e_*^{sH_*}{:}_{_{K}}$ 
is defined with double branched singularities on a
discrete set 

(c.f. \eqref{fundamenta2}). Thus, we have to 
prepare two sheets to consider ${:}e_*^{sH_*}{:}_{_{K}}$ for 
$s\in {\mathbb C}$. But, the origin $0$ of 

another sheet 
does not correspond to $1$, but $-1$.

\medskip
\noindent
(1.1) In general, there is no reflection symmetry in $s$ for the 
domain of existence of the solution of 

\eqref{starexp}. That is, the existence of ${:}e_*^{sH_*}{:}_{_{K}}$ does not 
necessarily imply that ${:}e_*^{-sH_*}{:}_{_{K}}$ exists: e.g.
$$
{:}e_*^{t\frac{1}{i\h}(u^2+v^2)}{:}_{_I}= 
\frac{1}{\cos t{-}\sin t}
e^{\frac{1}{i\h}\frac{\sin t}{\cos t{-}\sin t}(u^2+v^2)}\quad cf.  
\eqref{Delta}.
$$

\noindent
(1.2)  Moreover ${:}e_*^{sH_*}{:}_{_{K}}$ is double-valued 
holomorphic function in $K$ on an open connected dense 

domain,  i.e. double-valued holomorphic parallel section.

\medskip
\noindent
(2) If $H_*$, $G_*$ are quadratic forms, then the product 
${:}e_*^{tH_*}{*}e_*^{G_*}{:}_{_K}$ is defined as a double-valued 

holomorphic function of $(t,K)$ defined on 
an open connected dense 
domain containing $(0,0)$. 

\bigskip
For a given $K$, suppose that
\eqref{starexp} has  real analytic solutions in $t$ on some domain 
$D(K)$ including $0$ for the initial functions $1$  and $g$. 
We denote the solution of \eqref{starexp} 
with initial function $g$ by 
\begin{equation}
{:}e_*^{tH_*}{:}_{_K}{*_{_K}}g,\quad t{\in}D(K). 
\end{equation} 

\begin{prop}\label{exppoly}
If $H_*$ is a polynomial and ${:}e_*^{tH_*}{:}_{_K}$ is 
defined on a domain $D(K)$, then 
${:}e_*^{tH_*}{:}_{_K}{*_{_K}}p(\pmb u)$ is defined
for every polynomial $p(\pmb u)$ on the same domain $D(K)$.

If $p(\pmb u)=\sum A_\alpha(s){\pmb u}^{\alpha}$ 
is a polynomial whose coefficients depend smoothly on $s$, 
then the formula 
$$
\partial_s^{\ell}{:}e_*^{tH_*}{:}_{_K}{*_{_K}}p(\pmb u)=
{:}e_*^{tH_*}{:}_{_K}{*_{_K}}\partial_{s}^{\ell}p(\pmb u)
$$
holds for every $\ell$.
\end{prop}

\noindent
{\bf Proof}\, Multiplying 
the 
defining equation \eqref{starexp} by ${*}p(\pmb u)$ and applying the 
associativity in Proposition\,\ref{extholom}, we have  
\begin{equation}\label{eq-11}
\frac{d}{dt}f_t(K){*}p(\pmb u)=
{:}H_*{:}_{_K}{*_{_K}}(f_t(K){*}p(\pmb u)),
\quad f_0(K)= 1. 
\end{equation}
Since $f_t(K){*}p(\pmb u)$ is a real analytic solution, 
this is written in our notation as  $e_*^{tH_*}{*}p(\pmb u)$.
Applying $\partial^{\ell}_s$ to \eqref{eq-11} gives 
the second assertion by a similar argument. 
 \hfill $\Box$

\bigskip
For a quadratic form 
$\langle{\pmb u}A, {\pmb u}\rangle_*$,  
the $*$-exponential function 
$e_*^{\frac{t}{i\h}\langle{\pmb u}A, {\pmb u}\rangle_*}$ 
is given in a concrete form. 
For every $\alpha\in {\mathfrak{sp}}(m,{\mathbb C})$, we consider first the 
one parameter subgroup $e^{-2t\alpha}$ of $Sp(m,{\mathbb C})$, and 
consider the inverse image of twisted Cayley transform 
$C^{-1}_{\kappa}(e^{-2t\alpha})$:
We set 
\begin{equation}
\label{eq:kyakukei}
C_{\kappa}^{-1}(e^{-2t\alpha})=
\frac{1}{(I{-}{\kappa}){+}
e^{-2t\alpha}(I{+}{\kappa})}(I{-}e^{-2t\alpha})\\
=\frac{1}{\cosh t\alpha{-}(\sinh t\alpha){\kappa}}\sinh t\alpha.
\end{equation}
where $\frac{1}{X}$ stands for $X^{-1}$. 

The exponential function must lie on the integral manifold 
$\widetilde{\mathcal O}_{\kappa}$ through 
$(1;0)$, and the point of the 
integral manifold is determined by its phase part. Hence 
we have 
\begin{equation}
  \label{eq:starexp}
\exp_{*_{{\kappa}}}\frac{1}{i\h}t{\alpha}=
\Big(\big(\det(I+C_{\kappa}^{-1}(e^{-2t{\alpha}})
(I{{+}}{\kappa}))\big)^{\frac{1}{2}}
; C_{\kappa}^{-1}(e^{-2t{\alpha}})\Big).  
\end{equation}
In the original notation, we see 
$e_*^{s\frac{1}{i\h}\langle{\pmb u}
({\alpha}J),{\pmb u}\rangle_*}$
as follows by setting $\kappa{=}J\!K$: 
\begin{equation*} 
{:}e_*^{s\frac{1}{i\h}\langle{\pmb u}
({\alpha}J),{\pmb u}\rangle_*}{:}_{_{\kappa}}=
\big(\det(I{+}C_{\kappa}^{-1}
(e^{-2s{\alpha}})(I{{+}}{\kappa}))\big)^{\frac{1}{2}}
e^{\frac{1}{i\h}\langle{\pmb u}
(C_{\kappa}^{-1}(e^{-2s{\alpha}})J),{\pmb u}\rangle}.  
\end{equation*}
More precisely, for every $\alpha{\in}{\mathfrak{sp}}(m,{\mathbb C})$, 
the $K$-ordered expression of the $*$-exponential
function is given as follows:  
(Cf.\cite{ommy} \,\cite{OMMY6}\, \cite{OMMY7}
 for special cases.) 
\begin{equation}\label{fundamenta1}
\begin{aligned}
{:}e_*^{\frac{t}{i\h}
\langle{\pmb u}(\alpha J), {\pmb u}\rangle_*}{:}_{_K}
{=}
\frac{2^m}{\sqrt{\det(I{-}{\kappa}{+}e^{-2t\alpha}(I{+}\kappa))}}
e^{\frac{1}{i\h}\langle{\pmb u}
\frac{1}{I{-}{\kappa}{+}e^{-2t\alpha}(I{+}\kappa)}
(I{-}e^{-2t\alpha})J,{\pmb u}\rangle}
\end{aligned}
\end{equation}
where $\kappa{=}J\!K$. It is not hard to see that this is the 
real analytic solution of \eqref{starexp}. By this concrete form 
we see this is an element of ${\mathcal E}_{2+}({\mathbb C}^{2m})$ 
whenever this is defined. But it is remarkable that 
${:}e_*^{\frac{t}{i\h}
\langle{\pmb u}(\alpha J), {\pmb u}\rangle_*}{:}_{_K}$ remains 
in the space ${\mathbb C}e^{Q(u,v)}$ given in Theorem\,\ref{meromuseful}.

\subsection{Adjoint action to $V_{2m}$.}\label{adjoint}

$Sp_{\mathbb C}^{(\frac{1}{2})}(m)$ is not a genuine 
Lie group, as elements have double-valued nature in general, 
and it looks something like a double covering group of 
$Sp(m,{\mathbb C})$. But, because of this reason,  
$Sp_{\mathbb C}^{(\frac{1}{2})}(m)$ contains 
several genuine groups such as the metaplectic group which is not
contained in $Sp(m,{\mathbb C})$. Moreover,  
$Sp_{\mathbb C}^{(\frac{1}{2})}(m)$ contains 
$Spin(m)$ under the special ordered expression $K_s$ (cf. \cite{OMMY4}.)
In the case $m=1$, we have seen in \cite{OMMY3} some basic properties of Jacobi's 
$\theta$-functions by means of $*$-exponential functions of quadratic forms. 
\medskip

To avoid the vague issue of sign ambiguity, we first consider
adjoint representations of 
$Sp_{\mathbb C}^{(\frac{1}{2})}(m)$ on the linear space of 
generators, for the sign ambiguity 
disappears in adjoint representations, and it is 
independent of the expression parameter $K$.

For $\alpha\in {\mathfrak{sp}}(m;\mathbb C)$,  
the quadratic form 
$\langle{\pmb u}(\alpha J),{\pmb u}\rangle$ 
acts on the space of linear functions: 
$$
[\langle{\pmb u}(\frac{1}{2i\h}\alpha J),{\pmb u}\rangle, 
 \langle{\pmb a},{\pmb u}\rangle]={-}\langle{\pmb a\alpha},{\pmb u}\rangle.
$$ 

Hence, the Lie algebra ${\mathfrak{sp}}(m;\mathbb C)$ is obtained by 
the adjoint representation of quadratic forms 
$$
{\rm{ad}}(\langle{\pmb u}(\frac{1}{2i\h}\alpha J),{\pmb u}\rangle)
={-}\alpha \in {\mathfrak{sp}}(m;\mathbb C).
$$
It follows that for every $*$-function such as $*$-polynomials or 
$f_*(\pmb u)=\int_{{\mathbb R}^n}{\hat f}(\xi)
e_*^{\frac{1}{i\h}\langle\xi,{\pmb u}\rangle}\dbar\xi$, 
$$
e^{t{\rm{ad}}
(\langle{\pmb u}
(\frac{1}{2i\h}\alpha J),{\pmb u}\rangle)}f_*({\pmb u})
=f_*(e^{{-}t\alpha}{\pmb u}),
$$
where $e^{{-}t\alpha}$ is a linear transformation 
$e^{-t\alpha}\in Sp(m;\mathbb C)$.

\medskip 
A concrete form for the case $m=1$ is given 
by using the transposed matrices as follows, 
\begin{equation}
  \label{eq:ad-quad}
{\text{ad}}(\frac{i}{2\hbar}(au^2+bv^2+2cuv))
\begin{bmatrix}
u\\v  
\end{bmatrix}
=
\begin{bmatrix}
-c& -b\\
 a& c   
\end{bmatrix}
\begin{bmatrix}
u\\v  
\end{bmatrix}
\end{equation}

Let $V_{2m}{=}
\{\langle{\pmb\xi},{\pmb u}\rangle;
{\pmb\xi}{\in}{\mathbb C}^{2m}\}$. 
For every quadratic form 
$\frac{1}{2i\h}\langle{\pmb u}A,{\pmb u}\rangle_*$, 
${\rm{ad}}(\frac{1}{2i\h}\langle{\pmb u}A,{\pmb u}\rangle_*)$ 
is welldefined as a linear mapping independent of expression parameters.  
$$
{\rm{ad}}(\frac{1}{2i\h}\langle{\pmb u}A,{\pmb u}\rangle_*):
V_{2m}\to V_{2m},\quad 
{\rm{ad}}(\frac{1}{2i\h}\langle{\pmb u}A,{\pmb u}\rangle_*):
H\!ol({\mathbb C}^{2m})\to H\!ol({\mathbb C}^{2m})
$$ 
It is easy to see that
${\text{ad}}(\langle{\pmb u}
(\frac{1}{2i\h}\alpha J),{\pmb u}\rangle)
{=}{-}\alpha{\in}{\mathfrak{sp}}(m,{\mathbb C})$, hence it extends
 as a ${*}$-derivation 
$$
{\rm{ad}}(\frac{1}{2i\h}\langle{\pmb u}A,{\pmb u}\rangle_*):
({\mathcal E}_2({\mathbb C}^{2m}),{*})\to
({\mathcal E}_2({\mathbb C}^{2m}),{*}). 
$$ 
Linear algebra on finite dimensional vector space 
gives linear isomorphisms
$$
e^{{\rm{ad}}(\frac{1}{2i\h}\langle{\pmb u}A,{\pmb u}\rangle_*)}:
V_{2m}\to V_{2m},\quad 
e^{{\rm{ad}}(\frac{1}{2i\h}\langle{\pmb u}A,{\pmb u}\rangle_*)}:
H\!ol({\mathbb C}^{2m})\to H\!ol({\mathbb C}^{2m})
$$ 
and a ${*}$-isomorphism 
$$
e^{{\rm{ad}}(\frac{1}{2i\h}\langle{\pmb u}A,{\pmb u}\rangle_*)}:
({\mathcal E}_2({\mathbb C}^{2m}),{*})\to
({\mathcal E}_2({\mathbb C}^{2m}),{*}). 
$$

Set $A{=}\alpha J$. Since 
${:}e_*^{\frac{t}{2i\h}
\langle{\pmb u}A,{\pmb u}\rangle_*}{:}_{_K}$ 
is defined as a multi-valued holomorphic 
mapping from an open connected dense domain $D$ containing the origin into 
${\mathcal E}_{2+}({\mathbb C}^{2m})$. and 
the first associativity Theorem\,\ref{assocthm} applied 
to $t{=}\h s$  shows the following:

\begin{lem}\label{lem15}
Both sides are well-defined and associativity 
$$
{:}(e_*^{\frac{t}{2i\h}
\langle{\pmb u}A,{\pmb u}\rangle_*}
{*}\langle{\pmb\xi},{\pmb u}\rangle)
{*}e_*^{{-}\frac{t}{2i\h}
\langle{\pmb u}A,{\pmb u}\rangle_*}{:}_{_K}{=}
{:}e_*^{\frac{t}{2i\h}
\langle{\pmb u}A,{\pmb u}\rangle_*}
{*}(\langle{\pmb\xi},{\pmb u}\rangle
{*}e_*^{{-}\frac{t}{2i\h}
\langle{\pmb u}A,{\pmb u}\rangle_*})
{:}_{_K}
$$
holds for every $t{\in}D$. 
\end{lem}
Differentiating the identity of Lemma\,\ref{lem15} 
by using Theorem\,\ref{assocthm} 
several times, gives that 
\begin{equation*}
\frac{d}{dt}{:}e_*^{\frac{t}{2i\h}
\langle{\pmb u}A,{\pmb u}\rangle_*}
{*}\langle{\pmb\xi},{\pmb u}\rangle
{*}e_*^{{-}\frac{t}{2i\h}
\langle{\pmb u}A,{\pmb u}\rangle_*}{:}_{_K}{=}
{:}{\rm{ad}}(\frac{1}{2i\h}\langle{\pmb u}A,{\pmb u}\rangle_*)
(e_*^{\frac{t}{2i\h}
\langle{\pmb u}A,{\pmb u}\rangle_*}
{*}\langle{\pmb\xi},{\pmb u}\rangle
{*}e_*^{{-}\frac{t}{2i\h}\langle{\pmb u}A,{\pmb u}\rangle_*})
{:}_{_K}.
\end{equation*}

Uniqueness of the real analytic solution gives 
that the matrix obtained is independent of expression parameters:
\begin{equation*}
{:}e_*^{\frac{t}{2i\h}\langle{\pmb u}A,{\pmb u}\rangle_*}
{*}\langle{\pmb\xi},{\pmb u}\rangle
{*}e_*^{{-}\frac{t}{2i\h}
\langle{\pmb u}A,{\pmb u}\rangle_*}{:}_{_K}{=}
e^{t{\rm{ad}}(\frac{1}{2i\h}\langle{\pmb u}A,{\pmb u}\rangle_*)}
\langle{\pmb\xi},{\pmb u}\rangle{=}
\langle{\pmb\xi}e^{{-}t\alpha},{\pmb u}\rangle 
\,\,({=}\langle{\pmb\xi}, e^{{-}t\alpha}{\pmb u}\rangle), 
\end{equation*}
where $A{=}\alpha J, \alpha{\in}{sp}(m,{\mathbb C})$. 

\begin{thm}\label{The basic formula} 
If ${:}e_*^{\frac{t}{2i\h}
\langle{\pmb u}A,{\pmb u}\rangle_*}{:}_{_K}$ is defined, then
$$
e_*^{t\langle{\pmb u}(\frac{1}{2i\h}\alpha J),{\pmb u}\rangle}
    *{\langle{\pmb a},{\pmb u}\rangle}{*}
e_*^{-t\langle{\pmb u}(\frac{1}{2i\h}\alpha J),{\pmb u}\rangle}
=\langle {\pmb a}e^{-t\alpha},{\pmb u}\rangle.
$$
\end{thm}
The proof is based on the fact that 
$e_*^{\frac{t}{2i\h}
\langle{\pmb u}A,{\pmb u}\rangle_*}{*}
{\langle{\pmb a},{\pmb u}\rangle}{*}
e_*^{-t\langle{\pmb u}(\frac{1}{2i\h}\alpha J),{\pmb u}\rangle}$ 
is defined and real analytic on an open dense connected domain of $t$ containing 
$0$. Hence, one may replace 
${\langle{\pmb a},{\pmb u}\rangle}$ by any polynomial.

Since $\{e^{\alpha}, \alpha{\in}\mathfrak{sp}(m,{\mathbb C})\}$ 
generates $Sp(m,{\mathbb C})$, 
the following is easy to see:
\begin{prop}\label{ajoint222} 
As linear transformation of $V_{2m}$, we have 
${\rm{Ad}}(e_*^{t\langle{\pmb u}
(\frac{1}{2i\h}\alpha J),{\pmb u}\rangle})
=e^{t{\rm{ad}}(\langle{\pmb u}
(\frac{1}{2i\h}\alpha J),{\pmb u}\rangle)}.$
Hence,   
${\rm{Ad}}(e_*^{t\langle{\pmb u}
(\frac{1}{2i\h}\alpha J),{\pmb u}\rangle})$ has no singular point and 
generates the group $Sp(m,{\mathbb C})$.
\end{prop}

This identity holds in spite of the ambiguity of the amplitude of
$e_*^{t\langle{\pmb u}(\frac{1}{2i\h}\alpha J),{\pmb u}\rangle}$, 
because the ambiguity of amplitude disappears in the adjoint
formula. Hence, 
\begin{center}
\fbox{\parbox[c]{.7\linewidth}
{${}$\hfill${\rm{Ad}}(e_*^{t\langle{\pmb u}(\frac{1}{2i\h}\alpha J),{\pmb u}\rangle})$ 
generates the group $Sp(m,{\mathbb C})$.\hfill${}$\\
${\rm{Ad}}: Sp^{(\frac{1}{2})}_{\mathbb C}(m)\to Sp(m,{\mathbb C})$ is
a $2$-to-$1$ ``surjective homomorphism''.}}    
\end{center}
The blurred Lie group $Sp^{(\frac{1}{2})}_{\mathbb C}(m)$ 
generated by 
$e_*^{t\langle{\pmb u}(\frac{1}{i\h}\alpha J),{\pmb u}\rangle}$ 
looks like a double covering group of $Sp(m,{\mathbb C})$ 
which is known to be simply connected. 

\subsection{Several point set pictures for blurred subgroups} 
Recall the surjective ``homomorphism'' 
$$
{\rm{Ad}}: Sp^{(\frac{1}{2})}_{\mathbb C}(m)\to Sp(m,\mathbb C). 
$$
For every subgroup $G$ of $Sp(m,{\mathbb C})$, the full inverse  
 ${\rm{Ad}}^{-1}G$ may be viewed as a {\it blurred} covering of $G$. 
However, it is often possible that 
${\rm{Ad}}^{-1}G$ is a genuine Lie group under a suitable expression parameter.

\medskip 
Suppose we have a subgroup $G$ of $Sp(m,{\mathbb C})$. 
Take a simple open covering 
$\{V_{\alpha}\}_\alpha$ of $Sp(m,{\mathbb C})$ such that 
$\{V_{\alpha}{\cap}\,G\}_\alpha$ is also a simple open 
covering of $G$, and each $V_\alpha$ is contained in some 
$C_{\kappa}({\mathcal O}_{\kappa})$. (Cf. Lemma\,\ref{crucial00}.) 
For every $\alpha, \beta, \gamma$ we denote 
$$
V_{\alpha\beta}{\cap}\,G{=}
V_{\alpha}{\cap}V_{\beta}{\cap}G, \quad
V_{\alpha\beta\gamma}{\cap}\,G{=}
V_{\alpha}{\cap}V_{\beta}{\cap}V_{\gamma}
{\cap}\,G,\quad e.t.c.
$$ 
Although $Sp^{(\frac{1}{2})}_{\mathbb C}(m)$ is a blurred 
double covering, the simplicity of $V_{\alpha}{\cap}\,G$ 
ensures that \\ 
${\rm{Ad}}^{-1}(V_{\alpha}{\cap}\,G)
{=}(V_{\alpha}{\cap}\,G, {\mathbb Z}_2)$, and the patching 
diffeomorphisms
$\phi_{\alpha\beta}: V_{\alpha\beta}{\cap}\,G\to{\mathbb Z}_2 $ 
satisfies the cocycle condition 
$$
\phi_{\alpha\beta}\phi_{\beta\gamma}\phi_{\gamma\alpha}{=}\pm 1
$$
as $2$-to-$2$ mappings. These 
$2$-to-$2$ patching diffeomorphisms give on each 
$(\alpha,\beta)$ 
two choices of  patching diffeomorphisms, say 
$\pm\phi_{\alpha\beta}$. 
In a certain case, we can select one of these sign to 
clear the cocycle condition to obtain a genuine subset.  

\begin{thm}\label{dblecover}
For a connected subgroup $G$ of $Sp(m,{\mathbb C})$, if  
we can select patching diffeomorphisms so that 
they satisfies the cocycle condition, then 
there is a group $\tilde G$ contained in 
${\rm{Ad}}^{-1}(G)$ such that ${\rm{Ad}}:
\tilde G \to G$ is a surjective homomorphism.
\end{thm}

\noindent
{\bf Proof}. Since patching diffeomorphisms are so
adjusted that the cocycle condition is satisfied, 
we have a genuine point set. But it is 
easy to see that these satisfies the condition of 
covering group of $G$. Note that such a point 
set picture may not be unique.  \hfill ${\square}$

\bigskip
We have already in \cite{OMMY4} an example that 
$Sp^{(\frac{1}{2})}_{\mathbb C}(m)$ contains $Spin(m)$ under a special
ordered expression $K_s$. 
Here, we give a simplest example. Note that 
$$
\frac{i}{2\h}[\sum(u_i^2+v_i^2), 
\binom{\pmb u}{\pmb v}]=
\begin{bmatrix}
  0&-I_m\\
  I_m& 0\\
\end{bmatrix}\binom{\pmb u}{\pmb v}.
$$
We see that $Sp(m, \mathbb C)$ contains  $U(1)$ in the form 
\begin{equation}
  \label{eq:U(1)}
U(1)=
\left\{\begin{bmatrix}
\cos\theta I_m&-\sin\theta I_m\\
\sin\theta I_m&\,\,\,\cos\theta I_m   
\end{bmatrix}; \theta \in\mathbb R \right\}.   
\end{equation} 
Hence we see that
$\{{\rm{Ad}}(e_{*}^{\frac{i\theta}{2\h}\sum_m(u_i^2+v_i^2)})\}=U(1)$
and the full inverse ${\tilde U}(1)={\rm{Ad}}^{-1}(U(1))$ is a 
double covering group of $U(1)\subset Sp(m,{\mathbb C})$. 
In the next section, we see that there are open subsets 
${\mathfrak  K}_1$, ${\mathfrak  K}_2$ of expression
parameters such that 
$$
{:}{\tilde U}(1){:}_{_K}=
\left\{
\begin{matrix}
U(1){\times}{\mathbb Z}_2 & K\in {\mathfrak  K}_1\\ 
\text{the connected double cover of }\,\,U(1)&K\in {\mathfrak  K}_2
\end{matrix}
\right.
$$
Then, noting that $Sp(m,\mathbb R)\supset U(1)$, the full inverse   
${\rm{Ad}}^{-1}(Sp(m,\mathbb R))$ is a genuine connected double covering 
group of $Sp(m,\mathbb R)$ under the $K$-ordered expression such that
$K\in {\mathfrak  K}_2$. This is called the  
{\bf  metaplectic group} and denoted by $M\!p(m)$.  
The metaplectic group is the connected double covering group of 
$Sp(m,{\mathbb R})$, 
which appears naturally as patching diffeomorphisms of 
the symbols of the group of invertible 
Fourier integral operators. 
It is known that $M\!p(m)$ has no complexification as 
Lie groups. Thus $Sp^{(\frac{1}{2})}_{\mathbb C}(m)$ 
is viewed as its complexification  as blurred Lie groups.

\bigskip

For concrete computation, note that the adjoint mapping ${\rm{Ad}}$ gives 
$$
e_*^{\frac{r}{\h}u^2}\rightarrow 
\left[\begin{array}{cc}
 1,& 0\\
 -ri,& 1
\end{array}\right], \quad
e_*^{\frac{s}{\h}iuv}\rightarrow
\left[\begin{array}{cc}
 e^{-s},& 0\\
 0,& e^{s}
\end{array}\right], \quad 
e_*^{\frac{t}{\h}v^2}\rightarrow
\left[\begin{array}{cc}
 1,& ti\\
 0,& 1
\end{array}\right] 
$$
$$
e_*^{\frac{\theta}{2\h}(u^2+v^2)}\rightarrow
\left[\begin{array}{cc}
 \cosh\theta,& i\sinh\theta\\
 -i\sinh\theta,& \cosh\theta
\end{array}\right], \quad 
e_*^{\frac{s}{2\h}(u^2-v^2)}\rightarrow
\left[\begin{array}{cc}
 \cos s, & i\sin s\\
 i\sin s,& \cos s
\end{array}\right] 
$$
In particular, $Sp(1,\mathbb C)=S\!L(2,\mathbb C)$ contains
$S\!L(2,\mathbb R)$ and  
$$
\begin{aligned}
SU(2)=&\Big\{
\begin{bmatrix}
  \alpha & \beta\\
  -\bar\beta &\bar\alpha\\
\end{bmatrix}; 
|\alpha|^2{+}|\beta|^2=1\Big\}\cong 
S^3, \\
SU(1,1)=&\Big\{
\begin{bmatrix}
 \alpha & \beta\\
 \bar\beta &\bar\alpha\\
\end{bmatrix}; 
|\alpha|^2{-}|\beta|^2=1\Big\}.
\end{aligned}
$$ 
Through these subgroups we take the full inverse ${\rm{Ad}}^{-1}(G)$. 
Hence, for $S\!L(2,\mathbb R)$ we see 
$$
\begin{aligned}
\{{\rm{Ad}}&(e_*^{\frac{r}{2\h}i(u^2+v^2)}
 *e_*^{\frac{s}{\h}iuv}*e_*^{\frac{ti}{\h}u^2}); r, s, t\in \Bbb R\}\\
=&\left\{
\left[\begin{array}{cc}
 \cos r, & -\sin r\\
 \sin r,& \cos r
\end{array}\right]
\left[\begin{array}{cc}
 e^{-s},& 0\\
 0,& e^{s}
\end{array}\right]
\left[\begin{array}{cc}
 1,& 0\\
 t,& 1
\end{array}\right]; r,s,t\in \Bbb R \right\}
\end{aligned}
$$
Hence, ${\rm{Ad}}^{-1}(S\!L(2,{\mathbb R}))$ is the connected double
covering  of $S\!L(2,{\mathbb R} )$ under the 
$K$-ordered expression such that $K{\in}{\mathfrak K}_2$. 
 
\medskip
Similarly, under the 
$K$-ordered expression such that $K{\in}{\mathfrak K}_2$, we see 
${\widetilde{SU}}(1,1)={\rm{Ad}}^{-1}(SU(1,1))$ is the connected 
double covering group of $SU(1,1)$.

\medskip
Next, consider
$$
{\widetilde{SU}}(2)={\rm{Ad}}^{-1}(SU(2)). 
$$
Indeed, this is the simplest toy model of 
blurred covering group.  
More precisely, decompose $SU(2)$ as 
$$
\left\{
\left[\begin{array}{cc}
 \cos\theta, & -\sin\theta e^{i\psi}\\
 \sin\theta e^{-i\psi},& \cos\theta
\end{array}\right]
\left[\begin{array}{cc}
 e^{i\rho},& 0\\
 0,& e^{-i\rho}
\end{array}\right]
; \theta, \psi, \rho\in \Bbb R, \,\,|\theta|<\frac{\pi}{2}\right\}
$$
with singular points at $\cos\theta=0$, where $\theta, \psi$ may be
viewed as the latitude and longitude respectively. 
Under a suitable expression parameter, we have a double covering 
group of the group $\{e^{i\rho}\}$. Hence, we have a covering space by  
replacing $\rho$ by $\rho/2$ for each decomposition. 

By this observation we see also 
\begin{prop}\label{byprod22}
There is no expression parameter $K$ under which 
all one parameter subgroup are not $2\pi$-periodic but $4\pi$-periodic. 
\end{prop}  

\subsection{Several remarks on $*$-exponential functions}

By noting that $\det(e^{t\alpha}){=}1$ for every 
$\alpha{\in}{\mathfrak{sp}}(m,{\mathbb C})$, 
\eqref{fundamenta1} is rewritten as 
\begin{equation}\label{fundamenta2} 
\begin{aligned}
{:}e_*^{\frac{t}{i\h}
\langle{\pmb u}(\alpha J), {\pmb u}\rangle_*}{:}_{_K}
{=}
\frac{2^m}{\sqrt{\det(e^{t\alpha}(I{-}{\kappa})
{+}e^{-t\alpha}(I{+}\kappa))}}
e^{\frac{1}{i\h}\langle{\pmb u}
\frac{1}{e^{t\alpha}(I{-}{\kappa}){+}e^{-t\alpha}(I{+}\kappa)}
(e^{t\alpha}{-}e^{-t\alpha})J,{\pmb u}\rangle}
\end{aligned}
\end{equation}
In spite of the sign ambiguity of $\sqrt{\,\,}$, 
the exponential law 
\begin{equation}\label{explaw11}
\begin{aligned}
{:}e_*^{(s{+}t)\frac{1}{i\h}
\langle{\pmb u}(\alpha J), {\pmb u}\rangle_*}{:}_{_K}=&
{:}e_*^{s\frac{1}{i\h}
\langle{\pmb u}(\alpha J), {\pmb u}\rangle_*}{:}_{_K}
{*_{_K}}{:}e_*^{t\frac{1}{i\h}
\langle{\pmb u}(\alpha J), {\pmb u}\rangle_*}{:}_{_K}\\
{:}e_*^{s(a+\frac{1}{i\h}
\langle{\pmb u}(\alpha J), {\pmb u}\rangle_*)}{:}_{_K}=&
{:}e^{as}e_*^{s\frac{1}{i\h}
\langle{\pmb u}(\alpha J), {\pmb u}\rangle_*}{:}_{_K}
\end{aligned}
\end{equation}
holds under computations such as $\sqrt{a}\sqrt{b}{=}\sqrt{ab}$. 
This is because that the exponential law and associativity 
holds on the group $Sp(m,{\mathbb C})$. Note however that 
$\sqrt{1}{=}\pm 1$.

\medskip
By this observation we have the following:
\begin{prop}\label{oneparam}
For every fixed $\alpha$, $\kappa$, a suitable choice of 
angle $\theta$ gives various real one parameter subgroups 
${:}e_*^{se^{i\theta}\frac{1}{i\h}
\langle{\pmb u}(\alpha J), {\pmb u}\rangle_*}{:}_{_K}$, 
$s{\in}{\mathbb R}$. Moreover, we can find  many complex semi-groups 
on various sectors. 
\end{prop}
\medskip
By the concrete formula \eqref{fundamenta2},
we have also the following:
\begin{prop}\label{hbaranalytic}
Replacing $t$ by $t\h,$ 
${:}e_*^{t
\langle{\pmb u}(\alpha J), {\pmb u}\rangle_*}{:}_{_K}$ 
is real analytic in $\h$ in an open connected domain 
containing $\h=0$.  
\end{prop}

As \eqref{fundamenta2} has double branched singular points, we have to use two sheets 
by setting {\it slits} in the complex plane  
to treat ${:}e_*^{tH_*}{:}_{_K}$ univalent way. 
Although there is no general rule to set the slits, 
it is natural to set {\bf the slits periodically, since 
the singular points are distributed periodically}.  
We adopt this rule throughout this series.

\medskip
Note that $J\in {\mathfrak{sp}}(m,{\mathbb C})$ and also  
$J\in {Sp}(m,{\mathbb C})=
\{g\in G\!L(2m,{\mathbb C}); gJ\,{}^t\!g=J\}$.
For every $g\in Sp(m,{\mathbb C})$, 
$\tilde J=gJg^{-1}$ is both an element of Lie algebra and 
 a group element satisfies $\tilde J^2{=}{-}I$ and 
$e^{t\tilde J}{=}\cos tI{+}(\sin t)\tilde J$. 
Recall the formula \eqref{fundamenta1}, which is rewritten as  
\begin{equation*}
\begin{aligned}
{:}e_*^{\frac{t}{i\h}
\langle{\pmb u}(\alpha J), 
{\pmb u}\rangle_*}{:}_{_{({-}\!J\kappa)}}
{=}
\frac{2^m}{\sqrt{\det(I{-}{\kappa}{+}e^{-2t\alpha}(I{+}\kappa))}}
e^{\frac{1}{i\h}\langle{\pmb u}
\frac{1}{I{-}{\kappa}{+}e^{-2t\alpha}(I{+}\kappa)}
(I{-}e^{-2t\alpha})J,{\pmb u}\rangle}, \quad \kappa{=}J\!K.
\end{aligned}
\end{equation*} 
Setting $\alpha=\tilde J$ and noting 
$\alpha J=gJg^{-1}J={-}g\,{}^t\!g$, we see first  
$$
\begin{aligned}
I{-}{\kappa}&{+}e^{-2t\alpha}(I{+}\kappa)
=I{-}{\kappa}{+}(\cos 2t I{-}(\sin2t)\tilde J)(I{+}\kappa)\\
&=2(\cos t I{-}(\sin t)\tilde J)
(\cos t I{-}(\sin t){\tilde J}\kappa)\\
&=2g(\cos t I{-}(\sin t)J)
(\cos t I{-}(\sin t){J}\tilde\kappa)g^{-1},\quad 
(\tilde\kappa=g^{-1}\kappa g).
\end{aligned}
$$
We have also that 
$$
(I-e^{-2t\alpha})J{=}J{-}(\cos 2t I{-}(\sin2t)\tilde J)J
{=}-2g\sin t(\cos t I-(\sin t)J){}^t\!g.
$$

Since $\det(\cos t I{-}(\sin t)J)=1$, it follows 
$$
\det(I{-}{\kappa}{+}e^{-2t\alpha}(I{+}\kappa))
=2^{2m}\det(\cos t I{-}(\sin t)J\tilde{\kappa}).
$$
Recalling that 
$K{=}{-}J\kappa, \,\,\kappa{=}g{\tilde\kappa}g^{-1}$, and plugging these, we have 
\begin{equation}\label{tildeKK}
{:}e_*^{-\frac{t}{i\h}
\langle{\pmb u}g, 
{\pmb u}g\rangle_*}{:}_{_{({-}\!J\kappa)}}
{=}
{:}e_*^{\frac{t}{i\h}
u\langle{\pmb u}(\alpha J), 
{\pmb u}\rangle_*}{:}_{_{({-}\!J\kappa)}}\\
{=}
\frac{1}{\sqrt{\det(\cos t I{-}(\sin t)J\tilde\kappa)}}
e^{\frac{1}{i\h}\langle{\pmb u}g
\frac{-\sin t}{\cos t I{-}(\sin t){J}\tilde\kappa},{\pmb u}g\rangle}
\end{equation}
where 
$\cos t I{-}(\sin t){J}\tilde\kappa$ is a symmetric matrix. 

\medskip
Now, one may assume in generic ordered expressions, 
$-J\tilde\kappa$ has disjoint $2m$ simple eigenvalues. 
Considering the diagonalization of $J\tilde\kappa$ 
in \eqref{tildeKK},  we easily see that 
\begin{lem}\label{singularpts00}
In a generic $($open dense$)$ ordered expression, 
the singular points distributed $\pi$-periodically along $2m$ lines parallel 
to the real axis, 
%
%
and the singular points are all simple double 
branched singular points.
Moreover, ${:}e_*^{-\frac{t}{i\h}
\langle{\pmb u}g, 
{\pmb u}g\rangle_*}{:}_{_{({-}\!J\kappa)}}$ 
is rapidly decreasing along lines parallel to 
the pure imaginary axis of the growth order $e^{-|t|^m}$, where $2m=n$. 
\end{lem}

\noindent
{\bf Generic assumption}\,\,Throughout this series, 
we suppose above properties for generic ordered expressions except otherwise
stated.

In addition to generic assumption, we may suppose the following:
\begin{prop}\label{1parameter}
In generic ordered expression $K$, one may assume that 
 ${:}e_*^{\frac{t}{i\h}
\langle{\pmb u}g, 
{\pmb u}g\rangle_*}{:}_{_{({-}\!J\kappa)}}$
has no singular point on the real line. Hence,   
the exponential law proved by the uniqueness in the left evolution
equation gives that 
 ${:}e_*^{\frac{t}{i\h}
\langle{\pmb u}g, 
{\pmb u}g\rangle_*}{:}_{_{({-}\!J\kappa)}},$ $t\in{\mathbb R}$ 
forms a one parameter subgroup of period $\pi$, or $2\pi$ depending on
the expression parameter $K$.   
\end{prop}

One of the remarkable feature of this concrete formula 
(\ref{tildeKK}) is 
that it shows several extraordinary properties of 
$*$-exponential functions. For instance, we will see 
in the next section the following 
(cf. \eqref{defpolar00},  Lemma\,\ref{singularpts00}): 
\begin{prop}\label{remakable}
If $\alpha=gJg^{-1}$ for some $g\in Sp(m,{\mathbb C})$, 
then the $*$-exponential function 
of quadratic form 
$\frac{1}{i\h}
\langle{\pmb u}(\alpha J), {\pmb u}\rangle_*$
in a {\bf generic} $($open dense$)$ 
ordered expression $\kappa$  is $2\pi$-periodic along 
real line $($in precise, $\pi$-periodic or alternating $\pi$-periodic$)$, 
and rapidly decreasing in both sides along the imaginary axis $i{\mathbb R}$ 
in the growth order $e^{-|t|^m}$. Hence such a $*$-exponential
function must have singular points by Liouville's theorem. 
\end{prop}

\medskip
By \eqref{fundamenta2}, we see also  that 
$
{:}e_*^{s\frac{1}{i\h}
\langle{\pmb u}(\alpha J), {\pmb u}\rangle_*}{:}_{\kappa}
$
has in general discrete branched
singularities on the complex space 
$s{\in}{\mathbb C}$ with some 
periodicity depending on the parameter $\kappa{=}JK$. 
To obtain the value without sign ambiguity, we have to fix the path from 
$0$. To stress this, we use sometimes the notation 
\begin{equation}\label{uninotation}
{:}e_*^{[0\sim s]\frac{1}{i\h}
\langle{\pmb u}(\alpha J), {\pmb u}\rangle_*}{:}_{\kappa}
\end{equation}
where $[0\sim s]$ indicates a path joining $0$ to $s$ avoiding singular points.

\bigskip
Replacing $-J\kappa$ by $K$ in \eqref{tildeKK},  
we have $J\tilde\kappa= -{}^tgKg$, and replacing $t$ by $-t$ we have   
 the formula \eqref{eq:KK2} again:
\begin{equation}\label{tildeKKK}
\begin{aligned}
{:}e_*^{\frac{t}{i\h}
\langle{\pmb u}g, 
{\pmb u}g\rangle_*}{:}_{_{K}}
&{=}
\frac{1}{\sqrt{\det(\cos t I{-}(\sin t){}^t\!gKg)}}
e^{\frac{1}{i\h}\langle{\pmb u}g 
\frac{\sin t}{\cos tI-\sin t \,{}^t\!g Kg},{\pmb u}g\rangle}
\end{aligned}
\end{equation}

\bigskip  
By requiring $1$ at $t=0$ and by using $\det g=1$, 
we have by setting $t=\pm\pi$, and $t=\pm\frac{\pi}{2}$, that 
\begin{equation}\label{defpolar00}
{:}e_*^{\pi\frac{\pm 1}{i\h}
\langle{\pmb u}g, {\pmb u}g\rangle_*}
{:}_{_{K}}=\sqrt{(-1)^{2m}}{=}\sqrt{1}, \quad 
{:}e_*^{\pi\frac{\pm 1}{2i\h}
\langle{\pmb u}g, {\pmb u}g\rangle_*}{:}_{_{K}}
=
\frac{1}{\sqrt{\det{K}}}
e^{{-}\frac{1}{i\h}\langle{\pmb u}\frac{1}{K},{\pmb u}\rangle}.
\end{equation}
Note that the r.h.s.\! of the first equality looks independent 
of $g$ and the expression parameters, and 
that the r.h.s.\! of the second equality looks independent of $g$.
Since $Sp(m,{\mathbb C})$ is 
connected, it looks the sign of $\sqrt{1}$ and $\sqrt{\det K}$ 
can be fixed. 
However, 
the sign of $\sqrt{1}$ depends both on the expression $K$ and on the path 
from $0$ to $\pi$ by which we choose the sign of 
${:}e_*^{\pi\frac{1}{i\h}
\langle{\pmb u}{\alpha}J,{\pmb u}\rangle_*}{:}_{\kappa}$
under the condition  
$e_*^{0\langle{\pmb u}{\alpha}J,{\pmb u}\rangle_*}{:}_{\kappa}=1$, 
where the path should be so chosen that there is no singular point on the path. 

In the case $K{=}0$ (the Weyl ordered expression), the r.h.s.\! of the 
second identity diverges and the first identity gives  
$$
{:}e_*^{[0\sim t]\frac{1}{i\h}
\langle{\pmb u}g, {\pmb u}g\rangle_*}{:}_{0}=
\sqrt{(\cos t)^{2m}}
e^{\frac{1}{i\h}\langle{\pmb u}g
\frac{\sin t}{\cos t}I,{\pmb u}g\rangle}
=(\cos t)^me^{\frac{1}{i\h}\langle{\pmb u}g
\frac{\sin t}{\cos t}I,{\pmb u}g\rangle}
$$ 
by requesting $1$ at $t=0$. Hence 
${:}e_*^{[0\sim\pi]\frac{1}{i\h}
\langle{\pmb u}g,{\pmb u}g\rangle_*}{:}_{0}=(-1)^m$.

In general the $\pm$-sign depends on the path from $0$ to
$\pi$ or $\pi/2$. It depends on which sheet the end point of the path
is sitting.  
By this observation, we see that 
\begin{equation}\label{signsheets}
{:}e_*^{[0\sim\pi]\frac{\pm 1}{i\h}
\langle{\pmb u}g, {\pmb u}g\rangle_*}{:}_{_K}=(-1)^m, \,\,\,(\text{resp.}  -(-1)^m)
\end{equation}
if  
${:}e_*^{\pi\frac{\pm 1}{i\h}
\langle{\pmb u}g, {\pmb u}g\rangle_*}{:}_{_K}$ is sitting in the same
(resp. opposite)
sheet as in ${:}e_*^{0\frac{\pm 1}{i\h}
\langle{\pmb u}g, {\pmb u}g\rangle_*}{:}_{_K}$.

\bigskip 
On the other hand, for a fixed $K$, the r.h.s.\! of the 
second equality \eqref{defpolar00} is independent of $g$. Since $Sp(m,{\mathbb C})$ 
is connected, it looks that  one can fix  the sign of $\sqrt{\det K}$ 
in the r.h.s. of the second equality. Here, we meet the strange phenomenon 
that we have already met in \cite{OMMY4}. 
We call $e_*^{\pi\frac{1}{2i\h}\langle{\pmb u}g, {\pmb u}g\rangle_*}$ the 
(total) {\bf polar element} and denote this by ${\e}_{00}$. The polar
element will be discussed in the next section more closely.   

\subsubsection{The case $m=1$}

In this section, we treat the case of two variables $u,v$ 
(i.e. the case $m{=}1$). Note first that 
$\{\langle{\pmb u}g, {\pmb u}g\rangle_*; g{\in}S\!L(2,{\mathbb C})\}$
is spanned by quadratic forms given by  
$$
\begin{bmatrix}
\cosh r& \sinh r\\
\sinh r&\cosh r
\end{bmatrix},\quad
\begin{bmatrix}
\cos r& i\sin r\\
i\sin r&\cos r
\end{bmatrix},\quad
\begin{bmatrix}
e^{is}& 0\\
0&e^{-is}
\end{bmatrix},\quad
\begin{bmatrix}
e^{s}& 0\\
0&e^{-s}
\end{bmatrix},\quad r, s\in{\mathbb R}.
$$
In particular, we treat $*$-exponential 
functions $e_*^{t\frac{1}{i\h}(u_*^2{+}v_*^2)}$, 
$e_*^{t\frac{1}{i\h}2u{\ctt}v}$ more closely. 

In \eqref{fundamenta1}, we set the expression parameter  
$K= -J\!\kappa{=} 
\left[\begin{smallmatrix}
a&c\\
c&b
\end{smallmatrix}\right]$, and we set the amplitude part of \eqref{fundamenta1}  
$\frac{1}{\sqrt{\Delta_{K}(t)}}$ where 
\begin{equation}\label{Delta} 
\Delta_{K}(t){=} 
\det((\cos t)I{+}(\sin t)K){=}
\cos^2t{-}(a{+}b)\sin t\cos t{+}(ab{-}c^2)\sin^2t 
\end{equation} 
Note that ${a{+}b}$ and ${ab{-}c^2}$ 
can be arbitrary complex numbers. 

$\Delta_{K}(t)$ and the phase part of \eqref{fundamenta1}
are both $\pi$-periodic, but the sign of 
$\sqrt{\Delta_{K}(t)}$ depends on the expression parameter $K$ and the
path from $0$ to $t$ in the complex plane. 
The sign ambiguity is removed by putting the 
initial condition $e_*^{0\frac{1}{i\h}H_*}{=}1$ at 
$t{=}0$ only in the case that $a=-b$ and
$c^2{+}a^2=1$, i.e. $\Delta_{K}(t)=1$, or the case that 
$(a{-}b)^2{+}4c^2{=}0$, i.e. 
$\Delta_{K}(t){=}\frac{1}{4}(2\cos{t}{+}(a{+}b)\sin{t})^2$. 

\medskip
Moreover, singular points depend on expression parameters (cf.\cite {OMMY4}).
The case $c{=}0$ where $a, b$ are arbitrary in $\mathbb C$ 
gives an overview how the singular points are moving:
$$
{:}e_*^{t\frac{1}{i\h}(u_*^2{+}v_*^2)}{:}_{_K}{=}
\frac{1}{\sqrt{(\cos t{-}a\sin t)(\cos t{-}b\sin t)}}
\exp\frac{1}{i\h}\Big(\frac{\sin t}
{\cos t{-}b\sin t}u^2{+}\frac{\sin t}{\cos t{-}a\sin t}v^2\Big).
$$ 
By these observations, we see that the singular points appear 
$\pi$-periodically in general on two lines parallel to the 
real axis and the ${*}$-exponential functions have  
$e^{-|t|}$-growth with the exponential decay on the 
line parallel to the pure imaginary axis when these 
do not hit singular points.

The observation here gives in addition the following: 
\begin{lem}\label{reflem}
Choosing the expression parameter $K$, we can make both 
${:}e_*^{[0{\to}\pi]\frac{1}{i\h}
(u_*^2{+}v_*^2)}{:}_{_{K}}=1,
\,\,\text{and}\,\,-1$. 
Moreover, multiplying $e^{t}$, we have an extremal point, called {\it vacuum},  
$$
\lim_{t\to\infty}e^t
{:}e_*^{it\frac{1}{i\h}
(u_*^2{+}v_*^2)}{:}_{_{K}}=
\frac{2}{\sqrt{(1{-}a)(1{-}b)}}
\exp\frac{1}{i\h}
\Big(\frac{1}{1{-}b}u^2{+}\frac{1}{1{-}a}v^2\Big).
$$
depending on the expression parameters. 
\end{lem}

We fix the expression parameter $K$ as follows: 
$$
K_{re}=
\begin{bmatrix}
\rho&ic'\\
ic'&\rho\\
\end{bmatrix},\,\,\,{or}\,\,\, 
K_{im}=
\begin{bmatrix}
i\rho&c\\
c&i\rho\\
\end{bmatrix},\,\,\rho,\,c,\,c'\,{\in}{\mathbb R}.
$$ 
The formula \eqref{Delta} is rewritten in this case as 
$$
\begin{aligned}
\Delta_{K_{re}}(t)&{=} 
\det((\cos t)I{+}(\sin t)K_{re}){=}
\cos^2t{+}2\rho\sin t\cos t{+}(\rho^2{+}{c'}^2)\sin^2t,\\ 
\Delta_{K_{im}}(t)&{=} 
\det((\cos t)I{+}(\sin t)K_{im}){=}
\cos^2t{+}2i\rho\sin t\cos t{-}({\rho}^2{+}c^2)\sin^2t. 
\end{aligned}
$$ 
The first one is obviously positive definite if 
$c'\not=0$ (i.e. Siegel ordered expression in the case $m=1$) 
and hence 
$\sqrt{\Delta_{K_{re}}(t)}$ does not change sign when 
$t$ moves $0$ to $\pi$ along real line.

\medskip
On the other hand,  
$$
\sqrt{\Delta_{K_{im}}(t)}
{=}
\frac{1}{2}e^{-it}
\sqrt{(1{+}\rho)^2{+}c^2}\sqrt
{(e^{2it}{+}\alpha)(e^{2it}{+}\bar\alpha)}, \quad 
\alpha=\frac{1{-}(\rho{+}ic)}{1{+}(\rho{+}ic)}.
$$
One may assume generically that $|\alpha|\not=1$.
Hence, $\sqrt{\Delta_{K_{im}}(t)}$ changes sign when 
$t$ moves from $0$ to $\pi$. Thus, we have 
\begin{lem}\label{111}
${:}e_*^{t\frac{1}{i\h}(u^2{+}v^2)}{:}_{_{K_{re}}}$ is 
$\pi$-periodic, and the two lines of singular 
points are sitting in both upper and lower half plane. 
The real line is between these. 

On the other hand, 
${:}e_*^{t\frac{1}{i\h}(u^2{+}v^2)}{:}_{_{K_{im}}}$ is 
alternating $\pi$-periodic, and the two lines of singular 
points are sitting in upper or lower half plane depending 
on the sign of $\rho$.  
\end{lem}
The expression parameter $K_{im}$ is the case $m=1$ of 
the special expression parameter $K_s$ used in \cite{OMMY4}. 

\bigskip
Next, we take our attention to the quadratic 
form $2u{\ctt}v$, but we take a general expression parameter  
$K{=}
\footnotesize
{\begin{bmatrix}
\delta&c\\
c&\delta'
\end{bmatrix}}$.
A little complicated calculation via intertwiner 
$I_{_{K_0}}^{^{K}}$ from the normal ordered expression 
gives by setting 
$\Delta{=}e^t{+}e^{-t}{-}c(e^t{-}e^{-t})$ that  
\begin{equation}\label{genericparam00}
{:}e_*^{t\frac{1}{i\h}2u{\ctt}v}{:}_{_{K}}{=}
\frac{2}{\sqrt{\Delta^2{-}(e^t{-}e^{-t})^2\delta\delta'}}
\,\,e^{\frac{1}{i\h}
\frac{e^t-e^{-t}}{\Delta^2{-}(e^t{-}e^{-t})^2\delta\delta'}
\big((e^t-e^{-t})(\delta' u^2{+}\delta v^2){+}2\Delta uv\big)}.
\end{equation}
The $q$-scalar and the polar element are obtained by setting
$t=\pm{\pi i}$ and $t=\pm\frac{\pi i}{2}$ respectively.　

\medskip
For the simplest case in \eqref{genericparam00}, that  
the case $c=\delta=\delta'=0$ is the Weyl ordered expression. 
This is not a generic ordered expression having singular 
points on the imaginary axis, and this is $\pi i$-alternating 
periodic. 
 
On the contrary, the unit ordered expression is given by 
$K=I$, i.e. $\delta=1, \delta'=1$, $c=0$. 
By \eqref{genericparam00}, we have 
$$
{:}e_*^{t\frac{1}{i\h}2u{\ctt}v}{:}_{_{I}}=
\frac{2}{\sqrt{4}}e^{\frac{1}{4i\h}
(e^{2t}{+}e^{-2t}{+}2)(u^2{+}v^2){+}2(e^{2t}{-}e^{-2t})uv}.
$$
This is $\pi i$-periodic and there is no singular point.

For the case $\delta=\delta'=0$ but $c\not=0$ 
which involves the normal ordered expression, we see that 
\begin{equation}\label{normalantinormal}
{:}e_*^{t\frac{1}{i\h}2u{\ctt}v}{:}_{_{K}}{=}
\frac{2}{\sqrt{\Delta^2}}
\,\,e^{\frac{1}{i\h}
\frac{e^t-e^{-t}}{\Delta^2}\big(2\Delta uv\big)}
=
\frac{2}{\Delta}
\,\,e^{\frac{1}{i\h}
\frac{e^t-e^{-t}}{\Delta}2uv}.
\end{equation} 
This is the case where the singular points are 
not branching ones and they are sitting 
$\pi i$ periodically on a single line parallel to 
the imaginary axis whose real part are given by 
$\log\big|\frac{c+1}{c-1}\big|$. 
We see also that ${:}e_*^{\frac{t}{i\h}2u{\ctt}v}{:}_{_K}$ is 
alternating $\pi i$-periodic along the imaginary axis. 

\medskip

Suppose in \eqref{genericparam00} that $K{=}K_{re}$;  
\begin{equation}\label{genkkk}
\delta{=}\delta'{=}\rho,\,\,c=ic' \quad 
\rho,\,\,c'\in{\mathbb R},\,\,
 c'{\not=}0.
\end{equation}
By setting $\beta{=}\rho+ic'$, we have that    
\begin{equation}\label{valueatpi77}
\frac{1}{2}\sqrt{\Delta^2{-}(e^t{-}e^{-t})^2\delta\delta'}{=}
e^{-t}\sqrt{(1{-}\beta)(1{+}\bar\beta)}
\sqrt{e^{2t}{-}\frac{1{+}\beta}{1{-}\beta}}
\sqrt{e^{2t}{-}\frac{1{-}\bar\beta}{1{+}\bar\beta}}.
\end{equation}
Obviously 
$|\frac{1{+}\beta}{1{-}\beta}|
 |\frac{1{-}\bar\beta}{1{+}\bar\beta}|=1$, but 
one may assume in generic ordered expression that 
$|\frac{1{+}\beta}{1{-}\beta}|{\not=}1$. 
Hence, $\sqrt{e^{2t}{-}\frac{1{+}\beta}{1{-}\beta}}
\sqrt{e^{2t}{-}\frac{1{-}\bar\beta}{1{+}\bar\beta}}$ 
changes the sign when $t$ moves $0$ to $\pi i$. Thus we see 
$\frac{1}{2}\sqrt{\Delta^2{-}(e^t{-}e^{-t})^2\delta\delta'}$ 
does not change the sign on the interval $[0,\pi i]$. Hence 
${:}e_*^{t\frac{1}{i\h}2u{\ctt}v}{:}_{_{K_{re}}}$ is 
$\pi i$-periodic.
Remark now this is the case in 
${:}e_*^{\frac{it}{i\h}\langle{\pmb u}g,
{\pmb u}g\rangle}{:}_{_K}$ where  
$$
g=
\frac{1}{\sqrt{2}}
\begin{bmatrix}
1&i\\
i&1
\end{bmatrix}, \quad 
K_{re}=
\begin{bmatrix}
\rho&ic'\\
ic'&\rho
\end{bmatrix}.
$$ 
The concrete expression of polar element is   
\begin{equation}\label{polar400}
{:}{\e}_{00}{:}_{_{K_{re}}}=
{:}e_*^{\pi i\frac{1}{i\h}u{\ctt}v}{:}_{_{K_{re}}}{=}
\frac{1}{\sqrt{(\rho^2{+}{c'}^2)}}
e^{-\frac{1}{i\h}\frac{1}{\rho^2{+}{c'}^2}
\rho(u^2{+}v^2){-}2c'iuv}.
\end{equation}

\bigskip

Note  that the quadratic form $u_*^2{+}v_*^2$ is 
a representative of general quadratic forms 
$au_*^2{+}bv_*^2{+}2cu{\ctt}v$ with the discriminant 
$c^2{-}ab={-}1$ via $S\!L(2;{\mathbb C})$-linear change of 
generators. 

Since $S\!L(2;{\mathbb C})=Sp(1;{\mathbb C})$, such a linear 
change is covered by a change of expression parameters by 
the formula \eqref{eq:KK2}. 
Thus, even if an expression parameter $K$ is fixed 
generically, these patterns for the quadratic form 
$u_*^2{+}v_*^2$ must appear for $au_*^2{+}bv_*^2{+}2cu{\ctt}v$ 
via changing coefficients. We shall show that this 
appears slightly different, more delicate shape. 

In general, we set  
$$
e_*^{itH_*}{=}
e_*^{\frac{it}{i\h}(au_*^2{+}bv_*^2{+}2cu{\ctt}v)}, 
\quad c^2{-}ab=1.
$$ 

We have then three patterns as follows: 

\noindent
$(Q(1))$: $e_*^{itH_*}$ is alternating $\pi$-periodic 
and the 2 lines of singular points are in the upper 

half-plane.

\noindent
$(Q(2))$: $e_*^{itH_*}$ is alternating $\pi$-periodic 
and the 2 lines of singular points are in the lower 

half-plane.

\noindent
$(Q(3))$: $e_*^{itH_*}$ is $\pi$-periodic and the real 
line are between  2 lines of singular points. 

\bigskip
$(Q(k))$ are open subsets of $\{(a,b,c); c^2{-}ab{=}1\}$ 
such that $(Q(1))\cup(Q(2))\cup(Q(3))$ is dense. 
Since the time reversing sends the line of singularities 
to the opposite side, we see that $(Q(k))$ has the property 
$$
(Q(1))^{-1}=(Q(2)),\quad (Q(3))^{-1}=(Q(3)).
$$
This means that if $e_*^{itH_*}\in (Q(1))$, then 
$e_*^{it(-H_*)}\in (Q(2))$. 

\medskip
\noindent
{\bf Remark}\,\,Alternating $\pi$-periodicity appears when 
no sheet changing occurs. Thus, 
$$
{:}e_*^{\pi\frac{1}{i\h}(u^2{+}v^2)}{:}_{_{K}}=-1
$$
always on the positive sheet, as far as requesting
$e_*^{0\frac{1}{i\h}(u^2{+}v^2)}=1$. 
On the other hand , 
$$
{:}e_*^{\pi\frac{1}{i\h}(u^2{+}v^2)}{:}_{_{K}}=1,
$$
when the sheet changing occurs on a path from $0$ to $\pi$. 
It is very easy to make a mistake.

\medskip
Recall first the anomalous phenomena mentioned in 
\cite{OMMY4} that a polar element is obtained 
not only by  $e_*^{\frac{\pi i}{i\h}u{\ctt}v}$ but also by   
$e_*^{\frac{\pi i}{2i\h}(au_*^2{+}bv_*^2{+}c2u{\ctt}v)}$, 
$c^2{-}ab=1$. This shows that a polar element is sitting 
on various one parameter subgroups. This is just like 
the longitude lines starting at the north pole meet again 
 at the South Pole. We show in the next section this is a generic  
phenomena of 
$e_*^{\frac{\pi i}{2i\h}(au_*^2{+}bv_*^2{+}2cu{\ctt}v)}$, 
$c^2{-}ab=1$. Thus, a polar element has infinitely many 
{\it square roots} sitting on the equator. 

Beyond the south pole the longitude lines come back again to 
the north pole, where we give the initial value $1$ to every 
one parameter subgroup parameterized by longitude. 
However, it is a little surprising that 
the periodicity of these periodic movement 
depends on expression parameters.


\subsubsection{Product structure}\label{prodstr}

The product formula \eqref{eq:prodfmla} shows that the space  
${\mathbb C}e^{Q(u,v)}$ of  exponential functions
of polynomial of degree 2 forms a very special subclass in the space
${\mathcal E}_{2+}({\mathbb C}^{2m})$.  It is useful to memorize the next
theorem:

\begin{thm}\label{meromuseful}
In a generic ordered expression $K$,
the $*_{_K}$-product 
$$
{\pi}_{_K}: {\mathbb C}e^{Q(u,v)}\times {\mathbb C}e^{Q(u,v)}\to 
{\mathbb C}e^{Q(u,v)}
$$
is a mapping given in the form 
${\pi}_{_K}(ae^{Q}, be^{R})= ab\sqrt{f(Q,R,K)}\,\,e^{g(Q,R,K)}$
where $f$ and $g$ are meromorphic functions of $Q, R, K$.  Hence the 
continuity 
$$
\lim_{(k,\ell)}\pi_{_K}(a_ke^{Q_k}, b_{\ell}e^{R_{\ell}})=
\pi_{_K}(\lim_ka_ke^{Q_k}, \lim_{\ell}b_{\ell}e^{R_{\ell}})
$$
holds whenever $\lim_ka_ke^{Q_k}$, $\lim_{\ell}b_{\ell}e^{R_{\ell}}$
are defined in the space ${\mathbb C}e^{Q(u,v)}$, and 
$$
\pi_{_K}(a_ke^{Q_k}, b_{\ell}e^{R_{\ell}}),\quad   
\pi_{_K}(\lim_ka_ke^{Q_k}, \lim_{\ell}b_{\ell}e^{R_{\ell}})
$$ 
are defined.
\end{thm}

\medskip
As for products, we know already that associativity 
holds always with sign ambiguity. However 
the following theorem is useful as a corollary of   
Proposition\,\ref{hbaranalytic} and the formal associativity theorem 
(cf. Theorem 1.3),
\begin{thm}\label{2ndassociative}
For quadratic forms $K_*$, $L_*$, $M_*$, associativity 
$$
(e_{*}^{[0\sim r]K_*}{*}e_*^{[0\sim s]L_*})
{*}e_*^{[0\sim t]M_*}{=}
 e_{*}^{[0\sim r]K_*}{*}(e_*^{[0\sim s]L_*}
{*}e_*^{[0\sim t]M_*})
$$
holds without sign ambiguity whenever both sides are defined, 
where paths in both left/right hand sides 
with same symbol should be same path 
$($synchronized path selecting$)$.    
\end{thm}

\medskip
We next consider the product $e_{*}^{sH_*}{*}e_*^{tK_*}$ for two 
quadratic forms $H_*, K_*$ such that  $[H_*,K_*]=0$. First of all, 
we show the following 
\begin{prop}\label{prodcomm}
If $e_{*}^{sH_*}{*}e_*^{tK_*}$ 
are defined on $(s,t){\in}[0,a]^2$, then 
$e_{*}^{sH_*}{*}e_*^{tK_*}{=}e_{*}^{tK_*}{*}e_*^{sH_*}$.  
\end{prop}

\noindent
{\bf Proof}\,\,\,Since $K_*{*}e_{*}^{sH_*}$ and $e_{*}^{sH_*}{*}K_*$
satisfies the equation $\frac{d}{ds}f_s=H_*{*}f_s$ with $f_0=K_*$, 
we have $K_*{*}e_{*}^{sH_*}=e_{*}^{sH_*}{*}K_*$. Hence, we have 
$$
\frac{d}{dt}e_{*}^{sH_*}{*}e_*^{tK_*}{=}K_*{*}e_{*}^{sH_*}{*}e_*^{tK_*},
\quad 
\frac{d}{dt}e_{*}^{tK_*}{*}e_*^{tH_*}{=}K_*{*}e_{*}^{tK_*}{*}e_*^{sH_*}
$$
with the same initial condition $e_{*}^{sH_*}$. The uniqueness gives
the proof. \hfill $\Box$

\medskip
If $(s,t){\in}{\mathbb C}^2$, then we have in general 
$e_{*}^{sH_*}{*}e_*^{tK_*}=\pm e_*^{tK_*}{*}e_{*}^{sH_*}$ 
with the sign ambiguity by the product formula \eqref{prodkappa}. 
This means in particular and the phase parts of 
both sides coincides (the sign ambiguity appears only in the amplitude
parts).
In general, 
$e_{*}^{sH_*}{*}e_*^{tK_*}$ has a singular set $S$ of complex 
codimension 1. We see that the origin $(0,0)$ is not contained in $S$. 
Since $S$ is a branched singularity, we have to prepare two 
sheets ${\mathbb C}_+^2$, ${\mathbb C}_-^2$ and ``slit'' $\Sigma$ 
of real codimension 1 to connect these two sheets. $\Sigma$ is set so that 
${\mathbb C}^2{\setminus}\Sigma$ is locally simply connected 
and there is no singular point. 

\bigskip
Now, restrict the parameter $(s,t)\in {\mathbb R}^2$
in $e_{*}^{sH_*}{*}e_*^{tK_*}$ and suppose 
$e_{*}^{sH_*}{*}e_*^{tK_*}$ has a singular point 
in $(s,t){\in}(0,a){\times}(0,b)$.
One may assume that ${\mathbb R}^2$ is transversal to $S$ 
in generic ordered expression. Hence if 
$S\cap{\mathbb R}^2\not=\emptyset$, then this is a 
discrete set and  $\Sigma\cap{\mathbb R}^2$ is a collection of 
(real one dimensional) curves starting at a singular point 
ending another singular point or $\infty$. Hence one may assume that 
the boundary $\partial([0,a]{\times}[0,b])$ cuts the slit just once  
for all. 

\begin{prop}\label{singularpt}
Under the assumption as above, we have  
$e_{*}^{sH_*}{*}e_*^{tK_*}={-}e_{*}^{sK_*}{*}e_*^{tH_*}$ 
\end{prop}

\noindent
{\bf Proof}\,\, 
As we have two sheets, there are two ``origin'', 
$(0,0){\in}{\mathbb C}_+^2$ and $(0,0){\in}{\mathbb C}_-^2$. 
Since $e_{*}^{0H_*}{*}e_*^{0K_*}$ is $1$ in the positive sheet 
${\mathbb C}_+^2$, the origin in the negative sheet must 
be treated as $-1$.  Now, consider $e_{*}^{aH_*}{*}e_*^{bK_*}$ 
and  $e_*^{bK_*}{*}e_{*}^{aH_*}$. The first one is defined by 
the solution of the evolution equation 
$$
\frac{d}{dt}f_t=H_*{*}f_t,\quad f_0=e_*^{bK_*}. 
$$ 
We indicate this by the notation 
$e_{*}^{[0\to a]H_*}{*}e_*^{bK_*}$. This is the clockwise tracing
from the origin. On the contrary,  
$e_*^{[0\to b]K_*}{*}e_{*}^{aH_*}$ means the anti-clockwise 
tracing from the origin. Now suppose there is a singular point 
$(s_0,t_0)$, then  one of the paths $e_{*}^{[0\to a]H_*}{*}e_*^{bK_*}$
and $e_*^{[0\to b]K_*}{*}e_{*}^{aH_*}$ is crossing the slit 
hence they are sitting mutually in the opposite sheet. 
By this way, the sign changes around a singular point.\hfill $\Box$

\section{Rule of setting slits and polar elements}
\label{GenPat} 

If it is an absolute scalar, then  $(\sqrt{1})^2{=}1$ is trivial. 
Recall first 
\begin{prop}\label{polar111} 
If $e^{2\pi\alpha}{=}I$ such as $\alpha{=}J$ 
$($e.g. $\alpha=gJg^{-1}$, $\forall g{\in}Sp(m,{\mathbb C})$$)$, then 
${:}e_*^{\pi\frac{1}{i\h} 
\langle{\pmb u}{\alpha}J,{\pmb u}\rangle_*}{:}_{\kappa}
=\sqrt{1}$ independent of $K$. 
\end{prop} 
Note that l.h.s. is not a {\it classic} element, for this identity does not 
hold for $\h=0$.

Hence, the strict exponential law might be failed, that is,    
${:}e_*^{2\pi\frac{1}{i\h}
\langle{\pmb u}{\alpha}J,{\pmb u}\rangle_*}{:}_{\kappa}
=1$ or 
$$
{:}e_*^{\pi\frac{1}{i\h} 
\langle{\pmb u}{\alpha}J,{\pmb u}\rangle_*}{:}_{\kappa}
{*_{\kappa}}{:}e_*^{\pi\frac{1}{i\h} 
\langle{\pmb u}{\alpha}J,{\pmb u}\rangle_*}{:}_{\kappa}
=1
$$ 
may not hold automatically. 
If ${:}e_*^{t\frac{1}{i\h}
\langle{\pmb u}{\alpha}J,{\pmb u}\rangle_*}{:}_{\kappa}$   
has a singular point on the interval $[0,2\pi]$, then it 
may occur 
$(e_*^{\pi\frac{1}{i\h}
\langle{\pmb u}{\alpha}J,{\pmb u}\rangle_*})^2{\not=}
e_*^{2\pi\frac{1}{i\h}
\langle{\pmb u}{\alpha}J,{\pmb u}\rangle_*}$,
although the equality holds modulo $\pm$ sign. 

To avoid such a strange nature, we give a general rule to set {\it slits}. 
Because of the double branching singular points, we have to use two sheets 
by setting {\it slits} in the complex plane  
to treat these $*$-exponential functions ${:}e_*^{tH_*}{:}_{_K}$ univalent way. 

\begin{center}
\fbox{\parbox[c]{.7\linewidth}
{$\clubsuit$ \,\,As it is discussed already, it is natural to set the
slits periodically, \\
${}$\hfill{since the singular points are distributed periodically.}\hfill${}$}}  
\end{center}

\bigskip

By virtue of this rule, we have 
\begin{prop}\label{period}
If $e^{2\pi\alpha}{=}I$ $($e.g. $\alpha=gJg^{-1}$, $\forall g{\in}Sp(m,{\mathbb C})$$)$, then  
${:}\big(e_*^{[0{\to}\pi]\frac{1}{i\h}
\langle{\pmb u}{\alpha}J,{\pmb u}\rangle_*})^2{:}_{\kappa}{=}1$ for every 
$\kappa$-ordered expression such that 
${:}e_*^{t\frac{1}{i\h}
\langle{\pmb u}{\alpha}J,{\pmb u}\rangle_*}{:}_{\kappa}$ has no 
singular point on the interval $[0,\pi]$. Moreover, we have  
$${:}e_*^{2\pi\frac{1}{i\h}
\langle{\pmb u}{\alpha}J,{\pmb u}\rangle_*}{:}_{\kappa}{=}
{:}(e_*^{\pi\frac{1}{i\h}
\langle{\pmb u}{\alpha}J,{\pmb u}\rangle_*})^2{:}_{\kappa}.
$$ 
\end{prop}

\noindent
{\bf Proof}. Note first that this is by no means trivial. It is
crucial that the assumption and the $\pi$-periodicity of singular points shows that 
${:}e_*^{t\frac{1}{i\h}
\langle{\pmb u}{\alpha}J,{\pmb u}\rangle_*}{:}_{\kappa}$ has   
no singular point on the interval $[\pi,2\pi]$, but 
 if there is no rule to set the slit, 
it may happen that path $[0{\to}2\pi]$ cross the slit only once. 

By the rule of setting slits $\clubsuit$, we see that   
the slits are set $\pi$-periodically. Thus, the 
line segment $[0,2\pi]$ must cross the slits even (possibly $0$)
times. It follows ${:}e_*^{2\pi\frac{1}{i\h}
\langle{\pmb u}{\alpha}J,{\pmb u}\rangle_*}{:}_{\kappa}{=}1$, 
since this is sitting in the positive sheet. 

\medskip
To confirm 
${:}(e_*^{\pi\frac{1}{i\h}
\langle{\pmb u}{\alpha}J,{\pmb u}\rangle_*})^2{:}_{\kappa}
={:}e_*^{2\pi\frac{1}{i\h}
\langle{\pmb u}{\alpha}J,{\pmb u}\rangle_*}{:}_{\kappa},$
we have to recall how the $*$-product 
$e_*^{\pi\frac{1}{i\h}
\langle{\pmb u}{\alpha}J,{\pmb u}\rangle_*}{*}g$ is defined. 
We use the definition which is given by 
the evolution equation \eqref{starexp}. 

Since 
${:}e_*^{\pi\frac{1}{i\h} 
\langle{\pmb u}{\alpha}J,{\pmb u}\rangle_*}{:}_{\kappa}
={\pm 1}$, 
one can define
$$
{:}e_*^{t\frac{1}{i\h} 
\langle{\pmb u}{\alpha}J,{\pmb u}\rangle_*}{:}_{\kappa}
{*}_{\kappa}{:}1{:}_{\kappa}, \quad 
{\text{or}}\quad 
{:}e_*^{t\frac{1}{i\h} 
\langle{\pmb u}{\alpha}J,{\pmb u}\rangle_*}{:}_{\kappa}
{*}_{\kappa}{:}({-}1){:}_{\kappa}
$$ 
by the solution of the evolution equation \eqref{starexp} with 
the initial condition ${\pm 1}$. 
By Proposition \ref{exppoly}, the solution is 
${:}e_*^{t\frac{1}{i\h} 
\langle{\pmb u}{\alpha}J,{\pmb u}\rangle_*}{:}_{\kappa}$ or 
$-{:}e_*^{t\frac{1}{i\h} 
\langle{\pmb u}{\alpha}J,{\pmb u}\rangle_*}{:}_{\kappa}$ 
respectively. This gives the result. \hfill $\square$

\subsection{General polar element as $q$-scalars}

It is interesting that polar element ${\e}_{00}$ behaves just like a
scalar, but it behaves various ways. Sometimes, it behaves as if it
were $-1$, 
and sometimes it looks as if $i$ depending
on $K$. We call such elements $q$-scalars. 
But, to treat this as a univalent element, 
we have to distinguish more strictly. 
 
The strange double-valued nature of the polar element ${\e}_{00}$ is caused by that 
$e_*^{\pi\frac{1}{2i\h}\langle{\pmb u}g, {\pmb u}g\rangle_*}$ is
moving discontinuously in both positive and negative sheets when $g$ moves in 
$Sp(m,{\mathbb C})$.

\bigskip
 In this section, we analyze this phenomenon more clearly. 
In particular, we investigate the generic patterns of 
periodicity and singularities of $*$-exponential functions 
of quadratic forms under the assumption $\clubsuit$. In particular, 
we are interested the behaviour of polar element.
In what follows, we use several notions for the path as follows:

\medskip
\noindent
$[0{\to}a]$ : the path starting from the origin $0$ ending at $a$ along the line segment, but 
the $*$-exponential is evaluated at $t{=}a$ by the continuous chase 
from $0$ to $a$ along the path $[0{\to}a]$. 

\medskip
\noindent
$[0{\sim}a]$ : a path starting from the origin $0$ ending at $a$
avoiding singular points, but evaluated at $a$.

\medskip
\noindent
$[0{\approx}a]$ : a path starting from the origin $0$ ending at $a$
avoiding singular points and slits so that the end point is sitting in
the same sheet as the origin.

\bigskip
For a fixed $g$, 
${\e}_{00}={:}e_*^{\pi\frac{1}{2i\h}\langle{\pmb u}g, {\pmb u}g\rangle_*}{:}_{_K}$ 
is always viewed as a double-valued single parallel section. 
If $K$ is fixed, ${\e}_{00}$ looks independent of $g$ with $\pm$ ambiguity. 
To distinguish the sign, we use the notation   
\begin{equation}\label{strictnotion}
{:}{\e}_{00}[g]{:}_{_K}=
{:}e_*^{[0{\to}\pi]
\frac{1}{2i\h}\langle{\pmb u}g, {\pmb u}g\rangle_*}{:}_{_K}=
\frac{1}{\sqrt{\det(\cos([0{\to}1]\frac{\pi}{2})I{-}(\sin([0{\to}1]\frac{\pi}{2}){}^t\!gKg)}}
e^{-\frac{1}{i\h}\langle{\pmb u}\frac{1}{K}, {\pmb u}\rangle}
\end{equation}
to fix the sign of ${\e}_{00}$, where $[0{\to}a]$ is the path along
the straight line segment. 
Note that ${:}{\e}_{00}[g]{:}_{_K}$  
may not be defined at some $g$, when a singular point appears in 
the interval $(0,\pi/2]$. Although ${\e}_{00}=\pm{\e}_{00}[g]$ and 
${\e}_{00}$ is independent of $g$, ${\e}_{00}[g]$ may not be
continuous w.r.t.\! $g$. The sign changes discontinuously at some $g$. 
For a generic $K$, there is $g{\in}Sp(m,{\mathbb C})$ such that 
${}^t\!gKg$ is a real diagonal matrix. Hence 
${:}e_*^{t\frac{1}{2i\h}\langle{\pmb u}g, {\pmb u}g\rangle_*}{:}_{_K}$
has a singular point.  

\medskip
Note that for every $g{\in}Sp(m,{\mathbb C})$ 
there is $k{\in}Sp(m,{\mathbb C})$ such that 
$-\langle{\pmb u}g,{\pmb u}g\rangle_*{=}
\langle{\pmb u}k,{\pmb u}k\rangle_*$. 
This is shown for instance 
\begin{equation}\label{sp-inverse}
g
\begin{bmatrix}
iI&0\\
0&{-i}I
\end{bmatrix}
\begin{bmatrix}
iI&0\\
0&{-i}I
\end{bmatrix}
\,{}^t\!g= -g\,{}^t\!g. 
\end{equation}

Recall the rule $\clubsuit$ of setting slits. As sheets are set
$\pi$-periodically we see the next result: 
\begin{lem}
In generic $K$-expression, 
${:}e_*^{[0\to\pi]\frac{1}{i\h}\langle{\pmb u}g,{\pmb u}g\rangle_*}{:}_{_K}$ 
and ${:}e_*^{[0\to\pi]\frac{-1}{i\h}\langle{\pmb u}g,{\pmb u}g\rangle_*}{:}_{_K}$
belong to the same sheet, and 
$$
{:}e_*^{[0\to\pi]\frac{1}{i\h}\langle{\pmb u}g,{\pmb u}g\rangle_*}{:}_{_K}=
{:}e_*^{[0\to\pi]\frac{-1}{i\h}\langle{\pmb u}g,{\pmb u}g\rangle_*}{:}_{_K}
= 1\quad {\text{or}}\quad -1.
$$
However, this may not belong to the same (positive) sheet as 
${:}e_*^{0\frac{1}{i\h}\langle{\pmb u}g,{\pmb u}g\rangle_*}{:}_{_K}$.
\end{lem}

\noindent
{\bf Proof}\,\,If the path $[0{\to}\pi]$ crosses the slit $\ell$-times,
then the end point ${:}e_*^{[0{\to}\pi]\frac{1}{i\h}\langle{\pmb u}g,{\pmb u}g\rangle_*}{:}_{_K}$
is sitting on the $(-1)^{\ell}$-sheet. Since sheets are set
$\pi$-periodically, the path $[0{\to}\pi]$ for 
${:}e_*^{[0\to\pi]\frac{-1}{i\h}\langle{\pmb u}g,{\pmb u}g\rangle_*}{:}_{_K}$ 
also crosses the slit $\ell$-times. \hfill $\Box$ 

\medskip
On the other hand, the second equality of \eqref{defpolar00} does not
necessarily imply that 
$$
{:}e_*^{[0{\to}\pi]\frac{1}{2i\h}
\langle{\pmb u}g, {\pmb u}g\rangle_*}{:}_{_{K}}=
{:}e_*^{[0{\to}\pi]\frac{-1}{2i\h}
\langle{\pmb u}g, {\pmb u}g\rangle_*}{:}_{_{K}}.
$$
The sheet change may occur in the continuous tracing of 
$\sqrt{\det(\cos t I{-}(\sin t){}^t\!gKg)}$ from $-\frac{\pi}{2}$ to 
$\frac{\pi}{2}$ if the path from $-\frac{\pi}{2}$ to 
$\frac{\pi}{2}$ crosses the slit odd-times. 
\eqref{defpolar00} shows  
\begin{lem}\label{Rm b}
${:}{\e}_{00}[g]{:}_{_{K}}={:}{\e}_{00}[g]^{-1}{:}_{_K}$ if and only if 
$$
\sqrt{\det(\cos([0{\to}\pi])I{-}(\sin([0{\to}\pi]){}^t\!gKg)}=1.
$$
\end{lem}

If there is no singular point on 
${:}e_*^{t\frac{1}{2i\h}\langle{\pmb u}g, {\pmb u}g\rangle_*}{:}_{_{K}}$, 
$t\in{\mathbb R}$, then this forms a one parameter group, and thus the equality above 
is equivalent with 
${:}e_*^{[0{\to}\pi]\frac{1}{i\h}
\langle{\pmb u}g, {\pmb u}g\rangle_*}{:}_{_{K}}=1$ 
by the exponential law. 

Lemma\,\ref{reflem} in the this section  shows that for a certain $K$
there are $g$, $g'$ such that  $\sqrt{(-1)^{2m}}=1$ and $-1$ respectively. 
Thus, even if $K$ is fixed, the sign may depend on $g$ and the path from $0$ to $\pi$. 
Since $Sp(m,{\mathbb C})$ is connected, the sign changes 
discontinuously when the path from $0$ to $\pi$ hits a singular
point. The sign changes by the changing sheet caused when the path
crossing the slit drawn from the set of the singular points.

\bigskip
In the argument above, paths were restricted in 
line segment to fix the ambiguous sign.    
In fact, we can relax this condition. 
The next lemma shows that the sign-changing is caused only when  
the path moves across the set $S$ of singular points.  
Take an open connected subset $U$ of $Sp(m,{\mathbb C})$ 
which may be $U\not=-U$.  Suppose we can fix path 
${:}e_*^{[0\sim\pi]\frac{1}{i\h}
\langle{\pmb u}g,{\pmb u}g\rangle_*}{:}_{_{K}}$ 
from $t{=}0$ to $t{=}\pi$ avoiding singular points but depending continuously in $g\in U$.  
By setting $t=\pi$, and $t=\frac{\pi}{2}$, we have 
the following :

\begin{lem}\label{madamada}
Under the assumption for $U$ mentioned above, the $*$-exponential function 
$$
{:}e_*^{[0{\sim}t]\frac{1}{i\h}
\langle{\pmb u}g, {\pmb u}g\rangle_*}
{:}_{_{K}}
$$ 
is defined uniquely without sign ambiguity by the continuous tracing  
from the identity, and we see  
\begin{equation}\label{defpolar}
{:}e_*^{[0{\sim}\pi]\frac{1}{i\h}
\langle{\pmb u}g, {\pmb u}g\rangle_*}
{:}_{_{K}}=\sqrt{(-1)^{2m}},
\end{equation}
where 
$\sqrt{(-1)^{2m}}=(-1)^m$, when the end point of path is sitting in
the same (positive) sheet as $0$, and $-(-1)^m$, when the end point of path is
sitting in the opposite (negative) sheet.

\medskip
On the other hand for the polar element, we have 
\begin{equation}\label{defpolar22}
{:}e_*^{[0\sim\pi]\frac{1}{2i\h}
\langle{\pmb u}g, {\pmb u}g\rangle_*}{:}_{_{K}}
=
\frac{1}{\sqrt{\det{K}}}
e^{{-}\frac{1}{i\h}\langle{\pmb u}K^{{-}1},{\pmb u}\rangle}.
\end{equation}
The sign of $\sqrt{\det{K}}$ is determined by the sheet on which the 
end point of the path $[0{\sim}\pi]$ is sitting. 
\end{lem}

Note that $(-1)^m$ in Lemma\,\ref{madamada} is $-1$ if $m=$odd, and
$1$ if $m=$even. Thus, the mathematical context depends on
$(-1)^m=\pm1$ in the next Proposition. 
\begin{prop}\label{strange2}
Suppose there is $g\in Sp(m,{\mathbb C})$ such that 
${:}e_*^{[0\to\pi]\frac{1}{i\h}
\langle{\pmb u}g,{\pmb u}g\rangle_*}{:}_{_{K}}=-1$. 
Then, there must exist 
$h\in Sp(m,{\mathbb C})$ such that 
${:}e_*^{[0\to\pi]\frac{1}{i\h}
\langle{\pmb u}h, {\pmb u}h\rangle_*}{:}_{_{K}}=1$, and 
$\hat{h}\in Sp(m,{\mathbb C})$ such that the path  
${:}e_*^{[0\to\pi]\frac{1}{2i\h}
\langle{\pmb u}\hat{h}, {\pmb u}\hat{h}\rangle_*}{:}_{_{K}}$
must hit a singular points.
\end{prop}
 
\noindent
{\bf Proof}\, Suppose ${:}e_*^{[0\to\pi]\frac{1}{i\h}
\langle{\pmb u}g,{\pmb u}g\rangle_*}{:}_{_{K}}=-1$  
for every $g{\in}Sp(m,{\mathbb C})$ and suppose there is no 
singular point on the path $[0{\to}\pi]$. 

As $Sp(m,{\mathbb C})$ is connected, \eqref{sp-inverse} and  
the second equality of \eqref{defpolar00} give that for the mid-point 
\begin{equation}\label{contrdict}
{:}e_*^{[0{\to}\pi]\frac{1}{2i\h}
\langle{\pmb u}g,{\pmb u}g\rangle_*}{:}_{_{K}}
{=} 
{:}e_*^{[0{\to}\pi]\frac{1}{2i\h}
\langle{\pmb u}k,{\pmb u}k\rangle_*}{:}_{_{K}} 
{=}{:}e_*^{[0{\to}\pi]\frac{-1}{2i\h}
\langle{\pmb u}g,{\pmb u}g\rangle_*}{:}_{_{K}}. 
\end{equation}
The exponential law gives 
$$
{:}e_*^{[0{\to}\pi]\frac{-1}{2i\h}
\langle{\pmb u}g,{\pmb u}g\rangle_*}{:}_{_{K}}=
{:}\big(e_*^{[0{\to}\pi]\frac{1}{2i\h}
\langle{\pmb u}g,{\pmb u}g\rangle_*}\big)^{-1}{:}_{_{K}}
$$
and therefor multiplying 
$e_*^{[0{\to}\pi]\frac{1}{2i\h}
\langle{\pmb u}g,{\pmb u}g\rangle_*}$ to both sides of 
\eqref{contrdict}, we have the contradiction 
$$
-1=e_*^{[0{\to}\pi]\frac{1}{i\h}
\langle{\pmb u}g,{\pmb u}g\rangle_*}=
e_*^{[0{\to}\pi]\frac{1}{2i\h}
\langle{\pmb u}g,{\pmb u}g\rangle_*}
{*}\big(e_*^{[0{\to}\pi]\frac{1}{2i\h}
\langle{\pmb u}g,{\pmb u}g\rangle_*}\big)^{-1}=1.
$$ 
As a result $Sp(m,{\mathbb C})$ is 
divided into three parts $D_+, D_-, D_{sing}$ such that 
$$
{:}e_*^{[0{\to}\pi]\frac{1}{i\h}
\langle{\pmb u}g, {\pmb u}g\rangle_*}{:}_{_{K}}
=
\left\{
\begin{matrix}
-1& g{\in} D_+\\
{:}e_*^{t\frac{1}{i\h}
\langle{\pmb u}g, {\pmb u}g\rangle_*}{:}_{_{K}}
{\text{has a singular point on}}\,\,(0,\pi)& 
g{\in} D_{sing}\\
 1& g{\in} D_-
\end{matrix}
\right.
$$
and $D_{+}\subsetneqq Sp(m,{\mathbb C})$. In particular this 
yields $D_{sing}\not=\emptyset$. 

Now, we show that $D_-\not=\emptyset$. 
Since the points of $D_{sing}$ are branched singular points,   
the value of 
${:}e_*^{t\frac{1}{i\h}
\langle{\pmb u}g, {\pmb u}g\rangle_*}{:}_{_{K}}$ 
 changes sign around branched singular point. Since we assumed as a 
generic assumption that the singular points distributed
$\pi$-periodically along $2m$ lines parallel to the real line, there is 
at most one singular point on $(0,\pi)$. 
Thus, 
${:}e_*^{[0\to 1]\frac{\pi}{i\h}
\langle{\pmb u}g, {\pmb u}g\rangle_*}{:}_{_{K}}$ must change 
sign at $g{\in}D_{sing}$. Hence we see $D_-{\not=}\emptyset$. 
\hfill $\Box$

\medskip
We note that $Sp(k,{\mathbb C})$ is 
naturally included in $Sp(m,{\mathbb C})$ for $m{>}k$.  
Apparently, the result mentioned in \cite{OMMY4}
is a special case for $m=1$,  
$g=
\frac{1}{\sqrt{2}}\begin{bmatrix}
1&i\\
i&1
\end{bmatrix}
$
and $K{=}K_0$ (normal ordered expression). 

\medskip
Proposition\,\ref{strange2} gives in particular that 
if $D_+{\not=}\emptyset$, then $D_-{\not=}\emptyset$ and 
$D_{sing}{\not=}\emptyset$.

\bigskip
Consider now whether it is possible $D_{+}=\emptyset$ in
Lemma\,\ref{madamada}.   
First we note the following:
\begin{lem}\label{madamadada}
If $D_{sing}{\not=}\emptyset$, then $D_{\pm}{\not=}\emptyset$.  
\end{lem}

\noindent
{\bf Proof}\,\,For $(t,g){\in}{\mathbb C}{\times}Sp(m,{\mathbb C})$,
the set $S$ of singular points of 
${:}e_*^{t\frac{1}{i\h}\langle{\pmb u}g,{\pmb u}g\rangle_*}{:}_{_{K}}$
is a closed subset of complex codimension $1$. The slit $\Sigma$ 
is set so that $({\mathbb C}{\times}Sp(m,{\mathbb C})){\setminus}{\sigma}$  
is locally simply connected. Hence, if 
${:}e_*^{[0\to\pi]\frac{1}{i\h}\langle{\pmb u}g,{\pmb u}g\rangle_*}{:}_{_{K}}$
hits a singular point for some $g$, then there are 
$h, h'\in Sp(m,{\mathbb C})$ in a neighborhood of $g$ such that 
${:}e_*^{[0\to\pi]\frac{1}{i\h}\langle{\pmb u}h,{\pmb u}h\rangle_*}{:}_{_{K}}$
hits $\Sigma$, but 
${:}e_*^{[0\to\pi]\frac{1}{i\h}\langle{\pmb u}h',{\pmb u}h'\rangle_*}{:}_{_{K}}$
does not. Hence, these two must have different sign. \hfill $\Box$. 

\medskip

Now note that the comment following \eqref{strictnotion} shows that 
$D_{sing}{\not=}\emptyset$. Thus, we have 
\begin{thm}\label{doesnotexist}
Suppose $K$ is a generic expression parameter. Then, 
$Sp(m,{\mathbb C})$ is 
divided into three non empty subsets $D_+, D_-, D_{sing}$.
\end{thm}

\medskip
\noindent
{\bf Remark 1}\,\,As singular points are distributed $\pi$-
periodically, if $g\in D_{sing}$, then 
${:}e_*^{t\frac{1}{i\h}
\langle{\pmb u}g, {\pmb u}g\rangle_*}{:}_{_{K}}$ has singular 
points not only in the interval $(0,\frac{\pi}{2}]$ but 
also in the interval $(-\pi,-\frac{\pi}{2}]$.

\bigskip
Theorem\,\,\ref{doesnotexist} shows a polar element ${\e}_{00}$ is a
member of various one parameter subgroups with different periodicity 
${\e}_{00}^2=1$, and ${\e}_{00}^2={-}1$. 

\bigskip
\noindent
{\bf Note}\,\,Sometimes, $D_-$ contains a compact subgroup of 
$Sp(m,{\mathbb C})$. Indeed, we will show in the next section that such a case exists. 
That is, in the case $m=1$ there is a class $K_{re}$ of expression parameters such that 
$$
{:}e_*^{[0{\to}\pi]\frac{1}{i\h}
\langle{\pmb u}g, {\pmb u}g\rangle_*}{:}_{_{K}}=1 \quad \text{for
every }\,\, g{\in}SU(2)\quad {\rm {cf.\,\,Proposition\,\ref{nicenice22}}}.
$$

\medskip
\noindent
{\bf Remark 2}\,\, 
A polar element is a double-valued single element. Thus 
even though 
${:}e_*^{\frac{\pi}{2i\h}
\langle{\pmb u}g, {\pmb u}g\rangle_*}{:}_{_{K}}
=
{:}e_*^{\frac{\pi}{2i\h}
\langle{\pmb u}h, {\pmb u}h\rangle_*}{:}_{_{K}}$, 
square of these may be different 
$$
{:}(e_*^{\pm\frac{\pi}{2i\h}
\langle{\pmb u}g, {\pmb u}g\rangle_*})^2{:}_{_{K}}{\not=} 
{:}(e_*^{\pm\frac{\pi}{2i\h}
\langle{\pmb u}h, {\pmb u}h\rangle_*})^2{:}_{_{K}} 
$$
if the paths from $0$ to $\pi$ have different 
numbers of crossing slits.

It is quite difficult to control the $\pm$ sign 
in the product formula. We have always to chase continuously 
from the identity. Even though 
${:}{\e}_{00}[g]{:}_{_K}=\pm {:}{\e}_{00}[h]{:}_{_K}$, 
it does not necessarily imply 
${:}{\e}_{00}[g]^2{:}_{_K}={:}{\e}_{00}[h]^2{:}_{_K}$. 
Furthermore, we do not have enough information 
in order to determine ${:}{\e}_{00}[g]{*}{\e}_{00}[h]{:}_{_K}$, 
though this is $\{\pm 1\}$ by the product formula with sign
ambiguity. 
In such a situation, we cannot use 
${\e}_{00}[g], {\e}_{00}[h]$ as elements of a system with 
binary operations.

\bigskip
By these observation, it seems to be better to treat 
every element always together with a path from the 
origin, and products are defined always by path connecting.   
However, this is sometimes too much to treat the detail, for 
the object turns out to be a groupoid. We have to seek an amenable 
object to treat which gives informations what we want to 
know. 

\medskip
\noindent{\bf Strict polar element} 

Let $[0\approx\pi]$ be a path from $0$ to $\pi$ avoiding 
singular points and slits so that 
$e_*^{\pi\frac{1}{2i\h}
\langle{\pmb u}g, {\pmb u}g\rangle_*}$ is sitting in the same 
sheet as in 
$e_*^{0\frac{1}{2i\h}\langle{\pmb u}g, {\pmb u}g\rangle_*}$.
Then $e_*^{[0\approx\pi]\frac{1}{2i\h}
\langle{\pmb u}g, {\pmb u}g\rangle_*}$ is determined 
without sign ambiguity. 

$e_*^{[0\approx\pi]\frac{1}{2i\h}
\langle{\pmb u}g, {\pmb u}g\rangle_*}$ is called the 
{\bf strict  polar element} by requesting that the path is
so chosen that  
$e_*^{\pi\frac{1}{2i\h}
\langle{\pmb u}g, {\pmb u}g\rangle_*}$ is sitting in the same sheet 
as in $e_*^{0\frac{1}{2i\h}\langle{\pmb u}g, {\pmb u}g\rangle_*}$,
and it will be denoted by $\hat{\e}_{00}$. In precise, 
\begin{equation}\label{strictpolar}
\hat{\e}_{00}=
e_*^{[0\approx\pi]\frac{1}{2i\h}
\langle{\pmb u}g, {\pmb u}g\rangle_*},\quad 
{:}\hat{\e}_{00}{:}_{_K}=
\frac{1}{\sqrt{\det(\cos([0{\approx}1]\frac{\pi}{2})I{-}(\sin([0{\approx}1]\frac{\pi}{2}){}^t\!gKg)}}
e^{-\frac{1}{i\h}\langle{\pmb u}\frac{1}{K}, {\pmb u}\rangle}
\end{equation} 
but a little care is required for the r.h.s., for the sheet is not
distinguished by the notation itself.    

\medskip
Since singular points and slits are not 
sitting $\pi/2$-periodically but only $\pi$-periodically, the square 
${\hat{\e}_{00}}^2=\hat{\e}_{00}{*}\hat{\e}_{00}$ is defined only with 
sign ambiguity (cf.\eqref{starexp}). That is, 
${\hat{\e}_{00}}^2=\pm 1$ and the sign depends 
on $g$ and $K$ discontinuously, while ${\hat{\e}_{00}}^4=1$ (cf. Proposition\,\ref{period}).

But recall here that change of generators is covered by change of expression
parameters. Hence the same phenomenon must occur in the change of
expression parameters even when $g$ is fixed.

%
%

\subsection{Sign-changing by the order of continuous tracing}\label{Adjoint}
Recall \S\,\ref{prodstr}. We  have discussed the product formula
$e_*^{sH_*}{*}e_*^{tK_*}$ for the case $[H_*,K_*]=0$ in 
Propositions\,\ref{prodcomm},\,\,\ref{singularpt}.
In this section, we consider the the case $[H_*,K_*]{\not=}0$ and  
we give the product formula corresponding to 
Propositions\,\ref{prodcomm},\,\,\ref{singularpt}.  

As it is mentioned before, the product  
$$
{:}e_*^{t\langle{\pmb u}(\frac{1}{2i\h}\alpha J),
{\pmb u}\rangle}{*}f{:}_{_K},\quad 
{:}f{*}e_*^{{-}t\langle{\pmb u}(\frac{1}{2i\h}\alpha J),
{\pmb u}\rangle}{:}_{_K}
$$
are defined by the left/right evolution equations 
$$
\frac{d}{dt}f_t=\langle{\pmb u}(\frac{1}{2i\h}\alpha J),
{\pmb u}\rangle{*}f_t, \quad 
\frac{d}{dt}f_t=f_t{*}\langle{\pmb u}(\frac{1}{2i\h}\alpha J),
{\pmb u}\rangle,
$$
with initial data $f$
Suppose $f$ is another $*$-exponential function  
${:}e_*^{t\langle{\pmb u}(\frac{1}{2i\h}\beta J),
{\pmb u}\rangle}{:}_{_K}$.

\medskip
$$  
{:}e_*^{t\langle{\pmb u}(\frac{1}{2i\h}\alpha J),{\pmb u}\rangle}
{*}(f{*}e_*^{{-}t\langle{\pmb u}(\frac{1}{2i\h}\alpha J),
{\pmb u}\rangle}){:}_{_K} 
{\quad }
{:}(e_*^{t\langle{\pmb u}(\frac{1}{2i\h}\alpha J),
{\pmb u}\rangle}
{*}f){*}e_*^{{-}t\langle{\pmb u}(\frac{1}{2i\h}\alpha J),
{\pmb u}\rangle}{:}_{_K}
\quad \text{e.t.c.}
$$
are defined holomorphically, but multi-valued in $t$ 
on an open connected domain 
$D$ containing the origin $0\in{\mathbb C}$. 

Even in such a case, we can fix 
the value by tracing along a real analytic path from $0$.
We have then the following synchronized associativity:
\begin{thm}\label{synchronized}
Whenever the same path is used to fix the value in both sides, associativity 
$$
{:}e_*^{t\langle{\pmb u}(\frac{1}{2i\h}\alpha J),{\pmb u}\rangle}
{*}(f{*}e_*^{{-}t\langle{\pmb u}(\frac{1}{2i\h}\alpha J),{\pmb u}\rangle}){:}_{_K}
={:}(e_*^{t\langle{\pmb u}(\frac{1}{2i\h}\alpha J),{\pmb u}\rangle}
{*}f){*}e_*^{{-}t\langle{\pmb u}(\frac{1}{2i\h}\alpha J),{\pmb u}\rangle}{:}_{_K}
$$
holds and differentiating this gives as in \S\,\,\ref{adjoint}
$$
{:}e_*^{t\langle{\pmb u}(\frac{1}{2i\h}\alpha J),{\pmb u}\rangle}
  {*}f{*}e_*^{{-}t\langle{\pmb u}(\frac{1}{2i\h}\alpha J),{\pmb u}\rangle}{:}_{_K}
=
{:}e^{t{\rm{ad}}(\langle{\pmb u}(\frac{1}{2i\h}\alpha J),{\pmb u}\rangle)}f{:}_{_K}
$$
\end{thm}

Using Theorem\,\ref{assocthm} and Theorem\,\ref{synchronized}, 
we see also the following:
\begin{cor}\label{quadexpquad} 
Suppose $e_*^{s\langle{\pmb u}
(\frac{1}{2i\h}\alpha J),{\pmb u}\rangle}$ and  
$e_*^{[0{\sim}t]\langle{\pmb u}
(\frac{1}{2i\h}\beta J),{\pmb u}\rangle}$ are defined,
where $[0{\sim}\pi]$ is a real analytic curve in ${\mathbb C}$ joining
$0$ to $\pi$ avoiding singular points.  

Since for every fixed $s$,  
$$
e_*^{s\langle{\pmb u}
(\frac{1}{2i\h}\alpha J),{\pmb u}\rangle} 
{*}e_*^{t\langle{\pmb u}
(\frac{1}{2i\h}\beta J),{\pmb u}\rangle}
{*}
e_*^{{-}s\langle{\pmb u}
(\frac{1}{2i\h}\alpha J),{\pmb u}\rangle}
$$ 
are defined as a multi-valued holomorphic element on an open connected 
neighbourhood of $[0{\sim}\pi]$, we see   
$$
{:}{\rm{Ad}}(e_*^{s\langle{\pmb u}
(\frac{1}{2i\h}\alpha J),{\pmb u}\rangle})
e_*^{[0{\sim}t]\langle{\pmb u}
(\frac{1}{2i\h}\beta J),{\pmb u}\rangle}{:}_{_{K}}{=}
{:}e_*^{[0{\sim}t]\langle{\pmb u}
(\frac{1}{2i\h}\tilde\beta(s)J),{\pmb u}\rangle}{:}_{_K}, \,\,
\tilde\beta(s){=}e^{s\alpha}\beta e^{-s\alpha}
$$
hold without sign ambiguity, where $[0{\sim}t]$ in the r.h.s. is the
path naturally given by the adjoint action for the path of the l.h.s.   

In particular,  
${:}e_*^{[0{\to}t]\langle{\pmb u}
(\frac{1}{2i\h}\beta J),{\pmb u}\rangle}{:}_{_{K}}$ and  
${:}e_*^{[0{\to}t]\langle{\pmb u}
(\frac{1}{2i\h}\tilde\beta(s)J),{\pmb u}\rangle}{:}_{_K}$ must have
the same periodicity. 
\end{cor} 
Here we used the same notation as in previous section  
to stress that 
$e_*^{[0{\to}t]\langle{\pmb u}
(\frac{1}{2i\h}\tilde\beta(s) J),{\pmb u}\rangle}$ 
is defined by solving the evolution equation along 
the real segment $[0,t]$:
\begin{equation}\label{adjoint020}
\frac{d}{dt}f_*(t)=
\langle{\pmb u}
(\frac{1}{2i\h}\tilde\beta(s)J),{\pmb u}\rangle{*}f_*(t),\quad 
f_*(0){=}1.
\end{equation}

\bigskip 
Consider now 
${\rm{Ad}}(e_*^{s\langle{\pmb u}
(\frac{1}{2i\h}\alpha J),{\pmb u}\rangle})
e_*^{t\langle{\pmb u}
(\frac{1}{2i\h}\beta J),{\pmb u}\rangle}$ of two variables 
$(s,t){\in}[0,\pi]{\times}[0,\pi]$.
Note that Corollary\,\ref{quadexpquad} holds even if 
there is a singular point $(s_0,t_0)$ in 
the open square $(0,\pi){\times}(0,\pi)$, but 
there happens another phenomenon of change sheets depending on the order 
of continuous tracing of values. 

\medskip
By the observation Proposition\,\ref{singularpt} in the previous
section, we see that the singular points in ${\mathbb C}^2$ 
forms a set $S$ of complex codimension 1, which is transversal 
to the real plane. One may assume that 
$S\cap{\mathbb R}^2$ is a discrete set.  
Suppose now there is a singular point $(s_0,t_0)$ in 
the open square $(0,\pi){\times}(0,\pi)$. 
Then, there must be a slit starting 
from $(s_0,t_0)$ going outside the square. 
In what follows, we see that $*$-exponential function  
$e_*^{t\langle{\pmb u}
(\frac{1}{2i\h}\tilde\beta(s_0)J),{\pmb u}\rangle}$
is discontinuous at $t=t_0$. 

\bigskip
Hence fixing $t$ as $t_0{<}t{<}\pi$ and   
tracing $e_*^{t\langle{\pmb u}
(\frac{1}{2i\h}\tilde\beta(s)J),{\pmb u}\rangle}$
by moving $s$ from $s{=}0$, the curve must hit the slit and 
changes the sheet.

As it mentioned in Proposition\,\ref{singularpt}, the sheet changing gives 
\begin{equation}\label{2sheets}
e_*^{t\langle{\pmb u}
(\frac{1}{2i\h}\tilde\beta([0,s])J),{\pmb u}\rangle}
={-}e_*^{[0{\to}t]\langle{\pmb u}
(\frac{1}{2i\h}\tilde\beta(s)J),{\pmb u}\rangle}
\end{equation}
where the l.h.s.
is the element obtained by tracing continuously from 
$e_*^{t\langle{\pmb u}
(\frac{1}{2i\h}\tilde\beta(0)J),{\pmb u}\rangle}$
${=}$
$e_*^{t\langle{\pmb u}(\frac{1}{2i\h}\beta J),{\pmb u}\rangle}$ 
to $e_*^{t\langle{\pmb u}(\frac{1}{2i\h}\beta(s)J),{\pmb u}\rangle}$
under a fixed $t$.

One may understand how the sign 
changes by noting the difference $([0,t],s)$ 
and $(t,[0,s])$ in the next picture.   

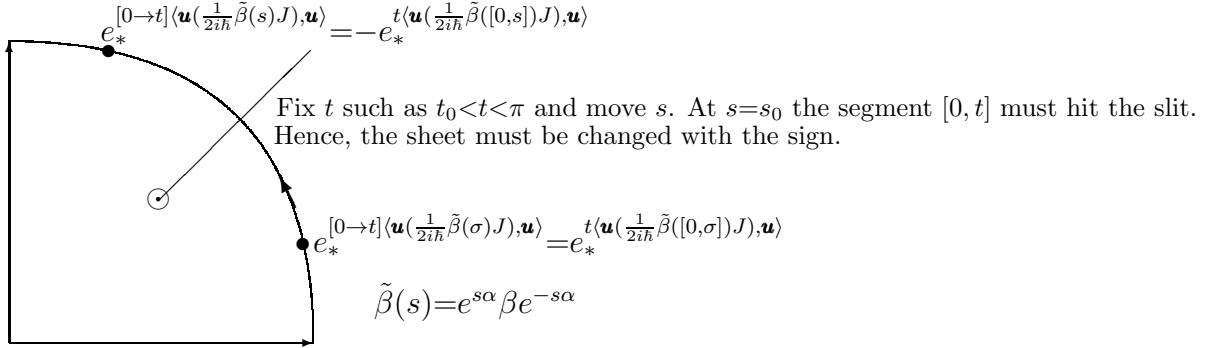
\setlength{\unitlength}{.4mm}
\begin{picture}(300,130)(0,-10)
\thinlines
\put(0,0){\vector(1,0){100}}
\put(0,0){\vector(0,1){100}}
\qbezier[400](0,100)(100,100)(100,0)
\put(45,45){$\odot$}
\put(49,47){\line(1,1){50}} 
\put(30,94){$\bullet$}
\put(30,100){$e_*^{[0{\to}t]\langle{\pmb u}
(\frac{1}{2i\h}\tilde\beta(s)J),{\pmb u}\rangle}{=}
{-}e_*^{t\langle{\pmb u}
(\frac{1}{2i\h}\tilde\beta([0,s])J),{\pmb u}\rangle}$}
\put(94,30){$\bullet$}
\put(100,30){$e_*^{[0{\to}t]\langle{\pmb u}
(\frac{1}{2i\h}\tilde\beta(\sigma)J),{\pmb u}\rangle}{=}
e_*^{t\langle{\pmb u}
(\frac{1}{2i\h}\tilde\beta([0,\sigma])J),{\pmb u}\rangle}$}
\put(85,75){\footnotesize{
Fix $t$ such as $t_0{<}t{<}\pi$ and move $s$. At $s{=}s_0$ 
the segment $[0,t]$ must hit the slit}.}
\put(87,66){\footnotesize{Hence, the sheet must be changed 
with the sign.}}
\thicklines
\put(94,45){\vector(-1,2){5}}
\put(120,10){$\tilde\beta(s){=}e^{s\alpha}\beta e^{-s\alpha}$}
\end{picture}

\bigskip
Now the formal associativity Theorem\,\ref{assocthm} gives 
the translation identity from the right 
evolution equation into 
the left evolution equation: 
\begin{equation}\label{master01}
{:}e_*^{s\langle{\pmb u}(\frac{1}{2i\h}\alpha J),{\pmb u}\rangle}
{*}
e_*^{[0{\to}t]\langle{\pmb u}(\frac{1}{2i\h}\beta J),
{\pmb u}\rangle}{:}_{_K}
={:}e_*^{[0{\to}t]\langle{\pmb u}
(\frac{1}{2i\h}\tilde\beta(s)J),{\pmb u}\rangle}{*}
 e_*^{s\langle{\pmb u}(\frac{1}{2i\h}\alpha J),{\pmb u}\rangle}{:}_{_K}.
\end{equation}

On the other hand, note that $e_*^{t\langle{\pmb u}
(\frac{1}{2i\h}\tilde\beta([0,s])J),{\pmb u}\rangle}$ 
is the solution of 
\begin{equation}\label{Adjoint111}
\frac{d}{d\eta}f_*(\eta)= 
[\langle{\pmb u}(\frac{1}{2i\h}\alpha J),
{\pmb u}\rangle,f_*(\eta)],\quad f_*(0){=}
e_*^{t\langle{\pmb u}(\frac{1}{2i\h}\beta J),{\pmb u}\rangle},
 \quad 
{\text{$t$ is fixed.}}
\end{equation} 
Since this is real analytic, 
the formal associativity theorem gives for fixed $t$ that  
\begin{equation}\label{master02}
{:}e_*^{[0{\to}s]\langle{\pmb u}
(\frac{1}{2i\h}\alpha J),{\pmb u}\rangle}
{*}e_*^{t\langle{\pmb u}
(\frac{1}{2i\h}\beta J),{\pmb u}\rangle}{:}_{_K}
={:}e_*^{t\langle{\pmb u}
(\frac{1}{2i\h}\tilde\beta([0,s])J),{\pmb u}\rangle}{*}
 e_*^{[0{\to}s]\langle{\pmb u}
(\frac{1}{2i\h}\alpha J),{\pmb u}\rangle}{:}_{_K}.
\end{equation}
If we use the tracing \eqref{2sheets}, then we have 
$$
\begin{aligned}
{:}
e_*^{[0{\to}s]\langle{\pmb u}(\frac{1}{2i\h}\alpha J),{\pmb u}\rangle}
{*}e_*^{t\langle{\pmb u}(\frac{1}{2i\h}\beta J),{\pmb u}\rangle}
{:}_{_K}
=&-{:}e_*^{[0{\to}t]\langle{\pmb u}
(\frac{1}{2i\h}\tilde\beta(s)J),{\pmb u}\rangle}{*}
 e_*^{[0{\to}s]\langle{\pmb u}(\frac{1}{2i\h}\alpha J),
{\pmb u}\rangle}{:}_{_K}\\
=&-{:}e_*^{[0{\to}t]\langle{\pmb u}
(\frac{1}{2i\h}\tilde\beta(s)J),{\pmb u}\rangle}{*}
 e_*^{s\langle{\pmb u}(\frac{1}{2i\h}\alpha J),
{\pmb u}\rangle}{:}_{_K}
\end{aligned}
$$ 
for $e_*^{[0{\to}s]\langle{\pmb u}(\frac{1}{2i\h}\alpha J),
{\pmb u}\rangle}$ on the right hand side can be replaced simply by 
$e_*^{s\langle{\pmb u}(\frac{1}{2i\h}\alpha J),
{\pmb u}\rangle}$ without 
changing meaning. 
It follows a little tricky result as follows:
\begin{thm}\label{master01}
If the square $[0,s]{\times}[0,t]$ contains no singular 
point, then the identify 
$$
{:}e_*^{[0{\to}s]\langle{\pmb u}(\frac{1}{2i\h}\alpha J),
{\pmb u}\rangle}
{*}e_*^{t\langle{\pmb u}
(\frac{1}{2i\h}\beta J),{\pmb u}\rangle}{:}_{_K}.
={:}e_*^{[0{\to}t]\langle{\pmb u}
(\frac{1}{2i\h}\tilde\beta(s)J),{\pmb u}\rangle}{*}
 e_*^{s\langle{\pmb u}(\frac{1}{2i\h}\alpha J),{\pmb u}\rangle}
{:}_{_K}
$$
holds, but if the square $[0,s]{\times}[0,t]$ 
contains a singular point $(s_0,t_0)$ in the interior, 
then  
$$
{:}e_*^{[0{\to}s]\langle{\pmb u}
(\frac{1}{2i\h}\alpha J),{\pmb u}\rangle}
{*}e_*^{t\langle{\pmb u}
(\frac{1}{2i\h}\beta J),{\pmb u}\rangle}{:}_{_K}.
=-{:}e_*^{[0{\to}t]\langle{\pmb u}
(\frac{1}{2i\h}\tilde\beta(s)J),{\pmb u}\rangle}{*}
 e_*^{s\langle{\pmb u}(\frac{1}{2i\h}\alpha J),{\pmb u}\rangle}
{:}_{_K}
$$
since the sheet is exchanged. 
\end{thm}  
\begin{cor}\label{trickymaster}
Suppose $\tilde\beta({\pi}){=}
e^{{\pi}\alpha}\beta e^{-{\pi}\alpha}{=}{-}\beta$. 
If there is no singular point in $(0,\pi){\times}(0,\pi)$, then 
$$
{:}e_*^{[0{\to}\pi]\langle{\pmb u}
(\frac{1}{2i\h}\alpha J),{\pmb u}\rangle}
{*}e_*^{\pi\langle{\pmb u}
(\frac{1}{2i\h}\beta J),{\pmb u}\rangle}{:}_{_K}.
={:}e_*^{[0{\to}\pi]\langle{\pmb u}
(\frac{-1}{2i\h}\beta J),{\pmb u}\rangle}{*}
 e_*^{\pi\langle{\pmb u}(\frac{1}{2i\h}\alpha J),
{\pmb u}\rangle}{:}_{_K}
$$
but if there is a singular point in $(0,\pi){\times}(0,\pi)$, then
$$
{:}e_*^{[0{\to}\pi]\langle{\pmb u}
(\frac{1}{2i\h}\alpha J),{\pmb u}\rangle}
{*}e_*^{\pi\langle{\pmb u}
(\frac{1}{2i\h}\beta J),{\pmb u}\rangle}{:}_{_K}.
=-{:}e_*^{[0{\to}\pi]\langle{\pmb u}
(\frac{-1}{2i\h}\beta J),{\pmb u}\rangle}{*}
 e_*^{\pi\langle{\pmb u}(\frac{1}{2i\h}\alpha J),
{\pmb u}\rangle}{:}_{_K}
$$
\end{cor}
 
\bigskip
The relation such as 
$e^{{\pi}\alpha}\beta e^{-{\pi}\alpha}{=}{-}\beta$ appears naturally
in the next section, 
but the relation in Corollary\,\ref{trickymaster}
is {\it not} a classical relation, 
for such a relation does not hold in the limit $\h\to 0$. 
%
%
\subsubsection{Formula obtained by adjoint relations}\,\,

In this subsection, we apply these results to the case $m=1$.  
First of all, we recall   
\begin{prop}\label{polar}
In a generic ordered expression $K$,  
${:}e_*^{\frac{t}{i\h}(au^2{+}bv^2{+}2cu{\ctt}v)}{:}_{_K}$ has   
no singular point on the real line and the pure imaginary line. 

Providing $c^2{-}ab=1$, polar element 
${:}e_*^{\frac{\pi i}{2i\h}(au^2{+}bv^2{+}2cu{\ctt}v)}{:}_{_K}$ 
depends only on $K$ and the path from $0$ to $\pi$.
\end{prop}

Except otherwise stated, the path is chosen as the 
segment $[0{\to}\pi]$.

\bigskip
\noindent
Note first that 
$\frac{1}{i\h}[u{\ctt}v, 
\begin{bmatrix}
u\\v
\end{bmatrix}]=
\begin{bmatrix}
1&0\\
0&{-}1
\end{bmatrix}
\begin{bmatrix}
u\\v
\end{bmatrix}$. 
It follows that 
$e^{\frac{it}{i\h}{\rm{ad}}(u{\ctt}v)}
\begin{bmatrix}
u\\v
\end{bmatrix}=
\begin{bmatrix}
e^{it}&0\\
0&e^{-it}
\end{bmatrix}
\begin{bmatrix}
u\\v
\end{bmatrix}$, and hence for any $*$-function such as  
$f_*(u,v,\h)=
\frac{1}{2\pi}\int_{{\mathbb R}^2}{\hat f}(s,t,\h)
e_*^{\frac{1}{i\h}(su{+}tv)}dsdt$ depending  
real analytically on $\h$ in some interval involving $\h=0$, 
we have 
$$
e_*^{\frac{is}{i\h}u{\ctt}v}{*}f_*(u,v,\h)
{*}e_*^{{-}\frac{is}{i\h}u{\ctt}v}{=}
f_*(e^{is}u, e^{-is}v, \h)
$$
by the formal associativity theorem. Furthermore, we have by 
the same reason that
\begin{equation}\label{adjointuv}
e_*^{\frac{is}{i\h}u{\ctt}v}{*}f_*(u,v,\h){=}
f_*(e^{is}u, e^{-is}v, \h){*}e_*^{\frac{is}{i\h}u{\ctt}v}.
\end{equation}

\medskip
$$
\frac{1}{2i\h}[u^2{-}v^2, 
\begin{bmatrix}
u\\v
\end{bmatrix}]=
\begin{bmatrix}
0&-1\\
-1&0
\end{bmatrix}
\begin{bmatrix}
u\\v
\end{bmatrix},\quad 
\frac{1}{2i\h}[u^2{+}v^2, 
\begin{bmatrix}
u\\v
\end{bmatrix}]=
\begin{bmatrix}
0&1\\
-1&0
\end{bmatrix}
\begin{bmatrix}
u\\v
\end{bmatrix}.
$$ 
It follows that 
$$
e^{\frac{it}{2i\h}{\rm{ad}}(u^2{-}v^2)}
\begin{bmatrix}
u\\v
\end{bmatrix}=
\begin{bmatrix}
\cos t&{-}i\sin t\\
{-}i\sin t&\cos t
\end{bmatrix}
\begin{bmatrix}
u\\v
\end{bmatrix},\quad 
e^{\frac{t}{2i\h}{\rm{ad}}(u^2{+}v^2)}
\begin{bmatrix}
u\\v
\end{bmatrix}=
\begin{bmatrix}
\cos t&\sin t\\
{-}\sin t&\cos t
\end{bmatrix}
\begin{bmatrix}
u\\v
\end{bmatrix}.
$$

Now, even if $f_*$ is a $*$-exponential function of 
quadratic form 
${:}e_*^{t\langle{\pmb u}(\frac{1}{2i\h}\beta J),
{\pmb u}\rangle}{:}_{_K}$, we can make several 
commutation relations  by using the product formula \eqref{prodkappa}. 
But for that purpose, we have to use {\it synchronized} path in both sides.

\bigskip
\noindent
{\bf Polar elements are splitting}

We next compute the case 
 $c{=}0$ and $\delta\delta'{\not=}\pm 1$ 
in \eqref{genericparam00}
i.e. $K=diag\{\delta,\delta'\}$. Then,  
$$
\begin{aligned}
&\sqrt{\Delta^2{-}(e^t{-}e^{-t})^2\delta\delta'}\\
&{=}
\sqrt{1{-}\delta\delta'}e^{-t}
\sqrt{e^{4t}{+}2\frac{1{+}\delta\delta'}
{1{-}\delta\delta'}e^{2t}{+}1}
{=}\sqrt{1{-}\delta\delta'}e^{-t}
\sqrt{(e^{2t}{+}\beta)(e^{2t}{+}\beta^{-1})}
\end{aligned}
$$
where 
$\beta=
\frac{1{+}\sqrt{\delta\delta'}}{1{-}\sqrt{\delta\delta'}}$.
If $|\beta|\not=1$ i.e. 
$\delta\delta'\not\in {\mathbb R}_{<0}$ 
then only one of 
$\sqrt{e^{2t}{+}\beta}$ or  
$\sqrt{e^{2t}{+}\beta^{-1}}$ changes sign when $t$ moves 
from $0$ to $\pi i$. Thus, this is the case 
where the singular points are distributed 
$\pi i$-periodically along two lines 
parallel to the imaginary axis both positive and negative 
real parts, whose real parts are  
$\pm\log\big|
\frac{\sqrt{\delta\delta'}+1}
{\sqrt{\delta\delta'}-1}\big|$, and 
${:}e_*^{\frac{t}{i\h}2u{\ctt}v}{:}_{_K}$ is 
$\pi i$-periodic along the imaginary axis. 

\begin{lem}\label{unitlike}
Suppose 
$
K{=}
\begin{bmatrix}
\delta&0\\
0&\delta'
\end{bmatrix}
$ 
such that $\delta\delta'\not=0, 1$, and $\delta\delta'$ is not a 
negative real.  
Then, ${:}e_*^{\frac{t}{i\h}2u{\ctt}v}{:}_{_K}$ 
is $\pi i$-periodic along the pure imaginary axis and  
singular points distributed 
$\pi i$-periodically along two lines 
parallel to the imaginary axis both positive and negative 
real parts.

If $\delta\delta'=1$, 
then ${:}e_*^{\frac{t}{i\h}2u{\ctt}v}{:}_{_K}$ 
is $\pi i$-periodic along the pure imaginary axis and there is 
no singular point. 
\end{lem}

However, if $|\beta|=1$ i.e. 
$\delta\delta'$ is negative real,  
then 
${:}e_*^{t\frac{1}{i\h}2u{\ctt}v}{:}_{_K}$, 
$K=diag\{\delta,\delta'\}$ has two branching singular points 
on the open interval $i(0,\pi)$. Even if this is the 
case, one may change $\delta, \delta'$ slightly  
so that $\delta\delta'$ avoids negative real and $1$.
By this procedure, we have the same periodical nature and the 
pattern of singularities as above.

\bigskip
We next change the generator by 
$(u,v)
\frac{1}{\sqrt{2}}
\begin{bmatrix}
1&-1\\
1&1
\end{bmatrix}$, 
and the expression parameters by two different ways: 
\begin{equation}\label{genchange}
\begin{aligned}
K_{re}&=
\frac{1}{2}
\begin{bmatrix}
1&1\\
-1&1
\end{bmatrix}
\begin{bmatrix}
\rho{-}ic'&0\\
\vspace{.3cm}
0&\rho{+}ic'
\end{bmatrix}
\begin{bmatrix}
1&-1\\
1&1
\end{bmatrix}=
\begin{bmatrix}
\rho&ic'\\
ic'&\rho
\end{bmatrix}\\
K_{im}&=
\frac{1}{2}
\begin{bmatrix}
1&1\\
-1&1
\end{bmatrix}
\begin{bmatrix}
i\rho{-}c&0\\
0&i\rho{+}c
\end{bmatrix}
\begin{bmatrix}
1&-1\\
1&1
\end{bmatrix}=
\begin{bmatrix}
i\rho&c\\
c&i\rho
\end{bmatrix}.
\end{aligned}
\end{equation}
Then for both cases, we see by \eqref{eq:KK2} 
that 
$$
{:}e_*^{\frac{t}{i\h}(u^2{-}v^2)}{:}_{_{K_{re}}}=
{:}e_*^{\frac{t}{i\h}2u'v'}{:}_{_{\hat{K}_0}},\quad
{:}e_*^{\frac{t}{i\h}(u^2{-}v^2)}{:}_{_{K_{im}}}=
{:}e_*^{\frac{t}{i\h}2u'v'}{:}_{_{K_0'}},
$$
where $u'=\frac{1}{\sqrt{2}}(u-v),\,v'=
\frac{1}{\sqrt{2}}(u+v)$,  
$\hat{K}_0=diag\{\rho{-}ic,\rho{+}ic\}$, 
$K_0'= diag\{i\rho{-}c,i\rho{+}c\}$.

Note now that 
$$
\begin{bmatrix}
\cos r& i\sin r\\
i\sin r&\cos r
\end{bmatrix} \subset 
\begin{bmatrix}
\rho& ic\\
ic&\rho
\end{bmatrix},\quad \rho, c\in{\mathbb R},
$$
and 
$$
\begin{bmatrix}
\cos r& i\sin r\\
i\sin r&\cos r
\end{bmatrix},\quad
\begin{bmatrix}
\cos s& -\sin s\\
sin s&\cos s
\end{bmatrix},\quad 
\begin{bmatrix}
e^{i\theta}& 0\\
0&e^{-i\theta}
\end{bmatrix}
$$ 
generate $SU(2)$.

\medskip
By these observation, we have first the following:
\begin{prop}\label{nicenice22}
If 
$K_{re}{=}
\begin{bmatrix}
\rho&ic'\\
ic'&\rho
\end{bmatrix}$ with $c', \rho \in {\mathbb R}$, satisfies 
$|\frac{1{+}\rho{+}ic'}{1{-}\rho{-}ic'}|{\not=}1$, 
then $K_{re}$ ordered expressions of those three $*$-exponential 
functions  
$$
e_*^{\frac{it}{i\h}2u{\ctt}v}, \quad 
e_*^{\frac{t}{i\h}(u^2+v^2)},\quad
e_*^{\frac{it}{i\h}(u^2-v^2)}, 
$$
have no singular point on the real axis and $\pi$-periodic, 
but each of them has singular points sitting $\pi$-periodically along 
two lines parallel to the real axis on both upper and  
lower half plane.

Hence, the polar element ${\e}_{00}$ may be written in the  
$K_{re}$-expression by 
$$
{:}{\e}_{00}{:}_{_{K_{re}}}=
{:}e_*^{\frac{\pi i}{i\h}u{\ctt}v}{:}_{_{K_{re}}}=
{:}e_*^{\frac{\pi i}{2i\h}(u^2{-}v^2)}{:}_{_{K_{re}}}=
{:}e_*^{-\frac{\pi}{2i\h}(u^2+v^2)}{:}_{_{K_{re}}}.
$$
and ${\e}_{00}^2=1$. Therefore, we have three square roots 
$$
e_1=e_*^{\frac{\pi i}{2i\h}u{\ctt}v}, \quad   
e_2=e_*^{\frac{\pi}{4i\h}(u^2{+}v^2)},\quad 
e_3=e_*^{\frac{\pi i}{4i\h}(u^2-v^2)} 
$$
such that $e_i^2={\e}_{00}$.
\end{prop}
To avoid possible confusion, we restrict the expression parameter in
the class $K_{re}$ in what follows.

For every $s$, \eqref{adjointuv} and Corollary\,\,\ref{quadexpquad}
gives in generic ordered expression $K$ that     

\begin{equation}\label{pmidenty}
\begin{aligned}
&{:}e_*^{\frac{\pi i}{4i\h}(u{\ctt}v)}
{*}e_*^{[0{\sim}s]\frac{1}{i\h}(u_*^2+v_*^2)}
{*}e_*^{{-}\frac{\pi i}{4i\h}(u{\ctt}v)}{:}_{_K}
{=}{:}e_*^{[0{\sim}s]\frac{i}{i\h}(u_*^2-v_*^2)}{:}_{_K},\\
&{:}e_*^{\frac{\pi i}{2i\h}(u{\ctt}v)}
{*}e_*^{[0{\sim}s]\frac{1}{i\h}(u_*^2+v_*^2)}
{*}e_*^{-\frac{\pi i}{2i\h}(u{\ctt}v)}{:}_{_K}
{=}{:}e_*^{[0{\sim}s]\frac{-1}{i\h}(u_*^2+v_*^2)}{:}_{_K}
\end{aligned}
\end{equation}
without sign ambiguity, where $[0{\sim}s]$ in the l.h.s. is a path from $0$ to $s$ 
in a complex plane on which there is no singular point, and  
$[0{\sim}s]$ in the r.h.s. is the path given naturally by the adjoint transformation.

We have also 
\begin{equation}\label{partI}
\begin{aligned}
&e_*^{\frac{\pi i}{4i\h}u{\ctt}v}
{*}e_*^{\frac{i[0\sim s]}{i\h}(u_*^2-v_*^2)}{*}
e_*^{{-}\frac{\pi i}{4i\h}u{\ctt}v}
{=}e_*^{-\frac{[0\sim s]}{i\h}(u_*^2+v_*^2)},\\
&e_*^{\frac{\pi i}{4i\h}u{\ctt}v}{*}
e_*^{\frac{i[0\sim s]}{i\h}(u_*^2-v_*^2)}
{=}e_*^{-\frac{[0\sim s]}{i\h}(u_*^2+v_*^2)}
{*}e_*^{\frac{\pi i}{4i\h}u{\ctt}v}
\end{aligned}
\end{equation}

Taking the synchronized use of path, we have 
\begin{equation}\label{partII}
\begin{aligned}
&e_*^{\frac{\pi}{4i\h}(u_*^2{+}v_*^2)}
{*}e_*^{\frac{i[0\sim s]}{i\h}u{\ctt}v}{=}
e_*^{-\frac{i[0\sim s]}{i\h}u{\ctt}v}{*}
e_*^{\frac{\pi}{4i\h}(u_*^2{+}v_*^2)},\\
&e_*^{\frac{\pi i}{4i\h}(u_*^2{-}v_*^2)}{*}
e_*^{\frac{i[0\sim s]}{i\h}u{\ctt}v}
{=}e_*^{-\frac{i[0\sim s]}{i\h}u{\ctt}v}
{*}e_*^{\frac{\pi i}{4i\h}(u_*^2{-}v_*^2)}.
\end{aligned}
\end{equation}

In these notations, we have also 
$$
e_*^{\frac{\pi i}{4i\h}u{\ctt}v}
{*}e_*^{\frac{[0\sim\pi]}{4i\h}(u_*^2+v_*^2)}
{*}e_*^{{-}\frac{\pi i}{4i\h}u{\ctt}v}
{=}e_*^{\frac{i[0\sim\pi]}{4i\h}(u_*^2-v_*^2)},
$$
Applying the second equality of \eqref{partII} to 
the part $e_*^{\frac{[0\sim\pi]}{4i\h}(u^2+v^2)}
{*}e_*^{{-}\frac{\pi i}{4i\h}u{\ctt}v}$, we have 
$$
e_*^{\frac{\pi i}{4i\h}u{\ctt}v}
{*}e_*^{\frac{[0\sim\pi]}{4i\h}(u^2+v^2)}
{*}e_*^{{-}\frac{\pi i}{4i\h}u{\ctt}v}{=}
e_*^{\frac{\pi i}{4i\h}u{\ctt}v}
{*}e_*^{\frac{\pi i}{4i\h}u{\ctt}v}
{*}e_*^{\frac{[0\sim\pi]}{4i\h}(u^2+v^2)}
{=}e_*^{\frac{i[0\sim\pi]}{4i\h}(u^2-v^2)}
$$
hold. This may be written simply by 
\begin{equation}\label{123}
e_*^{\frac{\pi i}{2i\h}u{\ctt}v}
{*}e_*^{\frac{\pi}{4i\h}(u^2+v^2)}
{=}e_*^{\frac{\pi i}{4i\h}(u^2-v^2)}
\end{equation}

Note also that \eqref{partII} yields a tricky result as follows:
\begin{prop}\label{rootsdouble}
The polar element $e_*^{\frac{i[0\sim \pi]}{i\h}u{\ctt}v}$ 
satisfies the equality 
$$
e_*^{\frac{\pi}{4i\h}(u_*^2{+}v_*^2)}
{*}e_*^{\frac{i[0\sim \pi]}{i\h}u{\ctt}v}{=} 
e_*^{-\frac{i[0\sim\pi]}{i\h}u{\ctt}v}{*}
e_*^{\frac{\pi}{4i\h}(u_*^2{+}v_*^2)}.
$$
Hence, such a polar element commutes with another 
square root of a polar element, if and only if 
$e_*^{-\frac{i[0\sim\pi]}{i\h}u{\ctt}v}=
e_*^{\frac{i[0\sim\pi]}{i\h}u{\ctt}v}$.
\end{prop}

Recall first \eqref{123}. This gives 
$e_1{*}e_2=e_3$ in the $K_{re}$-ordered 
expression. 

Generally, adjoint relations of quadratic forms give 
the following master relations for elements of square roots 
of the polar element.
\begin{lem}\label{master}
Let $H_*$ be a quadratic form with the discriminant $1$. 
Then,  
$e^{\pi i{\rm{ad}}(H_*)}e_j{=}e_j^{-1}$. 
This implies that $e_i{*}e_j{*}e_i^{-1}{=}e_j^{-1}$ by 
Theorem \ref{synchronized}. These relations  
hold without sign ambiguity. 
\end{lem}

\noindent
{\bf Proof}\,\, The first equality is easy to see. 
The second identity is a special case of the identity 
which is proved by using formal associativity theorem. 
\hfill $\Box$

By the master relation, we have in general
$$
e_i{*}e_j=e_j^{-1}{*}e_i
={\e}_{00}{*}e_j{*}e_i.
$$
By the identity $e_3=e_1{*}e_2$, we 
have 
$$
e_2{*}e_3=
e_2{*}e_1{*}e_2=
e_2{*}e_2^{-1}{*}e_1=e_1. 
$$  
Similarly, 
$$
e_3{*}e_1=
e_3{*}e_2{*}e_3=
e_3{*}e_3^{-1}{*}e_2=e_2.
$$

\bigskip
Note that all $e_i$ are elements of 
${\mathcal E}_{2+}({\mathbb C}^{2})$. Hence, we have 
\begin{thm}\label{surprise0}
In the $K_{re}$-ordered expression such that 
$|\frac{1{+}\rho{+}ic'}{1{-}\rho{-}ic'}|{\not=}1$,  
$\{{\e}_{00}, e_1, e_2, e_3\}$ 
generates an algebra ${\mathcal A}$ where exist  
two idempotent elements $\frac{1}{2}(1{+}{\e}_{00})$,  
$\frac{1}{2}(1{-}{\e}_{00})$ such that 
$$
1=\frac{1}{2}(1{+}{\e}_{00}){+}\frac{1}{2}(1-{\e}_{00}), \quad 
\frac{1}{2}(1{+}{\e}_{00}){*}\frac{1}{2}(1-{\e}_{00})=0.
$$

The subalgebra 
$\frac{1}{2}(1{-}{\e}_{00}){*}{\mathcal A}$ is naturally isomorphic to 
the complexification ${\mathbb C}{\otimes}{\mathbb H}$ 
of the quaternion field $\mathbb H$ such that by denoting $\hat 1=\frac{1}{2}(1{-}{\e}_{00})$  
$$
\hat{\e}_{00}=\frac{1}{2}(1{-}{\e}_{00}){*}{\e}_{00}=-\hat 1, \quad
\hat{e}_i^2=-\hat 1, \quad  
\hat{e}_i{*}\hat{e}_j={-}\hat 1{*}\hat{e}_j{*}\hat{e}_i,\quad 1\leq i,j\leq 3,
$$
where $\hat{e}_i=\frac{1}{2}(1{-}{\e}_{00}){*}{e}_i$, 
and the subalgebra 
$\frac{1}{2}(1{+}{\e}_{00}){*}{\mathcal A}$ is the group ring over
${\mathbb C}$ of the Klein's four group.
\end{thm}

\subsection{Independence of ordering principle and its failure}

In differential geometry, it is widely accepted 
that geometrical notion should have coordinate free expression. 
Obviously, algebraic structure of 
$({\mathbb C}[{\pmb u}], {*}_{\Lambda})$ 
depends only on the skew part of ${\Lambda}$.  
It seems reasonable to accept the independence  of ordering principle as 
a basic principle that the physical implication should be 
independent of ordered expressions. Theorem \ref{main01} supports 
this principle for elements in a class ${\mathcal E}_{2}({\mathbb C}^{n})$. 

By a direct calculation of intertwiner, we see that     
\begin{equation}
  \label{eq:intwin}
I_{_K}^{^{K'}}(e^{\frac{1}{i\h}\langle{\pmb a},{\pmb u}\rangle})
=e^{\frac{1}{4i\h}\langle{\pmb a}(K'{-}K),{\pmb a}\rangle}
e^{\frac{1}{i\h}\langle{\pmb a},{\pmb u}\rangle}.   
\end{equation}
Hence, $\{e^{\frac{1}{4i\h}\langle{\pmb a}K,{\pmb a}\rangle}
e^{\frac{1}{i\h}\langle{\pmb a},{\pmb u}\rangle}; 
K\in{\mathfrak S}_{\mathbb C}(2m)\}$ is a parallel section of 
$\coprod_{K\in{\mathfrak S}_{\mathbb C}(2m)}
{H{\!o}l}({\Bbb C}^{2m}).$

We denoted this collection symbolically by 
$e_*^{\frac{1}{i\h}\langle{\pmb a},{\pmb u}\rangle}$ and   
we regard each member 
\begin{equation}
  \label{eq:tempexp}
:e_*^{\frac{1}{i\h}\langle{\pmb a},{\pmb u}\rangle}:_{_K}
=e^{\frac{1}{4i\h}\langle{\pmb a}K,{\pmb a}\rangle}
e^{\frac{1}{i\h}\langle{\pmb a},{\pmb u}\rangle}
=e^{\frac{1}{4i\h}\langle{\pmb a}K,{\pmb a}\rangle
{+}\frac{1}{i\h}\langle{\pmb a},{\pmb u}\rangle}  
\end{equation}
as its $K$-expression.
Furthermore for every $K$, 
$e_*^{\frac{s}{i\h}\langle{\pmb a},{\pmb u}\rangle}$ 
is the solution of the evolution equation 
$$
\frac{d}{dt}{:}e_*^{\frac{s}{i\h}\langle{\pmb a},{\pmb u}\rangle}
{:}_{_{K}}
=\frac{1}{i\h}{:}\langle{\pmb a},{\pmb u}\rangle{:}_{_{K}}{*_{_K}}
{:}e_*^{\frac{s}{i\h}\langle{\pmb a},{\pmb u}\rangle}{:}_{_{K}}
\,\,{\text{with initial data}}\,\, {:}1{:}_{_{K}}=1.
$$
Note also that ${:}\langle{\pmb a},{\pmb u}\rangle{:}_{_{K}}
=\langle{\pmb a},{\pmb u}\rangle$. 
$e_*^{s\frac{1}{i\h}\langle{\pmb a},{\pmb u}\rangle}=
\{e^{s^2\frac{1}{4i\h}\langle{\pmb a}K\,{\pmb a}\rangle}
e^{s\frac{1}{i\h}\langle{\pmb a},{\pmb u}\rangle}; 
K\in{\mathfrak S}(2m)\}$ forms a one parameter 
group of parallel sections. 
The product formula in $K$-ordered expression gives 
the exponential law 
${:}e_*^{s\frac{1}{i\h}\langle{\pmb a},{\pmb u}\rangle}
{:}_{_{K}}{*_{_{K}}}
{:}e_*^{t\frac{1}{i\h}\langle{\pmb a},{\pmb u}\rangle}
{:}_{_{K}}=
{:}e_*^{(s{+}t)\frac{1}{i\h}\langle{\pmb a},{\pmb u}\rangle}
{:}_{_{K}}$ for every $K{\in}{\mathfrak S}(2m)$.    
Hence, this may be written by omitting the suffix $K$ as 
$e_*^{s\frac{1}{i\h}\langle{\pmb a},{\pmb u}\rangle}{*}
e_*^{t\frac{1}{i\h}\langle{\pmb a},{\pmb u}\rangle}=
e_*^{(s{+}t)\frac{1}{i\h}\langle{\pmb a},{\pmb u}\rangle}$.
The product formula may be written as
\begin{equation}\label{prodformula}
e_*^{\frac{1}{i\h}\langle{\pmb a},{\pmb u}\rangle}{*}
 e_*^{\frac{1}{i\h}\langle{\pmb b},{\pmb u}\rangle}
=e^{\frac{1}{2i\h}{\langle\pmb a}{J},{\pmb b}\rangle}
e_*^{\frac{1}{i\h}\langle({\pmb a}{+}{\pmb b}),
{\pmb u}\rangle}.
\end{equation}

\medskip
The main point is that we do not use operator theory, but 
instead various ordered expressions under the leading 
principle that a physical/mathematical object should be free from 
ordered expressions 
({\bf the independence of ordering principle}, (IOP) in short), 
just as geometrical objects are independent of local coordinate 
expressions. 

Recall this principle in geometry forced to accept 
the absolute abstract notion ``underlying topological space''
before a collection of local coordinate system. However, we saw 
in \cite{OMMY4} that the topology of a set depends on 
expression parameters. That is, 
$$
{\mathfrak P}^{(2)}_{K_0}\cong SO(m,{\mathbb C}){\times}{\mathbb Z}_2,
\quad \text{but} \quad 
{\mathfrak P}^{(2)}_{K_s}\cong Spin(m)\otimes{\mathbb C}.
$$ 

Furthermore,
if we apply this principle to our system, then it becomes a true 
nature that every linear form has two different inverses, 
for this holds for generic (open dense) expressions.
In general, parallel sections of 
$\coprod_{K{\in}{\mathfrak S}(n)}{\mathcal E}_{2+}({\mathbb C}^{n})$ 
are multi-valued with branched singular points depending 
on expression parameters. It is difficult to explain multi-valued 
parallel section in a picture of point set topology. 
Thus we have to think twice about the role of expression 
parameters in geometry and physics.

\medskip
\subsubsection{Philosophy of general dynamics}

It was widely accepted in classical physics that 
every dynamical movement must be caused 
by some Hamiltonian $H$. Another word, this is the definition 
of ``dynamical movement''. (IOP) is also widely accepted together 
with differential geometrical expressions, e.g. contact geometry, 
$G$-structures.

The philosophy was succeeded in non-relativistic quantum 
dynamics by replacing $H$ by a quantum Hamiltonian. 
This is given by the evolution equation of every quantum 
observable $f_t({\pmb u})$:
$$
\frac{d}{dt}f_t({\pmb u})=[H, f_t(\pmb u)]_*
$$
and the solution is given by $f_t=e^{t{\rm{ad}}(H)}f_0$, where 
$e^{t{\rm{ad}}(H)}$ acts on the space of quantum observables. 

In the relativity theory, ``time'' $t$ is never an absolute  
scalar, but a coordinate function of ``space-time''. Thus, 
the Hamiltonian $H$ which governs a relativistic movement  
is given by $H=H(t,e(t,\pmb u))$ involving $t$ and the quantum 
canonical conjugate $e$ of $t$. $e$ is called the ``energy'' 
variable, relating each other by $e=e(t,\pmb u)$, or 
$t=t(e,{\pmb u})$. 

The equation of the relativistic movement is written by similar 
differential equation by using ``proper time'' $\tau$ viewed as 
the individual time of observer:
$$
\frac{d}{d\tau}\phi_\tau(e,t,{\pmb u})
=[H, \phi_\tau(e,t,\pmb u)], 
\quad 
\frac{d}{d\tau}e{*}t=[H,e{*}t]=0,
$$
where $\phi_{\tau}$ is any quantum observable, and the solution is
given by $\phi_{\tau}=e^{\tau{\rm{ad}}(H)}\phi_0$, 
where $e^{\tau{\rm{ad}}(H)}$ acts on the space of quantum observables.
If one forgets about physics by neglecting the positivity of energy, 
such equations can be treated as Fourier integral operators, and the
principle (IOP) remains safe.    

\medskip
However in the field theory, quantum observables $\phi_0$ are regarded as operators 
acting on some pre-Hilbert space ${\mathbb H}$, and we are requested
to have $e_*^{\tau H}$ acting on ${\mathbb H}$  with suitable
associativity such that $\phi_{\tau}$ may be written as   
$$
\phi_{\tau}=(e_*^{\tau H}{*}\phi_0){*}e_*^{-\tau H}
           =e_*^{\tau H}{*}(\phi_0{*}e_*^{-\tau H}).
$$
On the other hand, as it is seen throughout this series of papers, 
$*$-exponential functions such as $e_*^{\tau H}$ often has branched 
singular point and the periodicity depends on the expression parameters.
We have delicate problems of failing associativity related to 
{\bf moving branched singular points}, which depends on
expression parameters. Stone's theorem shows that there is no
essential selfadjoint operator $H$ such that 
$\int_{\mathbb R}e^{tH}dt$ is finite.

At a first glance it is natural to replace  
$e_*^{\tau H}$ by ${\rm{Ad}}(e_*^{\tau H})$ and operator
representation of ${\rm{Ad}}(e_*^{\tau H})$. But this can not be the
mathematical Messiah, because such strange phenomena are already
involved in the transcendently extended algebra of ordinary calculus.
Hence phisysists are required always the mathematical consistency.   
Strictly speaking, this means that physical
phenomenon in field theory depends on how the element is expressed.   
Nature of individual element, in particular the nature of periodicity
depends on expression parameters.


\begin{thebibliography}{OM}
 
\bibitem{AAR}{G. Andrews, R. Askey, R. Roy},\,\,
\newblock{\sc Special functions},
\newblock{Encyclopedia Math, Appl. 71, Cambridge, 2000.}


\bibitem{AW}{G. S. Agawal, E. Wolf},\,\,
\newblock{\it Calculus for functions of noncommuting 
operators and general phase-space method of functions},
\newblock{Physcal Review D, vol.2, no.10, 1970, 2161-2186.}


\bibitem{BF}{F.Bayen,\,M,Flato,\,C.Fronsdal,\,A.Lichnerowicz,\,D.Sternheimer,\,}
{\it Deformation theory and quantization I, II}, Ann. Phys. 111, (1977),
61-151.  


\bibitem{GS}{I.M.Gel'fand,\,G.E.Shilov,\,\,}
{Generalized Functions, 2}, Acad. Press, 1968. 


\bibitem{gus}{V. Guillemin, S. Sternberg},\,\, 
{\sc Geometric Asymptotics}. A.M.S. Mathematical surveys, 14, 1977. 


\bibitem{Hi}{N. Hitchin},
\newblock{\it Lectures on special Lagrangian submanifolds}, 
\newblock{arXiv:math.DG/9907034vl 6Jul, 1999.}


 
\bibitem{Mm}{M.Morimoto,\,\,}{\sc An introduction to Sato's hyperfunctions},
 AMS Trans. Mono.129, 1993.  


\bibitem{om6}{H. Omori},\,\,{\it Toward geometric quantum theory}, in 
From Geometry to Quantum Mechanics. Prog. Math. 252,
Birkh{\"a}user,2007, 213-251.    


{\bibitem{ommy}H. Omori, Y. Maeda, N. Miyazaki, A. Yoshioka}, 
\newblock{\it Deformation quantization of Fr\'echet-Poisson algebras, 
--{Convergence of the Moyal product--}},  
in  Conf\'erence Mosh\'e Flato 1999, Quantizations, Deformations, 
and Symmetries, Vol II,  Math. Phys. Studies 22, 
Kluwer Academic Press, 2000, 233-246.


{\bibitem{OMMY7}H. Omori, Y. Maeda, N. Miyazaki, A. Yoshioka}, 
\newblock{\it Strange phenomena related to ordering problems in 
quantizations},  Jour. Lie Theorey, Vol 13, no.2, (2003) 491-510.



\bibitem{OMMY6}
{H.Omori, Y.Maeda, N.Miyazaki and A.Yoshioka :}
\newblock{\em Star exponential functions as two-valued elements}, 
  \newblock{in The breadth of symplectic and Poisson geometry}, 
  \newblock{Progress in Math. 232, Birkh{\"auser},2004}, 483-492. 




\bibitem{OMMY3}
{H.Omori, Y.Maeda, N.Miyazaki and A.Yoshioka :}
\newblock{\it Deformation of expressions for elements of algebras  (I)}, 
{arXiv:1104.2109} 
 

\bibitem{OMMY4}
{H.Omori, Y.Maeda, N.Miyazaki and A.Yoshioka :}
\newblock{\it Deformation of expressions for elements of algebras  (II)}, 
{arXiv:1105.1218} 
 






 

\bibitem{R}{M. Rieffel,\,\,}{\it Deformation quantization for actions of
  $\Bbb R^n$}, Memoir. A.M.S. 106, 1993.
\end{thebibliography}
\end{document}